%% file: main.tex
\documentclass[twocolumn, dvipsnames, switch]{article}

\usepackage{preprint}

\usepackage{xcolor}
\usepackage{lineno,hyperref}
\modulolinenumbers[5]
\usepackage{adjustbox}
\usepackage{amsmath}
\usepackage{booktabs}
\usepackage[skip=4pt]{caption}
\usepackage{color}
\usepackage{doi}
\usepackage{here}
\usepackage[utf8]{inputenc}
\usepackage{natbib}
\bibpunct{(}{)}{;}{a}{}{,}
\usepackage{pgf}
\usepackage{url}
\usepackage{tikz}
\usetikzlibrary{positioning,calc,shapes,arrows,chains}
\usepackage{verbatim}
\usepackage{hyperref}
\hypersetup{
colorlinks=true, linktocpage=true, pdfstartpage=3, pdfstartview=FitV,
breaklinks=true, pdfpagemode=UseNone, pageanchor=true, pdfpagemode=UseOutlines,
plainpages=false, bookmarksnumbered, bookmarksopen=true, bookmarksopenlevel=1,
hypertexnames=true, pdfhighlight=/O, urlcolor=RoyalBlue, linkcolor=RoyalBlue, citecolor=RoyalBlue,
}

\newcommand{\mono}[1]{\texttt{#1}}
\newcommand{\arcsec}{\ensuremath{^{\prime\prime}}}
\newcommand{\arcmin}{\ensuremath{^{\prime}}}

\title{The ssos Pipeline: Identification of Solar System Objects in Astronomical Images}

\usepackage{authblk}

\author[1]{Max Mahlke}
\author[1]{Enrique Solano}
\author[2]{Hervé Bouy}
\author[3]{Benoit Carry}
\author[4]{Gijs A. Verdoes Kleijn}
\author[5]{Emmanuel Bertin}

\affil[1]{CAB (INTA-CSIC), Campus ESAC (ESA),
         Camino Bajo del Castillo s/n,
         28692 Villanueva de la Ca\~nada,
         Madrid, Spain}
\affil[2]{Laboratoire d'astrophysique de Bordeaux, Univ. Bordeaux, CNRS, B18N, Allée Geoffroy Saint-Hilaire, 33615 Pessac, France}
\affil[3]{Universit{\'e} C{\^o}te d'Azur, Observatoire de la C{\^o}te d'Azur, CNRS, Laboratoire Lagrange, France}
\affil[4]{Kapteyn Astronomical Institute, University of Groningen, P.O. Box 800, 9700 AV Groningen, The Netherlands}
\affil[5]{Sorbonne Université, CNRS, UMR 7095, Institut d’Astrophysique de Paris, 98bis Boulevard Arago, 75014 Paris, France}

\begin{document}

\twocolumn[
  \begin{@twocolumnfalse}
\maketitle

\begin{abstract}
Observatories and satellites around the globe produce tremendous amounts of imaging data to study many different astrophysical phenomena. The serendipitous observations of Solar System objects are a fortunate by-product which have often been neglected due to the lack of a simple yet efficient identification algorithm.
Meanwhile, the determination of the orbit, chemical composition, and physical properties such as rotation period and 3D-shape of Solar System objects requires a large number of astrometry and multi-band photometry observations.
Such observations are hidden in current and future astrophysical archives, and a method to harvest these goldmines is needed.

This article presents an easy-to-implement, light-weight software package which detects bodies of the Solar System in astronomical images and measures their astrometry and photometry. The \mono{ssos} pipeline is versatile, allowing for application to all kinds of observatory imaging products. The sole principle requirement is that the images observe overlapping areas of the sky within a reasonable time range. Both known and unknown Solar System objects are recovered, from fast-moving near-Earth asteroids to slow objects in the distant Kuiper belt.

The high-level pipeline design and two test applications are described here, highlighting the versatility of the algorithm with both narrow-field pointed and wide-field survey observations. In the first study, 2,828 detections of 204 SSOs are recovered from publicly available images of the GTC OSIRIS Broad Band DR1 (Cortés-Contreras, in preparation). The false-positive ratio of SSO detections ranges from 0\%\,-\,23\% depending on the pipeline setup. The second test study utilizes the images of the first data release of J-PLUS, a 12-band optical survey. 4,606 SSO candidates are recovered, with a false-positive ratio of (2.0\,$\pm$\,0.2)\%. A stricter pipeline parameter setup recovers 3,696 candidates with a sample contamination below 0.68\%.
\end{abstract}
\vspace{0.35cm}

  \end{@twocolumnfalse}
]

\section{Introduction}
The identification and characterisation of Solar System objects (SSOs) are key steps in the understanding of the formation of the Solar System. Early evolutionary processes like planetary migration left their imprints in the distributions of the orbital parameters, chemical compositions, and size distributions of the bodies of the Solar System. The dynamical properties of the different populations have given rise to numerous models of planetary formation and evolution \citep{tsiganis2005origin,morbidelli2005chaotic,gomes2005origin,walsh2011low,2017SciA....3E1138R}. Multi-wavelength data is required to determine the spectral properties of the SSO populations and thereby constrain and refine these scenarios \citep{demeo2014solar}.

Besides the science case, there are more practical implications of SSO detection and identification. Near-Earth Objects (NEOs) pose a threat to  Earth, as reminded by impacts such as over the Russian city of Chelyabinsk in 2013 \citep{brown2013500}. The cited event raised awareness for the problem that sub-kilometre NEOs can have devastating local impacts. While dedicated efforts like the Catalina Sky Survey  likely completely surveyed the population of NEOs larger than 1\,km in diameter, smaller NEOs remain a constant risk \citep{1998BAAS...30.1037L,2008Natur.453.1178H}. Early detection and a detailed compositional study is required to choose among different response strategies.

To derive model constraints and assess risks, observational data is necessary. Since the discovery of the first asteroid (1) Ceres in 1801 at the Palermo Observatory, numerous efforts haven been taken to survey the SSO populations (see \citet{2015aste.book..795J} for a review). Over 790,000 objects have been detected and their orbital parameters were determined, though 9\,\% carry ephemeris uncertainties equal to or larger than 1\,deg in January 2019, underlining the need for further observations.\footnote{\url{https://minorplanetcenter.net/}} The physical and compositional characterisation of these objects is lacking far behind, due to the large increase of both required effort and possible parameter space between the identification and physical characterisation of SSOs.

Meanwhile, astronomical observatories are producing more imaging data each night than can be analysed in real-time. The Large Synoptic Survey Telescope (LSST), the next-generation wide-field telescope currently being built in Chile, will produce on average 15\,TB of raw data each observation night \citep{juric2015lsst}. Serendipitous observations of SSOs are a well-known by-product in these images and their astrometry and photometry is sometimes recovered. Automatic SSO detection pipelines have been developed for surveys where the required time and funding are available.

The Pan-STARRS survey demonstrates the full use of SSO discovery potential through pipelines. Their Moving Object Processing System (MOPS) automatically identifies SSOs in the survey images and reports them to the Minor Planet Centre (MPC) \citep{panstarrs2013}.
In October 2017, Pan-STARRS observed the first interstellar object, 1I/2017 U1, named ‘Oumuamua \citep{meech2017}.
The photometric pipeline of the Sloan Digital Sky Survey (SDSS) automatically flags moving objects based on a $\chi^2$-fit of their coordinates if the object has been observed at least three times \citep{2000AJ....120.1579Y,2001ASPC..238..269L}. The fourth data release of the Moving Object Catalogue (MOC) contains 471,569 moving objects, with an estimated completeness of 95\,\% and a contamination of $\sim$6\,\% \citep{ivezic2002asteroids}. Many of these moving objects could only later be associated to asteroids with certainty thanks to the increased number of SSO observations, highlighting the potential of legacy archive science \citep{CARRY2016340}. A citizen-science projected led by \citet{solano2014precovery} recovered 2,351 new measurements of 551 NEOs in the images of SDSS. 73\,\% of these measurements were recovered from images taken before May 2007, meaning the objects were not identified by the automated SDSS pipeline.

In space, the NASA \emph{WISE} mission \citep{wise} observed about 150,000 small bodies while surveying the entire celestial sphere in the mid-infrared, and reported their albedo \citep{masiero2011}.
The ESA \emph{Gaia} mission \citep{gaia}, repeatedly surveying the sky down to magnitude 20 in the visible, has already reported over 2,000,000 observations of 14,099 small bodies \citep{spoto2018}. The final catalog will contain multi-epoch astrometry, visible photometry and spectra for about 300,000 small bodies \citep{2007-EMP-101-Mignard}.
The ESA \emph{Euclid} space survey telescope \citep{2011arXiv1110.3193L} with an expected launch in 2022 will provide both infrared and optical observations of around 150,000 SSOs, allowing for detailed compositional studies \citep{Carry2018}. An automated identification of serendipitous SSO observations is investigated, e.g. by means of a convolutional neural network \citep{2018arXiv180710912L}.

Yet, for smaller missions or PI observations on large facilities, the resources for the development of an automated detection pipeline may not be available. A template solution may lower this threshold sufficiently to make the identification of SSOs feasible. As the telescope and observation conditions differ vastly from observatory to observatory, a mission-agnostic pipeline can only have one basic requirement: at least three images need to cover an overlapping area of the sky within a reasonable amount of time to identify moving sources. This work introduces the \mono{ssos} pipeline for the identification of SSOs and the recovery of their astrometric and photometric properties from astronomical images. As versatile tool, it is intended to be a template pipeline for all kinds of observatories, with minimal set-up and hardware requirements.

The \mono{ssos} pipeline was developed from the Kilo-Degree Survey (KiDS) SSO identification pipeline described in \citet{Mahlke2017}.
This proof-of-concept, survey-specific software recovered more than 20,000 SSO candidates with a degree of contamination below 0.05\,\%. However, it was tailored specifically to the KiDS observation strategy, instrument characteristics, and data product file specifications.
By applying the alpha-version to images from other observatories, these dependencies were identified and removed or weakened.
Among other changes, the \mono{ssos} pipeline no longer relies on regular image cadence or specific FITS keywords in the header.
Common observation cases like SSOs blended with other sources, missed SSO detections, long temporal baselines of observations, and single- or multi-band observations are treated better, leading to more SSOs being identified.
The \mono{python} implementation runs faster, has improved error-handling, and the distribution via the Python Package Index\footnote{\url{https://pypi.org/project/ssos/}} offers a simple installation process.

In \autoref{sec:pipeline}, the \mono{ssos} pipeline is described from a conceptual point of view.
The pipeline implementation and computational performance are discussed in \autoref{sec:implementation}.
In \autoref{sec:test}, the application and results of two test studies are presented, highlighting the versatility of the \mono{ssos} pipeline. Conclusions and an outlook on future development goals are given in \autoref{sec:conclusion}.

\begin {figure*}
\centering
\begin {tikzpicture}[-latex ,auto,node distance =1.8cm and 5.5cm ,on grid,semithick,
input/.style ={densely dashed, rectangle, draw=black, thick, fill=white,
text width=12em, text centered, minimum height=2em, inner sep=6pt},
output/.style ={rectangle, draw=black, thick, fill=white,
text width=10em, text centered, minimum height=3em, inner sep=6pt},
process/.style ={rectangle, rounded corners, draw=black, thick, fill=white, text width=10em, text centered, minimum height=3em, inner sep=6pt},
coord/.style={coordinate, on grid, node distance=1.8cm and 5.5},]
\node[input] (data) {Images, (Weights, Masks)};
\node[process] (sextractor) [below =of data] {\emph{SExtractor}};
\node[input] (sextractor_settings) [right =of sextractor] {SExtractor Configuration};
\node[process] (scamp) [below =of sextractor, yshift=-0.5cm] {\emph{SCAMP}};
\node[input] (ref_cat) [left =of scamp] {Reference Catalogue\\\small from VizieR Service};
\node[input] (scamp_settings) [right =of scamp] {SCAMP Configuration};

\node[process] (filter) [below =of scamp, yshift=-1cm] {\mono{python} \emph{Filter Routine}};
\node[input] (filter_settings) [right =of filter, yshift=0.4cm] {\mono{ssos} Configuration};
\node[input] (hyg) [below =of filter_settings, yshift=1.0cm] {Bright-Sources Catalogue};

\node[output] (candidates) [below =of filter, yshift=-1] {SSO Candidates};
\node[output] (cutouts) [right =of candidates, yshift=-1] {Cutouts};
\node[input] (skybot) [left =of candidates] {Known-SSOs Catalogue \\ \small from IMCCE};

\path (data) edge [bend left =0] (sextractor);
\path (sextractor_settings) edge [bend right =0](sextractor);
\path (sextractor) edge [bend left =0] node[right = .3 cm] {Source Catalogues}(scamp);

\path (ref_cat) edge [bend left =0] (scamp);
\path (scamp_settings) edge [bend right =0](scamp);
\path (scamp) edge [bend left =0] node[right = .3 cm, align=left] {Source Catalogues\\Full and Merged Catalogue}(filter);

\path (filter) edge [bend left =0](candidates);
\path (candidates) edge [bend left =0](cutouts);
\path (skybot) edge [bend left =0](candidates);

\path (filter_settings) -- (filter)coordinate[pos=0.5](mm);
\draw[-latex] (filter_settings) --(filter_settings.west)-| (mm) |- (filter.east);
\path (hyg) -- (filter)coordinate[pos=0.5](mm);
\draw[-latex] (hyg) --(hyg.west)-| (mm) |- (filter.east);

\node [coord, above=of sextractor_settings, xshift=3cm]  (c4)  {};
\path (data.east) to node [xshift=2em] {} (c4);
\draw (data.east) -- (c4)  |- (cutouts.east);

{\tiny }
\end{tikzpicture}
\caption{Flowchart of the steps (rounded), dependencies (dashed), and outputs (solid) in the \mono{ssos} pipeline. Dependencies on the left-hand side are retrieved online during the analysis, while the ones on the right-hand side are stored locally and can be fully configured by the user.}\label{fig:flow}
\end{figure*}
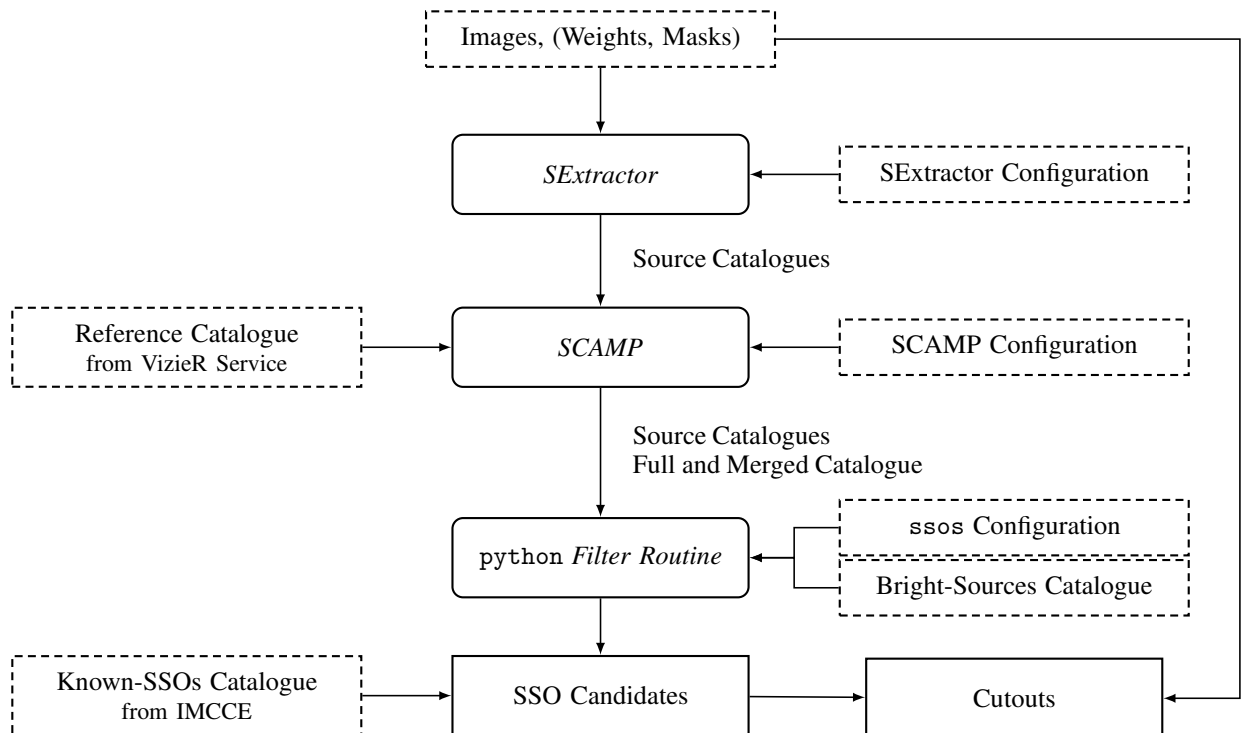

\section{Pipeline design}\label{sec:pipeline}
The focus of the pipeline design is placed on versatility. As a template solution for distinct telescope and observation set-ups, the requirements for the imaging data should be minimal. This versatility is achieved by relying on the widely used astr\emph{O}matic\footnote{\url{https://www.astromatic.net/}} software packages SExtractor and SCAMP for source detection and association \citep{1996A&AS..117..393B,2006ASPC..351..112B}. Both routines are robust, fast, and can handle many kinds of imaging data thanks to a large degree of configurability. After source detection with SExtractor, SCAMP is used to compute the global astrometric solution of the multi-epoch images. The resulting catalogues contain all stars, galaxies, imaging artefacts, and SSOs present in the images. A set of filter algorithms written in \mono{python} then identifies SSOs in this population primarily based on their apparent motion and trail appearance. The use of \mono{python} allows for simple automation, extensive error handling, and informative output to guide the user. The \mono{ssos} pipeline itself serves as a wrapper for these three steps and is written in \mono{python}.

All SSO selection filters are optional and configurable. Their parameters allow for different degrees of strictness in SSO candidate rejection. In general, the parameter selection steers the final set of SSO candidates either towards purity, i.e. no false-positive identifications, or completeness, i.e. all SSOs in the images are recovered. While in theory not mutually exclusive properties, in practice getting a complete and pure set of candidates is hardly feasible, and the user has to place the emphasis considering e.g. the final size of the output sample and the possibility of visual confirmation of the nature of all candidates. On the other hand, experience shows that artefacts are readily identified as outliers e.g. in colour-colour diagrams in the subsequent analysis of the SSO data.

This section steps through the pipeline processes and explains them on a physical, non-computational level. An overview of all steps and dependencies in the \mono{ssos} pipeline is given in \autoref{fig:flow}. Before executing the pipeline, the telescope metadata such as pixel scale, detector gain, and the FITS file properties such as the designation of the keywords of right ascension and declination coordinates in the headers must be entered in the configuration files of SExtractor, SCAMP, and the \mono{ssos} pipeline itself.\footnote{Until FITS header keywords are truly standardized, this tedious task cannot be reliably handled automatically.}

\subsection{Source detection: SExtractor}
SExtractor offers source detection in large astronomical images requiring only marginal setup. The source detection is described here briefly, the reader is referred to the official documentation\footnote{\url{https://sextractor.readthedocs.io/en/latest/}} and \citet{1996A&AS..117..393B} for details.

In agreement with the design principle of the \mono{ssos} pipeline, the minimal input for source detection with SExtractor are the FITS images. Corresponding weight images and pixel masks may be supplied, but the routine offers background estimation and several ways to reject artificial sources introduced by bad pixel apart from these auxiliary files.

A low-resolution, gridded map of the sky background is first computed by SExtractor by estimating the mode of the histogram of pixel values in each grid mesh. A full resolution background model is then derived through bi-cubic spline interpolation of this low-resolution map. After subtracting the background model, sources are detected by thresholding. By default, a source requires at least 5 contiguous pixel with more than 1.5 standard deviations of the local background to be extracted. Neighbouring sources that have merged in the image are deblended by re-thresholding the flux distribution in the source pixels in exponentially-spaced levels and evaluating the area and intensity of peaks at each level. Photometric measurements and morphometric parameters are extracted by calculating an adaptive elliptical aperture from the second order moments of the source's profile.

Each step of the source detection can be adapted by the user via a configuration file. For a reliable detection of SSOs, special focus should be placed on the source deblending parameters. As SSOs move through the field-of-view (FoV), they may get close in angular distance to other sources. If these mergers are not correctly deblended, the SSO is not detected in the image and the subsequent filter routine may reject all other detections of this SSO on the grounds of non-linear motion.

The source detection by SExtractor effectively converts the imaging pixel data to catalogue data, creating one catalogue of detected sources per input image. The next step aims to associate the detections of individual sources at distinct epochs by cross-matching these catalogues.

\subsection{Source association: SCAMP}\label{sec:scamp}
Standard procedure in astronomical observations suggests a dithering pattern when observing a part of the sky to bridge gaps in the camera CCDs and avoid biases e.g. introduced by hot pixel. The FoVs of the images are therefore shifted and possibly rotated and scaled with respect to each other. An astrometric calibration is required to trace source observations from one image to another, which is computed with SCAMP. Again, this computation is described in brevity, see \citet{2006ASPC..351..112B} and \citet{Bouy2013} for details.

SCAMP uses the centroid source positions retrieved by SExtractor to compute the astrometric solution. A reference catalogue is retrieved from the Vizier database \citep{2000A&AS..143...23O} and detected and reference sources are cross-correlated for each input image. Relative shifts, scaling, and rotations between the FoVs are thereby corrected for and the reprojected, detected sources are cross-matched in overlapping images. Finally, the astrometric solution is derived by a $\chi^2$-minimisation of the quadratic sum of differences in the positions of overlapping detected sources in pairs of images.

SCAMP computes the proper motions of all sources by computing a linear, weighted $\chi^2$-fit of the source coordinates over observation epoch. The reported uncertainty on the proper motions is computed from the covariance matrix of the best-fit parameters. Sources which display non-linear motion therefore have larger errors on their proper motion. Outlier detections in right ascension or declination space are identified by means of $\chi^2$-reduction. The source observation which increases the $\chi^2$ of the fit the most is seen as outlier and removed from the fit. This procedure is repeated until the reduced $\chi^2$ is below 6 or 20\,\% of the source detections were rejected.

As SExtractor, SCAMP offers a large degree of configuration. For reliable SSO detection, proper setting of the \mono{CROSSID\_RADIUS} is paramount. It defines the maximum distance in arcseconds between two source detections that SCAMP will consider for cross-matching.\footnote{While the \mono{CROSSID\_RADIUS} is defined in arcsecond in SCAMP, in the context of SSO identification it is more convenient to convert it to an equivalent proper motion limit by dividing by the time between the beginning of the two subsequent exposures. From here on, the \mono{CROSSID\_RADIUS} is therefore given in \arcsec/h.} Therefore, the cross-match radius sets an intrinsic upper limit on the proper motion of detected sources. For each pair of subsequent exposures, source detections which are displaced by more than this radius will not be regarded as belonging to the same source. Setting the \mono{CROSSID\_RADIUS} paramter too large, however, leads to an increase in mis-matched sources. A brief investigation of the optimal value is presented in \autoref{sec:gtc}.

\subsection{SSO identification: \mono{python} filter routine}
The difficult tasks of source detection and association across observation epochs are handled by the widely-used SExtractor and SCAMP packages, with the \mono{ssos} pipeline serving as a wrapper to handle the catalogue and image dependencies. The following filter routine is novel and has been developed specifically for the identification of SSO in source catalogues recovered from a series of images
\citep{Mahlke2017}.

The main characteristic of SSOs to differentiate them from other sources, astrophysical or artificial, is their apparent motion across the sky. Even over short observation times, they describe in first-order linear trails in images. This feature particularly sets them apart from stars and galaxies.
Meanwhile, artefacts like hot pixel or diffraction spikes can mimic the linear apparent motion if a linear dithering pattern is used. The differentiating feature here is the consistent appearance of the SSO trail in images taken under similar observation conditions.
The following six filter algorithms are based on these premises.

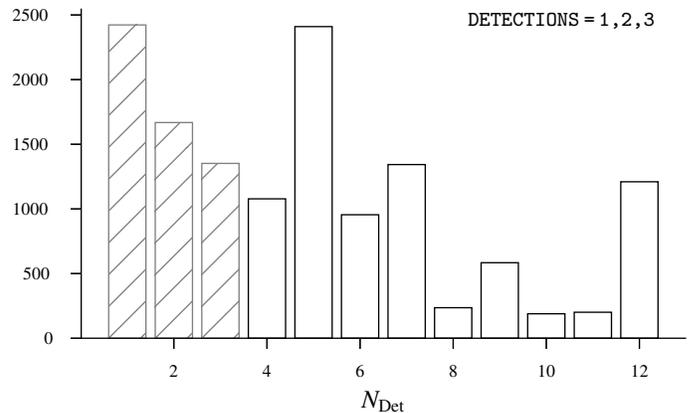
\begin{figure}
    \centering
    \input{number_of_detections.pgf}
    \caption{Depicted is the number of sources recovered in a randomly chosen tile of the J-PLUS survey, grouped by the number of detections per source. A large fraction of artefacts, especially cosmic rays, is filtered out by removing sources with only few detections across the images. The user can reject sources based on the number of detections via the \mono{DETECTIONS} filter parameter, here set to  3 or fewer detections and depicted by the grey, hatched bars.}
    \label{fig:number_of_detections}
\end{figure}

\paragraph{Number of Detections} The number of detections for each source is between 1 and \textit{N}, where \textit{N} is the total number of input images. Typically, stars and galaxies have close to $N$ detections as they do not leave the FoV unless the dithering pattern excludes them. Imaging artefacts like cosmic rays (CRs) make up the majority of sources with 1\,-\,3 detections as they only appear in singular images and may be associated by chance to another artefact by SCAMP. An example distribution of the number of detections per source in a series of 12 images is shown in \autoref{fig:number_of_detections}. The peak of sources with 5 detections is specific for this survey set-up (J-PLUS, see \autoref{sec:jplus}), which contains 5 broad-band in the set of 12 filters, increasing the chance of observing a source five times.

The number of detections for SSOs is largely dependent on the exposure time and observed bands of the survey. Filtering sources with few detections ($\leq$ 3) is still advisable to remove a large fraction of imaging artefacts. Furthermore, the ability to correctly evaluate a source's motion and appearance over time increases with each detection, strongly reducing the probability of a false-positive detection.

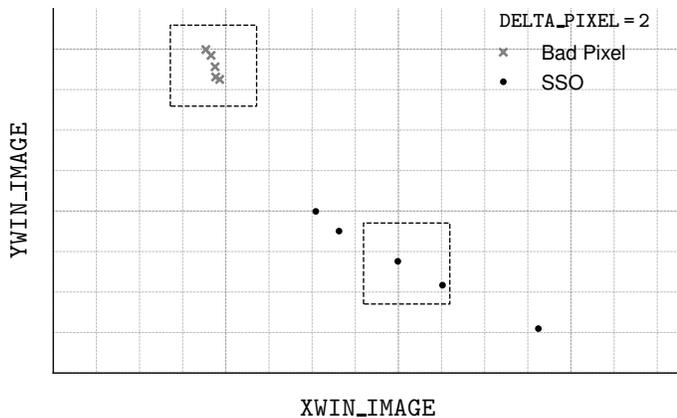
\begin{figure}
    \centering
    \input{bad_pixel_filter.pgf}
    \caption{Rejection of artificial sources introduced by bad pixel is done in CCD-space. Astrophysical sources are in general projected on a wide range of pixel (black dots), while artificial sources introduced by bad pixel will be localized (grey marks). The \mono{DELTA\_PIXEL} filter parameter sets the minimum required difference between the minimum and maximum pixel position in x- and y-dimension that sources have to cover to pass this filter. This is illustrated here by the dashed rectangles centred on the mean source positions.}
    \label{fig:bad_pixel_filter}
\end{figure}
\paragraph{Bad Pixel} Hot pixels in CCDs mimic linearly moving sources of constant appearance in case of a linear dithering pattern. While there are several ways to eliminate these artificial sources from the sample prior to this step,\footnote{Bad pixel masks and increasing the minimum number of pixel per source in the SExtractor configuration are effective ways to reject artificial sources caused by broken CCD pixel, but the former might not be available and the latter might impede the detection of faint SSOs.}  a simple yet effective method is provided at this point.

Bad-pixel sources are characterised by their fixed position in the CCD-space. The centroid position recovered by SExtractor lies on or adjacent to the bad pixel itself, while non-artificial sources should not fall onto the same pixel in every exposure. The \mono{DELTA\_PIXEL} filter parameter defines the minimum spread in pixel-space that sources have to exhibit to pass this filter. \autoref{fig:bad_pixel_filter} shows a rejected source, a hot pixel, in the upper left corner, while an accepted SSO projected onto CCD-space is shown below.

\paragraph{Proper Motion} Motion is the primary characteristic differentiating an SSO from extrasolar sources. A strong separation of non-SSOs and SSOs is the signal-to-noise ratio (SNR) of the source candidate's proper motion. As explained in Sec.~\ref{sec:scamp}, SCAMP computes the proper motion with a weighted, linear $\chi^2$-fit. A large uncertainty derived from the best fit parameters' covariance matrix indicates non-linear motion. A lower limit on the SNR therefore rejects sources exhibiting non-linear proper motion. The SNR is defined as

\begin{equation}
\mathrm{SNR}_\mu = \mu / \sigma_\mu,
\end{equation}

where $\mu$ is the absolute value of the proper motion and $\sigma_\mu$ the combined error of the proper motion components as derived by SCAMP.

\begin{figure}
    \centering
    \input{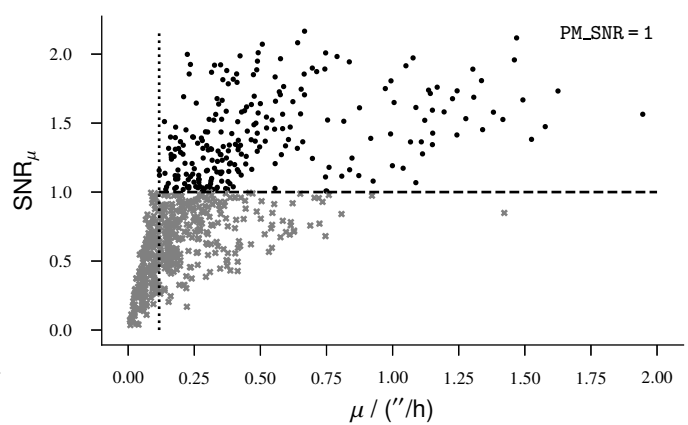}
    \caption{SCAMP computes proper motions by means of a weighted, linear $\chi^2$-fit. Non-linear motion leads to large uncertainties on the derived absolute value, which decreases the signal-to-noise ratio. Setting a threshold \mono{PM\_SNR} (black, dashed line) removes these non-SSO sources (grey marks). As the clustering towards zero SNR and proper motion shows, filtering sources based on the signal-to-noise ration  effectively introduces a lower limit on the proper motion (black, dotted line).}
    \label{fig:proper_motion_snr_filter}
\end{figure}

\autoref{fig:proper_motion_snr_filter} shows that setting the \mono{PM\_SNR} filter parameter is equivalent to introducing a lower limit on the proper motion. In general, sources with larger proper motion have a larger SNR. Extrasolar sources, however, will exhibit random coordinate fluctuations on the order of the seeing, leading to non-linear apparent motion and hence a small SNR in the proper motion calculation.

\begin{figure}
    \centering
    \input{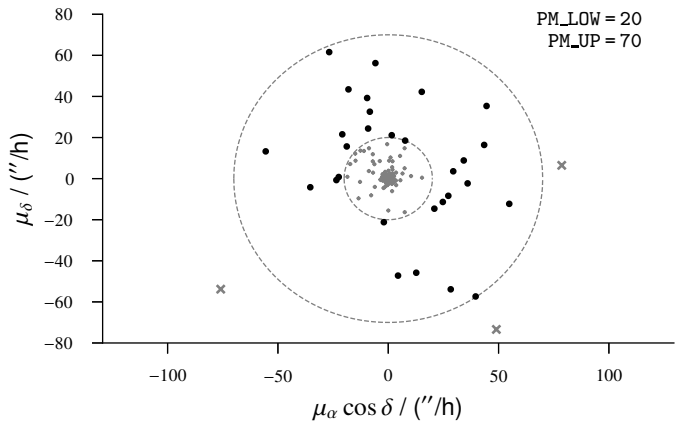}
    \caption{The distribution of sources in proper motion space in a randomly chosen tile of the J-PLUS survey. SSO candidates can be rejected on the basis of the absolute value of their proper motion. Sources with proper motions larger than \mono{PM\_UP} or smaller than \mono{PM\_LOW} are rejected (grey marks), while sources within the range are accepted (black dots). Setting \mbox{\mono{PM\_LOW}$\,\sim10\arcsec$/h} in this particular example removes all stars and galaxies from the candidates sample, however, distant, slower-moving SSOs are rejected as well. For clarity, only 50\% of the sources are shown.}
    \label{fig:proper_motion_filter}
\end{figure}

The absolute value of the apparent motion should be restricted with care. As shown e.g. in \citet{Carry2018}, the proper motion space of SSOs encompasses several orders of magnitudes, from a few hundredths of \arcsec/h for distant Kuiper-Belt objects (KBOs) to over 1000\,\arcsec/h for the close population of NEOs. Instead, the possibility to filter the proper motion values outside a user-specified range is intended for focusing either on fast- or slow-moving sources like NEOs or KBOs, or to overcome persistent spurious detections in highly contaminated images. \autoref{fig:proper_motion_filter} shows a typical distribution of proper motion values of sources for an input set of 12 images, split into the right-ascension and declination components. The cluster towards zero proper motion is made up of background stars and galaxies.

\paragraph{Linear Motion} The filter on the linearity of the apparent motion is the most effective and reliable step in the source selection. While the proper motion computation by SCAMP is already exploited in the previous filter, it excludes outlier detections in coordinate-space from the fit, weakening the rejection. Furthermore, SSO observations with large gaps between observations may fool the linear motion filter due to the long baseline of the fit. These cases need to be identified and handled differently and make this dedicated filter step necessary.

For each SSO candidate, the right ascension and declination coordinates are fit over epoch using a weighted, linear $\chi^2$-fit. The coefficient of determination parameter $R^2$ is then defined by regarding the total sum of the squares over the sum of the squared residuals,

\begin{equation}
    R^2 = 1-\frac{\sum_i(y_i-f_i)^2}{\sum_i (y_i-\bar{y})^2},
\end{equation}

where $f_i$ describes the linear model function evaluated at the observation epochs $i$, $y_i$ are the coordinate data points and $\bar{y}$ the mean coordinate value. The fraction compares the variance of data and model to the variance of the data. If the fit describes the data well, i.e. the source exhibits linear apparent motion in this coordinate over time, $R^2$ will be close to unity. By setting the \mono{R\_SQU\_M} filter parameter, sources with $R^2$ values smaller than the limit will be rejected. This is analogous to the proper motion computation by SCAMP, however, no source detections are dismissed during the fitting procedure.

Linear fits can be fooled by long temporal baselines. Any source candidate observed several times in the first hour of the observation night and one more during morning hours exhibits an $R^2$ value close to unity. To detect and treat these occurrences, source detections are checked for outliers in epoch-space. Outlier detections are defined using the median absolute deviation (MAD) of the observation epochs $E_i$,

\begin{equation}
\mathrm{MAD}=\mathrm{median}(|E_{i} - \mathrm{median}(E)|),
\end{equation}

describing the median time step between the single observation epochs and the median observation epoch. Using the \mono{OUTLIER\_THRESHOLD} filter parameter, the user sets the upper limit of observation time between two observations in units of MADs. If any gap in observation time is larger than the set threshold, the source detections are divided into two subgroups at this gap. This may occur at multiple epochs for a single source, specifically if the observation strategy includes frequent changes between narrow- and wide-band filter observations.

\begin{figure}
    \centering
    \input{outlier_motion_filter.pgf}
    \caption{Displayed are the right ascension coordinates of an SSO over epoch (black dots). A coincidental cosmic ray detection was erroneously added to the SSO detections by SCAMP (grey mark). Performing a linear fit on all 6 detections would result in the source being rejected ($R^2$\,=\,0.36, dotted line). By identifying the CR outlier in epoch space using the median absolute deviation of observation epochs, it can be dismissed for the fit, and the SSO correctly passes the filter step ($R^2$\,=\,1.00, black line). The CR detection is flagged as outlier in the output data. The coordinate error bars smaller than the marker size.}
    \label{fig:outlier_motion_filter}
\end{figure}
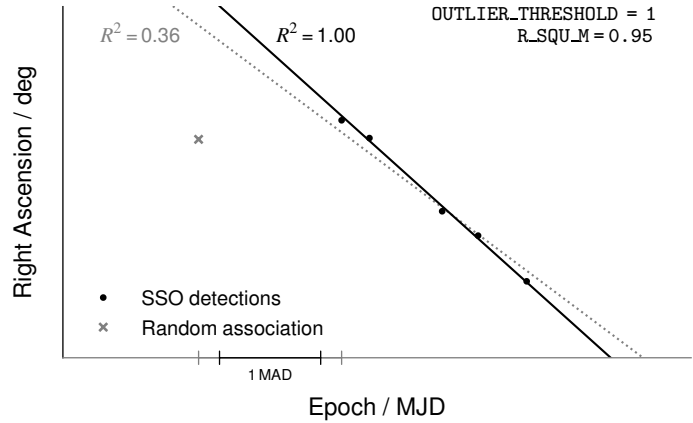
Each detection subgroup is then checked for linear motion in both coordinate dimensions by the fitting described above. If all linear fits pass the $R^2$ limit, the subgroup is accepted. If a subgroup does not pass or if it contains fewer than 3 detections, it is flagged as outlier detection. If all subgroups of a source are flagged as outlier detections, the source is rejected.

\autoref{fig:outlier_motion_filter} illustrates this for an SSO with 5 detections, which erroneously was associated with a CR detection by the SCAMP cross-match. After identifying the CR outlier in epoch-space, the motion of the SSO is correctly evaluated as linear and the source is accepted, with the CR detection flagged as outlier.

Sources with flagged outliers can then be accepted, visually confirmed, or, as a more conservative approach, only kept if matched with a known SSO.
Alternatively, all outlier detections can be rejected, while the remaining source detections are kept.

\paragraph{Trail consistency} SSOs imaged under similar observation conditions like seeing, binning, exposure times, and observation band should leave consistent trails in the images. Meanwhile, artefacts do not have any bounding conditions on their appearances, so they will likely fluctuate in size. The morpohmetric trail parameters can therefore be examined for consistency to differentiate these sources.

The consistency of the trail parameters with the hypothesis of a constant SSO imprint in the image is quantified using the semi-major and -minor axis lengths of the ellipse fit to the source by SExtractor, denoted by \textit{a} and \textit{b} respectively. If the observation conditions are constant, the standard deviation should be in the order of the seeing.

\begin{figure}
    \centering
    \input{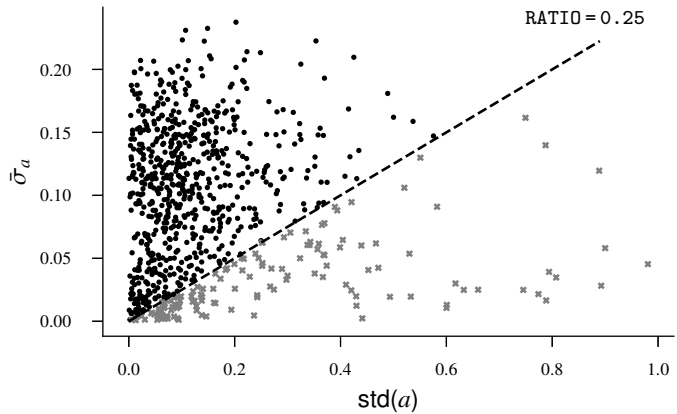}
    \caption{Weighted average uncertainty versus standard deviation of the semi-major axes of sources in single-band observations by the Gran Telescopio Canarias. The dashed black line depicts the \mono{RATIO} of 0.25 between the two quantities. Sources below this ratio are showing large size variations and are rejected (grey marks). Sources above this ratio are accepted (black dots).}
    \label{fig:trail_consistency_fullcat}
\end{figure}

For each axis, the average weighted uncertainty $\bar{\sigma}$ is computed and compared to the standard deviation std of the data.
The \mono{RATIO} filter parameter sets an upper limit on the observed standard deviation with respect to the uncertainties of the morphometric parameters,

\begin{align}
    \begin{split}
        \bar{\sigma}_{a} &= \frac{1}{\sqrt{\sum_i \frac{1}{\sigma_{a,i}^2}}}\\
        \mono{RATIO}_{a} &= \frac{\bar{\sigma}_{a}}{\mathrm{std}(a)},
    \end{split}
\end{align}
and analogously for the semi-minor axis $b$. A smaller \mono{RATIO} allows for larger dispersion in the ellipse parameters. \autoref{fig:trail_consistency_fullcat} depicts the relation between the weighted average uncertainty and the standard deviation of the semi-major axes of sources in a random set of single-band observations by the Gran Telescopio Canarias. The majority of sources is of astrophysical nature and exhibits small standard deviations of their axes parameters even as the average uncertainties increase. Artificial sources are located towards the bottom of the figure, indicating large size fluctuations with respect to their morphometric uncertainties.

In practice, the \mono{RATIO} value is set by examining the distribution of the values plotted in \autoref{fig:trail_consistency_fullcat} or by trial-and-error analysis of a set of known SSOs and artefacts in the images.

The trail filter should be applied to multi-band data with caution. SSOs display exhibit wavelength-dependent fluxes depending on their spectral energy distribution (SED), leading to a change in their trail appearance under constant exposure time if the observation band is changed.

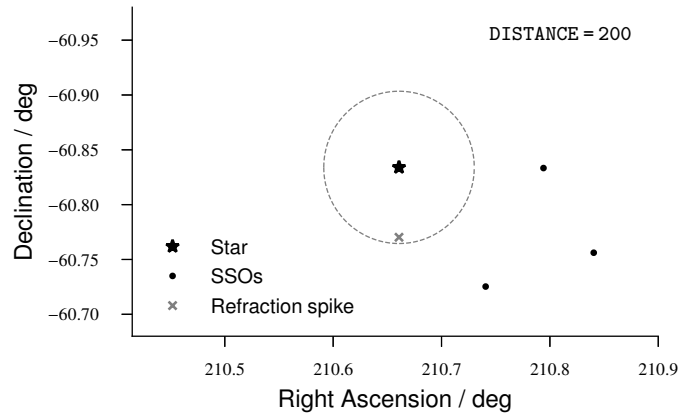
\begin{figure}
    \centering
    \input{stellar_catalogue.pgf}
    \caption{SSO candidates in close vicinity to bright sources are likely artefacts such as diffraction spikes or ghosts. The \mono{DISTANCE} parameter sets the lower limit in arcseconds that the mean source position must have to pass this filter step (grey circle). The black star is Alpha Centauri, the grey mark depicts a rejected candidate, the black circles are accepted candidates.}
    \label{fig:stellar_catalogue}
\end{figure}

\paragraph{Bright-Sources Catalogue} Diffraction spikes and ghosts introduced by bright sources are tough challenges for the filter pipeline. They can mimic linear motion given a linear dithering pattern. The trail consistency filter can mitigate this contamination, but in the case of multi-band data or when aiming for a more conservative approach, filtering the SSO candidates in close vicinity to bright sources is the most reliable way to reject these artefacts. The \mono{DISTANCE} parameter sets the minimum distance in arcseconds to sources that the mean candidate coordinates must have in order to pass the filter, as depicted in \autoref{fig:stellar_catalogue}.
The reference source catalogue can be either the same catalogue as retrieved by SCAMP, with optional lower- and upper magnitude limits applied on the reference sources via the \mono{MAG\_LIMITS} parameter, or a user-provided, local catalogue.

\subsection{Optional analyses}
The \mono{ssos} pipeline produces a catalogue of SSO candidates detections. Depending on the filter parameters and quality of the images, the output sample will be contaminated by false-positive detections to a certain percentage. To investigate the amount of artefacts surviving the pipeline, cutout images of every SSO detection can optionally be extracted using the SWARP software \citep{2002ASPC..281..228B}. These images allow for quick visual confirmation of the results and further aid in the identification of common artefacts and necessary filter parameter tuning.

Another standard analysis procedure is the cross-match of all SSO candidate detections with the SkyBoT service \citep{2006ASPC..351..367B}. The \mono{ssos} pipeline automatically queries SkyBoT providing the image FoV and observation epoch. The service returns the metadata of SSOs that are predicted to cross the FoV at the given epoch by comparing the observation metadata to pre-computed SSO ephemerides. If a known SSO is within a user-defined cross-match radius, it designation and other properties are added to the source candidate's metadata. In case there is more than one known SSO within the cross-match radius, the one exhibiting the closest co-linear motion to the SSO candidate is chosen by minimizing the difference in the proper motion angles between the candidate source and the known SSO.

Finally, an additional SExtractor run can be performed for all source detections to compute fixed-aperture magnitudes for each SSO,  required e.g. for colour indices. SExtractor is run in dual-image mode, with a user-specified filter as reference filter. The fixed-aperture flux and magnitude are added to the output data.

\section{Implementation and performance}\label{sec:implementation}

The \mono{ssos} pipeline is written in \mono{python3}.
The source code is open-source\footnote{\url{https://github.com/maxmahlke/ssos}} and extensive documentation is available online.\footnote{\url{https://ssos.readthedocs.io}} The code is further available from the Python Package Index, simplifying the install process and the handling of the \mono{python} dependencies. The astr\emph{O}matic software SExtractor, SCAMP, and SWARP need to be installed separately. As such, the pipeline can be installed on personal computers for the configuration setup and testing, as well as on computer grids for the batch-processing of images.

The development of the source code so far focused on simplicity of use. All pipeline steps are initiated using the \mono{ssos} executable from the command line. The pipeline settings are read from a configuration file. The user interface by design resembles closely that of SExtractor and SCAMP, to facilitate the application for users already familiar with these software.

Upon instantiation, the \mono{ssos} pipeline creates and manages all necessary directory structures, image and catalogue dependencies for each step. In the usual application mode, the user should not have to specify more than the directory of the FITS images and the path to the configuration file.

Some development effort has been placed on the execution speed. Improvements were achieved by profiling the python code and reducing the number of value lookups in catalogues as far as possible. Nevertheless, the main workload is handled by the efficient SExtractor and SCAMP runs, which are the analysis' bottlenecks in terms of speed and can be multi-threaded.

Execution time depends on a variety of factors, such as the number of images searched, the pixel dimensions of each image, the number of sources present, the chosen filter steps and parameters, and finally the computer hardware. As a reference, the complete analysis of 6,132 J-PLUS images with dimensions of 2\,deg$^2$ in 9,500$\times$9,500 pixel took $\sim$30 hours on an 2 GHz Intel Core i5 computer architecture with 8\,GB RAM (MacBookPro Late 2016), around 3.5 minutes for each of the 511 input fields made up of 12 images each.

\section{Test studies}\label{sec:test}
 Applying the alpha-version of the pipeline described in \citet{Mahlke2017}, 20,221 SSO candidates with an estimated false-positive content of less than 0.05\,\% were recovered in 346\,deg$^2$ of the KiDS survey. Since then, the pipeline has been re-structured to facilitate the application not only to surveys but all kinds of imaging data products. While the well-defined observation strategy of survey telescopes simplifies the application of the \mono{ssos} pipeline due to the set dithering patterns and exposure times, it is not required for successful identification of SSOs.
The versatility of the \mono{ssos} pipeline is now illustrated using two case studies: images from the GTC OSIRIS Broad Band DR1 Catalogue, made up of PI-lead observations, and survey data from the J-PLUS DR1.

\subsection{GTC OSIRIS Broad Band DR1 Catalogue}\label{sec:gtc}

\begin{figure}
    \centering
    \input{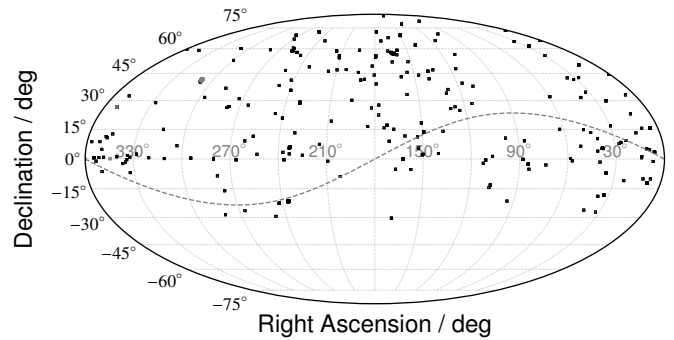}
    \caption{Footprints of the GTC OSIRIS Broad Band DR1 catalogue. The footprint scale was increased for readability. The grey dashed line represents the ecliptic plane.}
    \label{fig:gtc_moc_mollweide}
\end{figure}

The \mono{ssos} pipeline is applied to 8,096 images acquired with the Gran Telescopio Canarias (GTC) OSIRIS, which were used to derive the GTC OSIRIS Broad Band DR1 catalogue (Cortés-Contreras, in preparation). OSIRIS (Optical System for Imaging and low-Intermediate-Resolution Integrated Spectroscopy) is an optical imaging and spectroscopy instrument part of GTC with a vignetted FoV of 7.8$\arcmin\times$7.5$\arcmin$.
Observations in the DR1 catalogue are taken in the Sloan system broadband filters $g'$, $r'$, $i'$, and $z'$. The SSO identification is carried out on the astrometrically and photometrically corrected images provided by the Spanish Virtual Observatory. The observation footprints are depicted in \autoref{fig:gtc_moc_mollweide}.

Given the small FoV and in general large ecliptic latitude of the footprints, the output sample of SSO candidates is small, allowing for illustrative studies of the optimal pipeline filter configuration and the influence of the SCAMP cross-match radius parameter  \mono{CROSSID\_RADIUS}, as visual confirmation of the results is feasible.

The pointed observations apply varying observation strategies. A single observation run can include different time-spans, singular or a multitude of exposures, different exposure times, and filter configurations. To apply the \mono{ssos} pipeline, the 8,096 exposures are therefore divided into logical groups. Each logical group is made up of all exposures with overlapping FoVs taken in a single night. If a group is made up of 3 or fewer exposures, it is discarded, as the linear motion cannot be detected reliably.
712 exposures are rejected on the basis of this criterion. An additional 402 exposures had to be discarded due to a large contamination by imaging artefacts, introduced during the detector read-out or caused by telescope tracking failures.

In total, 6,982 footprints are searched, divided into 420 logical groups. The observation epochs span the time from April 2009 to July 2014. The smallest number of exposures per group is 4, the largest group consists of 218 images, while 316 groups have 20 or fewer exposures. The average exposure time per group ranges from 17\,s to 1\,h.

For each exposure, a weight image is computed from the flat field images available online and passed to SExtractor for source detection.

The small FoV presents a challenge for SCAMP, as the reference catalogues might contain too few sources to compute the astrometric solution. The standard reference catalogue is set to SDSS-R9. For 129 groups, the reference had to be switched to the GSC-2.3 catalogue due to coverage reasons \citep{2008AJ....136..735L,2012ApJS..203...21A}.

A further challenge arises due to the absence of an underlying observation strategy. The cross-match radius has to be chosen on a per-group basis as the cadence between single exposures for each observation night varies. The radius is computed on-the-fly so that the average cadence between exposures allows for the observation of SSOs with proper motions up to 50\,\arcsec/\,h, 100\,\arcsec/\,h, and 150\,\arcsec/\,h. These three values are chosen to test the hypotheses that choosing a small radius misses fast moving SSOs if there are gaps in the observation epochs, while slower SSOs are recovered more reliably. An incomplete but pure output sample is expected. On the other hand, larger \mono{CROSSID\_RADIUS} values should yield a higher chance of recovering a complete SSO set, however, the amount of contamination due to spurious detections should increase as well.

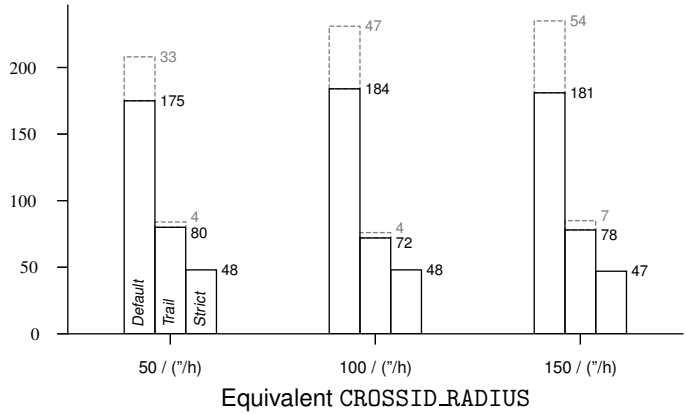
\begin{figure}
    \centering
    \input{gtc_runs.pgf}
    \caption{Results of the 9 pipeline runs on images of the GTC OSIRIS Broad Band DR1 Catalogue. The contamination and completeness for three different SCAMP cross-match radii and three different filter parameter settings is investigated. For each radius, the results of the \emph{default}, \emph{trail}, and \emph{strict} runs are given. The solid black bar gives the number of recovered SSOs, the dashed grey bar the number of artefacts in the output sample. Over all 9 analysis runs, 204 distinct SSOs are recovered. No pipeline run recovers a complete set of SSOs, while the degree of contamination increases with the cross-match radius.}
    \label{fig:gtc_runs}
\end{figure}

For each of the three runs with varying cross-match radii, three different filter parameter settings are set, again aiming to investigate the effect on the completeness versus purity relation. The \emph{default} pipeline settings are listed in \ref{app:settings}. Only sources with more than 3 detections are accepted and the trail consistency filter is off. The \emph{HYG} database is chosen as reference catalogue for bright sources in the FoV.  It contains all stars in the Hipparcos, Yale Bright Star, and Gliese catalogues.\footnote{\url{http://www.astronexus.com/hyg}}

In a second step, denoted \emph{trail}, the trail consistency filter is turned on and \mono{RATIO} is set to 0.25, to reject artefacts more conservatively. A final run, \emph{strict}, increases the \mono{RATIO} to 0.5, \mono{R\_SQU\_M} from 0.95 to 0.97, and rejects sources with 6 or fewer detections.

With the three chosen cross-match radii and three filter parameter settings, a total of 9 analysis runs is analysed and the output samples are visually inspected. Between the SSO candidates resulting from the \emph{default} runs, there is a large overlap as the input images remain the same, but the different cross-match radii include some sources and exclude others.

After visually confirming the nature of all SSO candidates, 204 distinct SSOs are identified. One SSO was recovered in 53 exposures. No \mono{CROSSID\_RADIUS} setting successfully recovered all 204 SSOs in the \emph{default} setting. \autoref{fig:gtc_runs} displays the results of the 9 analysis runs. The most complete output set is achieved with the \mono{CROSSID\_RADIUS} set to 100\,\arcsec/h, however, the difference between the sets is within a few percent. On the other hand, the number of recovered false-positives is proportional to the radius. The contamination increases from 19\% to 30\% between the 50\,\arcsec/h and 150\,\arcsec/h runs. For the two stricter parameter settings, the differences in the output samples disappear. Ultimately, a \mono{CROSSID\_RADIUS} of $\sim$100\,\arcsec/\,h is recommended for most use cases, as the majority of SSOs should not exhibit proper motions faster than this, and the degree of contamination is limited.

\begin{table}
\centering
\caption{The distribution of cross-matched SSOs over SSO classes. MB describes Main-Belt asteroids.}
\label{table:skybot_gtc}
    \begin{tabular}{lllll} \toprule
Comet  & Inner MB &Middle MB &Outer MB & Trojan\\
 \midrule
4  & 21 & 17 & 20 & 1\\ \bottomrule
 \end{tabular}
\end{table}
\begin{figure}
    \centering
    \input{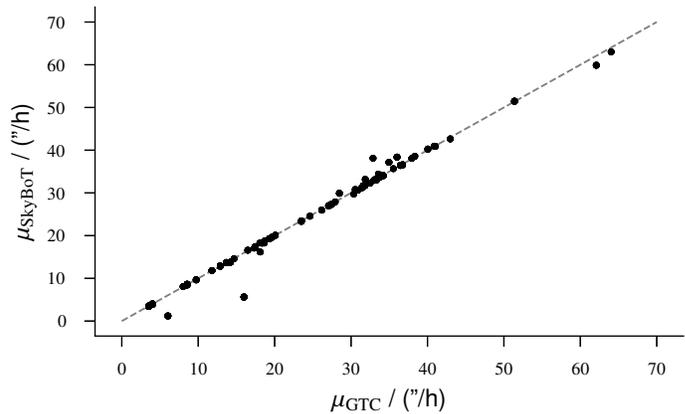}
    \caption{Comparison of predicted and recovered proper motion values of the cross-matched GTC SSOs (black dots). Ratios close to unity (dashed grey line) indicate a successful cross-match.}
    \label{fig:skybot_gtc_pm}
\end{figure}
Finally, the number of known SSOs in the output sample is determined. GTC images from early observation epochs may carry an error of up to 15 minutes on their \mono{DATE\_OBS} header keyword. While this presents no problem for the identification pipeline, the SkyBoT cross-matching may be inaccurate. Therefore, the \mono{CROSSMATCH\_RADIUS} parameter in the SkyBoT queries was increased from 10\arcsec~to 40\arcsec.

63 of the 204 SSOs are matched to known SSOs using the SkyBoT service. Their classes are given in \autoref{table:skybot_gtc}. The 4 observed comets are recovered from targeted observations, as well as one observation of (300163) 2006 VW139. All targeted observations of SSOs are successfully recovered, though the targets of some observation runs were left undefined by the observer. The vast majority of serendipitously observed SSOs are from the Main-Belt, with one Jupiter Trojan observed as well. \autoref{fig:skybot_gtc_pm} compares the absolute proper motion values of the cross-matched SSOs as computed by SkyBoT and as recovered by the \mono{ssos} pipeline. A successful cross-match is indicated by a ratio close to unity between these two values, which is the case for most matches. Ratios far from unity point to co-incidental cross-matches.

All 2,828 detections of the 204 detected SSOs were reported to the MPC. Further scientific exploitation of this dataset will be presented in Cortés-Contreras (in preparation).

\subsection{J-PLUS DR1}\label{sec:jplus}
\begin{figure}
    \centering
    \input{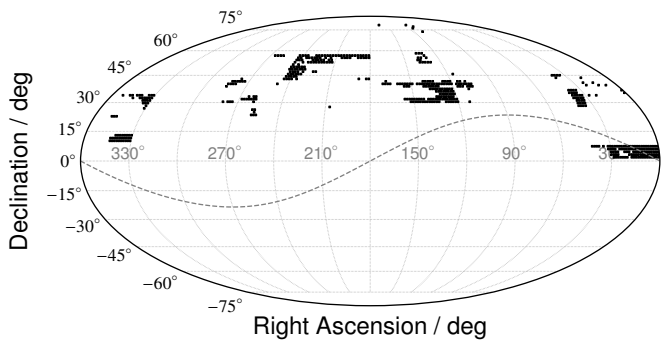}
    \caption{The footprints of the 511 fields of the J-PLUS DR1 cover approximately 1020\,deg$^2$. The grey dashed line represents the ecliptic plane.}
    \label{fig:jplus_moc_mollweide}
\end{figure}

The J-PLUS (Javalambre Photometric Local Universe Survey) is an on-going optical survey executed at the Observatorio Astrofísico de Javalambre (OAJ) \citep{cenarro2018j}. The observation strategy entails 36 observations in 12 bands taken within a few hours, using the JAST/T80 telescope with an FoV of 2\,deg$^2$. The whole survey aims to image 8,500\,deg$^2$. \autoref{fig:jplus_moc_mollweide} depicts the footprints of the J-PLUS first data release from July 2018 containing 511 fields  imaged between November 2015 and January 2018.\footnote{\url{http://archive.cefca.es/catalogues/jplus-dr1/}} The observations imaged 1022\,deg$^2$ of the sky with limiting magnitudes ranging from 20 to 23.

For each of the 511 J-PLUS fields, the 3 dithers per band are combined to 12 images, which are retrieved from the online repository, including their weights and image masks. When combining the 3-step dither pattern, a frame of zero value pixels is padded around a reference exposure. This frame is later removed with the supplied image masks.
The weight images are supplied to SExtractor during source detection.

At a first inspection of the images, it is found that the reduction pipeline successfully removes most imaging artefacts. The \mono{CROSSID\_RADIUS} is therefore set to 125\,\arcsec/h, as the amount of artificial sources is expected to remain low. Applying the \mono{ssos} pipeline on a test set of 6 tiles showed that the contamination of the final dataset is almost exclusively made up of stars. There is a probability that stars will mimic linear motion when in reality their centres are stochastically displaced due to seeing, which in combination with the large FoV of the JAST/T80 camera leads to this contamination.

The reference catalogue for the astrometric matching by SCAMP is the 2MASS All-Sky Catalogue of Point Sources \citep{2003yCat.2246....0C}.

The \mono{ssos} pipeline is applied with two filter settings. The \emph{default} setting is analogous to the one used in the GTC study, \ref{app:settings}, except that the bad pixel filter is turned off due to the low level of contamination. The \emph{strict} setting increases the required number of detections per source to 5 and the \mono{R\_SQU\_M} parameter to 0.97, to reject the stellar contaminants. The HYG database is used again as reference catalogue for bright sources in the FoV.

The results of the two applications are given in \autoref{table:jplus_results}. The \emph{default} sample contains 4,606 SSO candidates, with an estimated contamination of (2.0\,$\pm$\,0.2)\%. The \emph{strict} sample is made up of 3,696 candidates and has a false-positive ratio (FPR) below 0.68\,\%. The number of false-positive detections for each sample is derived from visual inspection of a random sub-sample of 500 (441) SSOs for the \emph{default} and the \emph{strict} sample respectively. 10 artefacts are found in the \emph{default} sub-sample, none of which is present in the \emph{strict} sub-sample.

\begin{figure}
    \centering
    \input{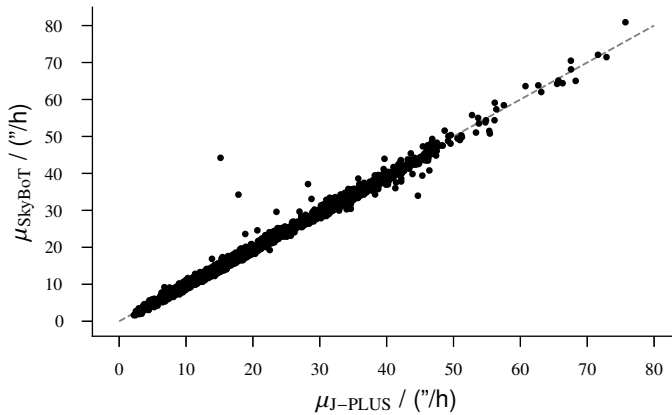}
    \caption{Comparison of predicted and recovered proper motion values of the cross-matched J-PLUS SSOs (black dots). Ratios close to unity (dashed grey line) indicate a successful cross-match.}
    \label{fig:skybot_jplus_pm}
\end{figure}

Of the 4,606 SSO candidates in the \emph{default} sample, 3,932 are matched within 10\,\arcsec with a known SSO by SkyBoT. The distribution of the cross-matched SSOs over the SSO populations is given in \autoref{table:skybot_jplus}. The majority of known SSOs is identified in the Main-Belt, while also members of the close NEAs and far KBOs are recovered. \autoref{fig:skybot_jplus_pm} displays the relation between the computed and recovered proper motion values of the cross-matched SSOs, showing a good agreement in general.

\begin{table}
\centering
\caption{The number of SSO candidates and the derived false-positive ratio (FPR) for the two pipeline runs on images of the J-PLUS survey. }
\label{table:jplus_results}
    \begin{tabular}{lll}       \toprule
          & SSO Candidates      & FPR    \\ \midrule
  \emph{default} & 4,606 & (2.0\,$\pm$\,0.2)\% \\
  \emph{strict}  & 3,696   & $\leq$0.68\,\%    \\ \bottomrule
 \end{tabular}
\end{table}

\begin{table}
\centering
\caption{The distribution of cross-matched J-PLUS SSOs over the SSO classes. NEA describes near-Earth asteroids, MCs are Mars-Crosser, MB are asteroids in the Main-Belt, while KBO are Kuiper-Belt objects.}
\label{table:skybot_jplus}
    \begin{tabular}{lllll} \toprule
NEA & MC & MB & Trojan & KBO\\
 \midrule
3 & 31 & 3,887 & 10 & 1\\ \bottomrule
 \end{tabular}
\end{table}

After photometric calibration of the recovered magnitudes, 34,014 detections of 4,596 SSOs were reported to the MPC.

Surveys like J-PLUS are prime candidates for the application of the \mono{ssos} pipeline, as the strict observation strategy definition offers precise settings of the filter parameters. Especially imaging campaigns covering areas close to the ecliptic will often yield a large number of SSO candidates, rendering visual confirmation of the nature of all candidates impossible. Therefore, a detailed discussion of the statistical derivation of the false-positive ratio is given in \ref{app:fpr}.

Upper limits for the completeness of the SSOs sample could be derived by regarding the fraction of recovered, known SSOs using SkyBoT. The large ephemeris uncertainties of SSOs with few recorded observations and the uncertainty on the predicted visual magnitude have to be accounted for.

\section{Conclusion and Outlook}\label{sec:conclusion}

Presented is an easy-to-implement, light-weight SSO recovery pipeline  which can be applied to increase the science-yield of any kind of imaging data. The \mono{ssos} pipeline offers a high degree of versatility thanks to the underlying SExtractor, SCAMP, and \mono{python} software. A flexible set of filter algorithms developed in various use cases offers a wide range of SSO identification methods.

The two presented test cases illustrate the potential of the pipeline. The observation strategies and telescope set-ups differ substantially, yet the \mono{ssos} pipeline is successfully applied in both cases with only minor changes in the set-up. The wide range of classes of recovered SSOs in the J-PLUS survey, from close NEAs to distant KBOs, displays that the method does not discriminate between the populations, as the only prerequisite are subsequent observations of an overlapping area in the sky.

The \mono{ssos} pipeline becomes more versatile with each new application case, as different observation strategies or data product formats are accounted for.
Future development will focus on easier filter parameter testing as well as increasing the science yield. The former may be achieved by automatically evaluating the number of source candidates in each step for different filter setups, and comparing them to a sample of previously identified artefacts and SSOs. The most promising improvement in terms of SSO science would be the successful cross-matching of SSOs recovered in subsequent observation nights, greatly increasing the recovered arc of the orbit. Finally, the prior knowledge of the ephemerides of known SSOs provided by SkyBoT can be exploited more by cross-matching sources before applying the filter pipeline, and extracting matches with co-linear movement right away.

\section*{Acknowledgements}
MM is funded by the European Space Agency under the research contract C4000122918.

MM would like to thank the EDAS Helpdesk at ESAC for their advice on the derivation of the false-positive sample ratios.

Based on data from the GTC Public Archive at CAB (INTA-CSIC).

Based on observations made with the JAST/T80 telescope for the J-PLUS project at the Observatorio Astrofísico de Javalambre, in Teruel, owned, managed and operated by the Centro de Estudios de Física del Cosmos de Aragón.

This research has made use of the VizieR catalogue access tool, CDS, Strasbourg, France.

H. Bouy acknowledges funding from the European Research Council (ERC) under the European Union’s Horizon 2020 research and innovation programme (grant agreement No 682903, P.I. H. Bouy), and from the French State in the framework of the "Investments for the future" Program, IdEx Bordeaux, reference ANR-10-IDEX-03-02 .

\bibliographystyle{aa_doi}
\bibliography{bib}
\clearpage

\appendix
\section{Estimating the false-positive ratio}
\label{app:fpr}
Applying the \mono{ssos} pipeline to a large set of imaging data often results in an amount of SSO candidates where visual confirmation of their true nature is no longer feasible. The number of false-positive detections has to be estimated by limiting the inspection to a smaller subset. In the following, the statistical foundation of this estimation is described in brevity. Further information can be found e.g. in \citet{jaynes2003probability,rice2003mathematical}.

The recovered SSO candidates form a dichotomous population of size $N$. They can be divided into $K$ artefacts and $(N-K)$ SSOs.\footnote{\emph{Artefacts} includes any kind of false-positive result, not necessarily only source detections caused by imaging artefacts.} A random sample of size $n$ drawn without replacement from this population forms a vector of Bernoulli random variables $\textbf{X}=(X_{1}, X_{2}, \dots, X_{n})$, where

\[
  X_i =
  \begin{cases}
                                   0 & \text{if SSO} \\
                                   1 & \text{if artefact} \qquad i\in\{1, \dots ,n\}
  \end{cases}
\]

The number of artefacts $k$ in the sample $n$ is then
\begin{equation}
    k = \sum_{i=1}^n X_i
\end{equation}

The probability density function of $k$ is described by the hypergeometric distribution
\begin{equation}
    h(k\,|\,N, K, n) = \frac{\binom{K}{k}\binom{N-K}{n-k}}{\binom{N}{n}}.
\end{equation}
The notation here emphasizes that the probability distribution $h(k\,|\,N, K, n)$ of the sample statistic $k$ depends, among others, on the unknown population parameter $K$. In the following, this notation is dropped and the dependency is implicit.

Each population member has a probability of $n/N$ of being included in the sample. The linearity of the expectation then yields the expected number of artefacts $E(k)$ in the sample, given by the contribution of the expectation of all $K$ artefacts;
\begin{equation}
    E(k) = K\frac{n}{N}.
\end{equation}

The derivation of the sample variance $s^2(k)$ is more challenging as the population members are dependent variables. While the result is given here, the reader is referred to the literature for detailed steps:
\begin{equation}\label{equ:vark}
    s^2(k) = \frac{nK(1-\frac{K}{N})(N-n)}{N(N-1)}.
\end{equation}

The sample statistic $k$ is now used to estimate the population parameter $K$. An intuitive estimator of $K$ is defined by using the method of moments; the ratio of artefacts in the sample is assumed to be equal to that in the population:
\begin{equation}
    \frac{k}{n} \approx \frac{K}{N} \implies K \approx k\frac{N}{n}
\end{equation}
The expectation of $K$ is then given by
\begin{equation}
    \mu(k\frac{N}{n}) = K.
\end{equation}
As the estimator is unbiased, its variance is equal to the mean square error of $K$, where \autoref{equ:vark} is used:
\begin{equation}\label{equ:sigma_K}
    \sigma^2(k\frac{N}{n}) = s^2(k)\cdot\Big(\frac{N}{n}\Big)^2= (N-K)\frac{K}{N}\frac{N-n}{N-1}
\end{equation}
It is apparent that, as the sample size $n$ increases towards $N$, the variance of $K$ decreases. The intuitive conclusion is that a larger test sample leads to a more accurate estimation of the population contamination.

The description above is valid for all population parameters $N, K$ and sample parameters $n$. However, special care should be taken  if the sample does not contain any artefact; i.e. $k=0$. In this case, \autoref{equ:sigma_K} yields an uncertainty of 0, which is underestimated unless $n=N$.

Instead, the upper limit of the the 95\% confidence interval on the false-positive ratio of the population can be estimated by approximating the hypergeometric distribution with the binomial distribution. This is valid for large $N/n$ values. The upper limit is then given by

\begin{equation}
    \mathrm{Pr}(k=0) = (1-p)^n \overset{!}{\geq}0.05
\end{equation}

If all inspected sources in a sample of size $n>30$ are SSOs, the false-positive ratio is therefore well approximated by $3/n$, which is known as the rule of three \citep{eypasch1995probability}.

To demonstrate the application, the derivation of the false-positive ratio of the J-PLUS survey pipeline runs presented in \autoref{sec:jplus} is given in \autoref{table:jplus_fpr}.

\begin{table}
\centering
\caption{Calculation of sample contamination for the two J-PLUS pipeline runs presented in \autoref{sec:jplus}. FPR refers to the false-positive ratio. Note that for the Default run, the values refer to the expectation and standard deviation, while for the Strict run, the rule of three was used to derive the upper limit of the 95\% confidence interval.}
\label{table:jplus_fpr}
    \begin{tabular}{llllll} \toprule
  & N & n & k & K & FPR\\
 \midrule
  Default & 4,606 & 500 & 10 & 92\,$\pm$\,9 & (2.0\,$\pm$\,0.2)\%\\
  Strict  & 3,696 & 441 & 0  & 0\,$\pm$\,0 & $\leq$\,0.68\%\\\bottomrule
 \end{tabular}
\end{table}
\clearpage
\section{\mono{ssos} pipeline settings file}\label{app:settings}
\begin{verbatim}
# ------
# Image Parameters

SCI_EXTENSION        All
WEIGHT_IMAGES        False
RA                   RA
DEC                  DEC
OBJECT               OBJECT
DATE-OBS             DATE-OBS
FILTER               FILTER
EXPTIME              TEXPTIME

# ------
# SExtractor File Settings

SEX_CONFIG           semp/ssos.sex
SEX_PARAMS           semp/ssos.param
SEX_FILTER           semp/gauss_2.5_5x5.conv
SEX_NNW              semp/default.nnw

# ------
# SCAMP File Settings

SCAMP_CONFIG         semp/ssos.scamp

# ------
# SWARP File Settings

SWARP_CONFIG         semp/ssos.swarp

# ------
# SSO Filter Settings

FILTER_DETEC         True
DETECTIONS           1,2,3

FILTER_PM            True
PM_LOW               0
PM_UP                300
PM_SNR               0.5

FILTER_PIXEL         True
DELTA_PIXEL          2

FILTER_MOTION        True
IDENTIFY_OUTLIER     True
OUTLIER_THRESHOLD    1.5
R_SQU_M              0.95

FILTER_TRAIL         False
RATIO                0.25

FILTER_REF_SOURCES   True
DISTANCE             200
BRIGHT_SOURCES_CAT   semp/hygdata_v3.csv
MAG_LIMITS           -99,99

# ------
# Optional analyses

CROSSMATCH_SKYBOT    True
CROSSMATCH_RADIUS    10
OBSERVATORY_CODE     500
FOV_DIMENSIONS       0.5x0.5

EXTRACT_CUTOUTS      True
CUTOUT_SIZE          256

FIXED_APER_MAGS      False
REFERENCE_FILTER     r
\end{verbatim}
\end{document}

%% file: number_of_detections.pgf
\begingroup%
\makeatletter%
\begin{pgfpicture}%
\pgfpathrectangle{\pgfpointorigin}{\pgfqpoint{3.611457in}{2.232003in}}%
\pgfusepath{use as bounding box, clip}%
\begin{pgfscope}%
\pgfsetbuttcap%
\pgfsetmiterjoin%
\definecolor{currentfill}{rgb}{1.000000,1.000000,1.000000}%
\pgfsetfillcolor{currentfill}%
\pgfsetlinewidth{0.000000pt}%
\definecolor{currentstroke}{rgb}{1.000000,1.000000,1.000000}%
\pgfsetstrokecolor{currentstroke}%
\pgfsetdash{}{0pt}%
\pgfpathmoveto{\pgfqpoint{0.000000in}{0.000000in}}%
\pgfpathlineto{\pgfqpoint{3.611457in}{0.000000in}}%
\pgfpathlineto{\pgfqpoint{3.611457in}{2.232003in}}%
\pgfpathlineto{\pgfqpoint{0.000000in}{2.232003in}}%
\pgfpathclose%
\pgfusepath{fill}%
\end{pgfscope}%
\begin{pgfscope}%
\pgfsetbuttcap%
\pgfsetmiterjoin%
\definecolor{currentfill}{rgb}{1.000000,1.000000,1.000000}%
\pgfsetfillcolor{currentfill}%
\pgfsetlinewidth{0.000000pt}%
\definecolor{currentstroke}{rgb}{0.000000,0.000000,0.000000}%
\pgfsetstrokecolor{currentstroke}%
\pgfsetstrokeopacity{0.000000}%
\pgfsetdash{}{0pt}%
\pgfpathmoveto{\pgfqpoint{0.386510in}{0.446667in}}%
\pgfpathlineto{\pgfqpoint{3.549235in}{0.446667in}}%
\pgfpathlineto{\pgfqpoint{3.549235in}{2.169781in}}%
\pgfpathlineto{\pgfqpoint{0.386510in}{2.169781in}}%
\pgfpathclose%
\pgfusepath{fill}%
\end{pgfscope}%
\begin{pgfscope}%
\pgfsetbuttcap%
\pgfsetroundjoin%
\definecolor{currentfill}{rgb}{0.000000,0.000000,0.000000}%
\pgfsetfillcolor{currentfill}%
\pgfsetlinewidth{0.501875pt}%
\definecolor{currentstroke}{rgb}{0.000000,0.000000,0.000000}%
\pgfsetstrokecolor{currentstroke}%
\pgfsetdash{}{0pt}%
\pgfsys@defobject{currentmarker}{\pgfqpoint{0.000000in}{-0.055556in}}{\pgfqpoint{0.000000in}{0.000000in}}{%
\pgfpathmoveto{\pgfqpoint{0.000000in}{0.000000in}}%
\pgfpathlineto{\pgfqpoint{0.000000in}{-0.055556in}}%
\pgfusepath{stroke,fill}%
}%
\begin{pgfscope}%
\pgfsys@transformshift{0.871396in}{0.446667in}%
\pgfsys@useobject{currentmarker}{}%
\end{pgfscope}%
\end{pgfscope}%
\begin{pgfscope}%
\pgftext[x=0.871396in,y=0.307778in,,top]{\sffamily\fontsize{6.664000}{7.996800}\selectfont \(\displaystyle 2\)}%
\end{pgfscope}%
\begin{pgfscope}%
\pgfsetbuttcap%
\pgfsetroundjoin%
\definecolor{currentfill}{rgb}{0.000000,0.000000,0.000000}%
\pgfsetfillcolor{currentfill}%
\pgfsetlinewidth{0.501875pt}%
\definecolor{currentstroke}{rgb}{0.000000,0.000000,0.000000}%
\pgfsetstrokecolor{currentstroke}%
\pgfsetdash{}{0pt}%
\pgfsys@defobject{currentmarker}{\pgfqpoint{0.000000in}{-0.055556in}}{\pgfqpoint{0.000000in}{0.000000in}}{%
\pgfpathmoveto{\pgfqpoint{0.000000in}{0.000000in}}%
\pgfpathlineto{\pgfqpoint{0.000000in}{-0.055556in}}%
\pgfusepath{stroke,fill}%
}%
\begin{pgfscope}%
\pgfsys@transformshift{1.358719in}{0.446667in}%
\pgfsys@useobject{currentmarker}{}%
\end{pgfscope}%
\end{pgfscope}%
\begin{pgfscope}%
\pgftext[x=1.358719in,y=0.307778in,,top]{\sffamily\fontsize{6.664000}{7.996800}\selectfont \(\displaystyle 4\)}%
\end{pgfscope}%
\begin{pgfscope}%
\pgfsetbuttcap%
\pgfsetroundjoin%
\definecolor{currentfill}{rgb}{0.000000,0.000000,0.000000}%
\pgfsetfillcolor{currentfill}%
\pgfsetlinewidth{0.501875pt}%
\definecolor{currentstroke}{rgb}{0.000000,0.000000,0.000000}%
\pgfsetstrokecolor{currentstroke}%
\pgfsetdash{}{0pt}%
\pgfsys@defobject{currentmarker}{\pgfqpoint{0.000000in}{-0.055556in}}{\pgfqpoint{0.000000in}{0.000000in}}{%
\pgfpathmoveto{\pgfqpoint{0.000000in}{0.000000in}}%
\pgfpathlineto{\pgfqpoint{0.000000in}{-0.055556in}}%
\pgfusepath{stroke,fill}%
}%
\begin{pgfscope}%
\pgfsys@transformshift{1.846042in}{0.446667in}%
\pgfsys@useobject{currentmarker}{}%
\end{pgfscope}%
\end{pgfscope}%
\begin{pgfscope}%
\pgftext[x=1.846042in,y=0.307778in,,top]{\sffamily\fontsize{6.664000}{7.996800}\selectfont \(\displaystyle 6\)}%
\end{pgfscope}%
\begin{pgfscope}%
\pgfsetbuttcap%
\pgfsetroundjoin%
\definecolor{currentfill}{rgb}{0.000000,0.000000,0.000000}%
\pgfsetfillcolor{currentfill}%
\pgfsetlinewidth{0.501875pt}%
\definecolor{currentstroke}{rgb}{0.000000,0.000000,0.000000}%
\pgfsetstrokecolor{currentstroke}%
\pgfsetdash{}{0pt}%
\pgfsys@defobject{currentmarker}{\pgfqpoint{0.000000in}{-0.055556in}}{\pgfqpoint{0.000000in}{0.000000in}}{%
\pgfpathmoveto{\pgfqpoint{0.000000in}{0.000000in}}%
\pgfpathlineto{\pgfqpoint{0.000000in}{-0.055556in}}%
\pgfusepath{stroke,fill}%
}%
\begin{pgfscope}%
\pgfsys@transformshift{2.333364in}{0.446667in}%
\pgfsys@useobject{currentmarker}{}%
\end{pgfscope}%
\end{pgfscope}%
\begin{pgfscope}%
\pgftext[x=2.333364in,y=0.307778in,,top]{\sffamily\fontsize{6.664000}{7.996800}\selectfont \(\displaystyle 8\)}%
\end{pgfscope}%
\begin{pgfscope}%
\pgfsetbuttcap%
\pgfsetroundjoin%
\definecolor{currentfill}{rgb}{0.000000,0.000000,0.000000}%
\pgfsetfillcolor{currentfill}%
\pgfsetlinewidth{0.501875pt}%
\definecolor{currentstroke}{rgb}{0.000000,0.000000,0.000000}%
\pgfsetstrokecolor{currentstroke}%
\pgfsetdash{}{0pt}%
\pgfsys@defobject{currentmarker}{\pgfqpoint{0.000000in}{-0.055556in}}{\pgfqpoint{0.000000in}{0.000000in}}{%
\pgfpathmoveto{\pgfqpoint{0.000000in}{0.000000in}}%
\pgfpathlineto{\pgfqpoint{0.000000in}{-0.055556in}}%
\pgfusepath{stroke,fill}%
}%
\begin{pgfscope}%
\pgfsys@transformshift{2.820687in}{0.446667in}%
\pgfsys@useobject{currentmarker}{}%
\end{pgfscope}%
\end{pgfscope}%
\begin{pgfscope}%
\pgftext[x=2.820687in,y=0.307778in,,top]{\sffamily\fontsize{6.664000}{7.996800}\selectfont \(\displaystyle 10\)}%
\end{pgfscope}%
\begin{pgfscope}%
\pgfsetbuttcap%
\pgfsetroundjoin%
\definecolor{currentfill}{rgb}{0.000000,0.000000,0.000000}%
\pgfsetfillcolor{currentfill}%
\pgfsetlinewidth{0.501875pt}%
\definecolor{currentstroke}{rgb}{0.000000,0.000000,0.000000}%
\pgfsetstrokecolor{currentstroke}%
\pgfsetdash{}{0pt}%
\pgfsys@defobject{currentmarker}{\pgfqpoint{0.000000in}{-0.055556in}}{\pgfqpoint{0.000000in}{0.000000in}}{%
\pgfpathmoveto{\pgfqpoint{0.000000in}{0.000000in}}%
\pgfpathlineto{\pgfqpoint{0.000000in}{-0.055556in}}%
\pgfusepath{stroke,fill}%
}%
\begin{pgfscope}%
\pgfsys@transformshift{3.308010in}{0.446667in}%
\pgfsys@useobject{currentmarker}{}%
\end{pgfscope}%
\end{pgfscope}%
\begin{pgfscope}%
\pgftext[x=3.308010in,y=0.307778in,,top]{\sffamily\fontsize{6.664000}{7.996800}\selectfont \(\displaystyle 12\)}%
\end{pgfscope}%
\begin{pgfscope}%
\pgftext[x=1.967872in,y=0.170033in,,top]{\sffamily\fontsize{9.600000}{11.520000}\selectfont \(\displaystyle N_{\mathrm{Det}}\)}%
\end{pgfscope}%
\begin{pgfscope}%
\pgfsetbuttcap%
\pgfsetroundjoin%
\definecolor{currentfill}{rgb}{0.000000,0.000000,0.000000}%
\pgfsetfillcolor{currentfill}%
\pgfsetlinewidth{0.501875pt}%
\definecolor{currentstroke}{rgb}{0.000000,0.000000,0.000000}%
\pgfsetstrokecolor{currentstroke}%
\pgfsetdash{}{0pt}%
\pgfsys@defobject{currentmarker}{\pgfqpoint{-0.055556in}{0.000000in}}{\pgfqpoint{0.000000in}{0.000000in}}{%
\pgfpathmoveto{\pgfqpoint{0.000000in}{0.000000in}}%
\pgfpathlineto{\pgfqpoint{-0.055556in}{0.000000in}}%
\pgfusepath{stroke,fill}%
}%
\begin{pgfscope}%
\pgfsys@transformshift{0.386510in}{0.446667in}%
\pgfsys@useobject{currentmarker}{}%
\end{pgfscope}%
\end{pgfscope}%
\begin{pgfscope}%
\pgftext[x=0.192258in,y=0.414550in,left,base]{\sffamily\fontsize{6.664000}{7.996800}\selectfont \(\displaystyle 0\)}%
\end{pgfscope}%
\begin{pgfscope}%
\pgfsetbuttcap%
\pgfsetroundjoin%
\definecolor{currentfill}{rgb}{0.000000,0.000000,0.000000}%
\pgfsetfillcolor{currentfill}%
\pgfsetlinewidth{0.501875pt}%
\definecolor{currentstroke}{rgb}{0.000000,0.000000,0.000000}%
\pgfsetstrokecolor{currentstroke}%
\pgfsetdash{}{0pt}%
\pgfsys@defobject{currentmarker}{\pgfqpoint{-0.055556in}{0.000000in}}{\pgfqpoint{0.000000in}{0.000000in}}{%
\pgfpathmoveto{\pgfqpoint{0.000000in}{0.000000in}}%
\pgfpathlineto{\pgfqpoint{-0.055556in}{0.000000in}}%
\pgfusepath{stroke,fill}%
}%
\begin{pgfscope}%
\pgfsys@transformshift{0.386510in}{0.785309in}%
\pgfsys@useobject{currentmarker}{}%
\end{pgfscope}%
\end{pgfscope}%
\begin{pgfscope}%
\pgftext[x=0.081532in,y=0.753192in,left,base]{\sffamily\fontsize{6.664000}{7.996800}\selectfont \(\displaystyle 500\)}%
\end{pgfscope}%
\begin{pgfscope}%
\pgfsetbuttcap%
\pgfsetroundjoin%
\definecolor{currentfill}{rgb}{0.000000,0.000000,0.000000}%
\pgfsetfillcolor{currentfill}%
\pgfsetlinewidth{0.501875pt}%
\definecolor{currentstroke}{rgb}{0.000000,0.000000,0.000000}%
\pgfsetstrokecolor{currentstroke}%
\pgfsetdash{}{0pt}%
\pgfsys@defobject{currentmarker}{\pgfqpoint{-0.055556in}{0.000000in}}{\pgfqpoint{0.000000in}{0.000000in}}{%
\pgfpathmoveto{\pgfqpoint{0.000000in}{0.000000in}}%
\pgfpathlineto{\pgfqpoint{-0.055556in}{0.000000in}}%
\pgfusepath{stroke,fill}%
}%
\begin{pgfscope}%
\pgfsys@transformshift{0.386510in}{1.123952in}%
\pgfsys@useobject{currentmarker}{}%
\end{pgfscope}%
\end{pgfscope}%
\begin{pgfscope}%
\pgftext[x=0.026169in,y=1.091835in,left,base]{\sffamily\fontsize{6.664000}{7.996800}\selectfont \(\displaystyle 1000\)}%
\end{pgfscope}%
\begin{pgfscope}%
\pgfsetbuttcap%
\pgfsetroundjoin%
\definecolor{currentfill}{rgb}{0.000000,0.000000,0.000000}%
\pgfsetfillcolor{currentfill}%
\pgfsetlinewidth{0.501875pt}%
\definecolor{currentstroke}{rgb}{0.000000,0.000000,0.000000}%
\pgfsetstrokecolor{currentstroke}%
\pgfsetdash{}{0pt}%
\pgfsys@defobject{currentmarker}{\pgfqpoint{-0.055556in}{0.000000in}}{\pgfqpoint{0.000000in}{0.000000in}}{%
\pgfpathmoveto{\pgfqpoint{0.000000in}{0.000000in}}%
\pgfpathlineto{\pgfqpoint{-0.055556in}{0.000000in}}%
\pgfusepath{stroke,fill}%
}%
\begin{pgfscope}%
\pgfsys@transformshift{0.386510in}{1.462594in}%
\pgfsys@useobject{currentmarker}{}%
\end{pgfscope}%
\end{pgfscope}%
\begin{pgfscope}%
\pgftext[x=0.026169in,y=1.430477in,left,base]{\sffamily\fontsize{6.664000}{7.996800}\selectfont \(\displaystyle 1500\)}%
\end{pgfscope}%
\begin{pgfscope}%
\pgfsetbuttcap%
\pgfsetroundjoin%
\definecolor{currentfill}{rgb}{0.000000,0.000000,0.000000}%
\pgfsetfillcolor{currentfill}%
\pgfsetlinewidth{0.501875pt}%
\definecolor{currentstroke}{rgb}{0.000000,0.000000,0.000000}%
\pgfsetstrokecolor{currentstroke}%
\pgfsetdash{}{0pt}%
\pgfsys@defobject{currentmarker}{\pgfqpoint{-0.055556in}{0.000000in}}{\pgfqpoint{0.000000in}{0.000000in}}{%
\pgfpathmoveto{\pgfqpoint{0.000000in}{0.000000in}}%
\pgfpathlineto{\pgfqpoint{-0.055556in}{0.000000in}}%
\pgfusepath{stroke,fill}%
}%
\begin{pgfscope}%
\pgfsys@transformshift{0.386510in}{1.801236in}%
\pgfsys@useobject{currentmarker}{}%
\end{pgfscope}%
\end{pgfscope}%
\begin{pgfscope}%
\pgftext[x=0.026169in,y=1.769120in,left,base]{\sffamily\fontsize{6.664000}{7.996800}\selectfont \(\displaystyle 2000\)}%
\end{pgfscope}%
\begin{pgfscope}%
\pgfsetbuttcap%
\pgfsetroundjoin%
\definecolor{currentfill}{rgb}{0.000000,0.000000,0.000000}%
\pgfsetfillcolor{currentfill}%
\pgfsetlinewidth{0.501875pt}%
\definecolor{currentstroke}{rgb}{0.000000,0.000000,0.000000}%
\pgfsetstrokecolor{currentstroke}%
\pgfsetdash{}{0pt}%
\pgfsys@defobject{currentmarker}{\pgfqpoint{-0.055556in}{0.000000in}}{\pgfqpoint{0.000000in}{0.000000in}}{%
\pgfpathmoveto{\pgfqpoint{0.000000in}{0.000000in}}%
\pgfpathlineto{\pgfqpoint{-0.055556in}{0.000000in}}%
\pgfusepath{stroke,fill}%
}%
\begin{pgfscope}%
\pgfsys@transformshift{0.386510in}{2.139879in}%
\pgfsys@useobject{currentmarker}{}%
\end{pgfscope}%
\end{pgfscope}%
\begin{pgfscope}%
\pgftext[x=0.026169in,y=2.107762in,left,base]{\sffamily\fontsize{6.664000}{7.996800}\selectfont \(\displaystyle 2500\)}%
\end{pgfscope}%
\begin{pgfscope}%
\pgfpathrectangle{\pgfqpoint{0.386510in}{0.446667in}}{\pgfqpoint{3.162725in}{1.723114in}}%
\pgfusepath{clip}%
\pgfsetbuttcap%
\pgfsetmiterjoin%
\pgfsetlinewidth{0.501875pt}%
\definecolor{currentstroke}{rgb}{0.501961,0.501961,0.501961}%
\pgfsetstrokecolor{currentstroke}%
\pgfsetdash{}{0pt}%
\pgfpathmoveto{\pgfqpoint{0.530270in}{0.446667in}}%
\pgfpathlineto{\pgfqpoint{0.725199in}{0.446667in}}%
\pgfpathlineto{\pgfqpoint{0.725199in}{2.087728in}}%
\pgfpathlineto{\pgfqpoint{0.530270in}{2.087728in}}%
\pgfpathclose%
\pgfusepath{stroke}%
\end{pgfscope}%
\begin{pgfscope}%
\pgfsetbuttcap%
\pgfsetmiterjoin%
\pgfsetlinewidth{0.501875pt}%
\definecolor{currentstroke}{rgb}{0.501961,0.501961,0.501961}%
\pgfsetstrokecolor{currentstroke}%
\pgfsetdash{}{0pt}%
\pgfpathrectangle{\pgfqpoint{0.386510in}{0.446667in}}{\pgfqpoint{3.162725in}{1.723114in}}%
\pgfusepath{clip}%
\pgfpathmoveto{\pgfqpoint{0.530270in}{0.446667in}}%
\pgfpathlineto{\pgfqpoint{0.725199in}{0.446667in}}%
\pgfpathlineto{\pgfqpoint{0.725199in}{2.087728in}}%
\pgfpathlineto{\pgfqpoint{0.530270in}{2.087728in}}%
\pgfpathclose%
\pgfusepath{clip}%
\pgfsys@defobject{currentpattern}{\pgfqpoint{0in}{0in}}{\pgfqpoint{1in}{1in}}{%
\begin{pgfscope}%
\pgfpathrectangle{\pgfqpoint{0in}{0in}}{\pgfqpoint{1in}{1in}}%
\pgfusepath{clip}%
\pgfpathmoveto{\pgfqpoint{-0.500000in}{0.500000in}}%
\pgfpathlineto{\pgfqpoint{0.500000in}{1.500000in}}%
\pgfpathmoveto{\pgfqpoint{-0.416667in}{0.416667in}}%
\pgfpathlineto{\pgfqpoint{0.583333in}{1.416667in}}%
\pgfpathmoveto{\pgfqpoint{-0.333333in}{0.333333in}}%
\pgfpathlineto{\pgfqpoint{0.666667in}{1.333333in}}%
\pgfpathmoveto{\pgfqpoint{-0.250000in}{0.250000in}}%
\pgfpathlineto{\pgfqpoint{0.750000in}{1.250000in}}%
\pgfpathmoveto{\pgfqpoint{-0.166667in}{0.166667in}}%
\pgfpathlineto{\pgfqpoint{0.833333in}{1.166667in}}%
\pgfpathmoveto{\pgfqpoint{-0.083333in}{0.083333in}}%
\pgfpathlineto{\pgfqpoint{0.916667in}{1.083333in}}%
\pgfpathmoveto{\pgfqpoint{0.000000in}{0.000000in}}%
\pgfpathlineto{\pgfqpoint{1.000000in}{1.000000in}}%
\pgfpathmoveto{\pgfqpoint{0.083333in}{-0.083333in}}%
\pgfpathlineto{\pgfqpoint{1.083333in}{0.916667in}}%
\pgfpathmoveto{\pgfqpoint{0.166667in}{-0.166667in}}%
\pgfpathlineto{\pgfqpoint{1.166667in}{0.833333in}}%
\pgfpathmoveto{\pgfqpoint{0.250000in}{-0.250000in}}%
\pgfpathlineto{\pgfqpoint{1.250000in}{0.750000in}}%
\pgfpathmoveto{\pgfqpoint{0.333333in}{-0.333333in}}%
\pgfpathlineto{\pgfqpoint{1.333333in}{0.666667in}}%
\pgfpathmoveto{\pgfqpoint{0.416667in}{-0.416667in}}%
\pgfpathlineto{\pgfqpoint{1.416667in}{0.583333in}}%
\pgfpathmoveto{\pgfqpoint{0.500000in}{-0.500000in}}%
\pgfpathlineto{\pgfqpoint{1.500000in}{0.500000in}}%
\pgfusepath{stroke}%
\end{pgfscope}%
}%
\pgfsys@transformshift{0.530270in}{0.446667in}%
\pgfsys@useobject{currentpattern}{}%
\pgfsys@transformshift{1in}{0in}%
\pgfsys@transformshift{-1in}{0in}%
\pgfsys@transformshift{0in}{1in}%
\pgfsys@useobject{currentpattern}{}%
\pgfsys@transformshift{1in}{0in}%
\pgfsys@transformshift{-1in}{0in}%
\pgfsys@transformshift{0in}{1in}%
\end{pgfscope}%
\begin{pgfscope}%
\pgfpathrectangle{\pgfqpoint{0.386510in}{0.446667in}}{\pgfqpoint{3.162725in}{1.723114in}}%
\pgfusepath{clip}%
\pgfsetbuttcap%
\pgfsetmiterjoin%
\pgfsetlinewidth{0.501875pt}%
\definecolor{currentstroke}{rgb}{0.501961,0.501961,0.501961}%
\pgfsetstrokecolor{currentstroke}%
\pgfsetdash{}{0pt}%
\pgfpathmoveto{\pgfqpoint{0.773931in}{0.446667in}}%
\pgfpathlineto{\pgfqpoint{0.968860in}{0.446667in}}%
\pgfpathlineto{\pgfqpoint{0.968860in}{1.576378in}}%
\pgfpathlineto{\pgfqpoint{0.773931in}{1.576378in}}%
\pgfpathclose%
\pgfusepath{stroke}%
\end{pgfscope}%
\begin{pgfscope}%
\pgfsetbuttcap%
\pgfsetmiterjoin%
\pgfsetlinewidth{0.501875pt}%
\definecolor{currentstroke}{rgb}{0.501961,0.501961,0.501961}%
\pgfsetstrokecolor{currentstroke}%
\pgfsetdash{}{0pt}%
\pgfpathrectangle{\pgfqpoint{0.386510in}{0.446667in}}{\pgfqpoint{3.162725in}{1.723114in}}%
\pgfusepath{clip}%
\pgfpathmoveto{\pgfqpoint{0.773931in}{0.446667in}}%
\pgfpathlineto{\pgfqpoint{0.968860in}{0.446667in}}%
\pgfpathlineto{\pgfqpoint{0.968860in}{1.576378in}}%
\pgfpathlineto{\pgfqpoint{0.773931in}{1.576378in}}%
\pgfpathclose%
\pgfusepath{clip}%
\pgfsys@defobject{currentpattern}{\pgfqpoint{0in}{0in}}{\pgfqpoint{1in}{1in}}{%
\begin{pgfscope}%
\pgfpathrectangle{\pgfqpoint{0in}{0in}}{\pgfqpoint{1in}{1in}}%
\pgfusepath{clip}%
\pgfpathmoveto{\pgfqpoint{-0.500000in}{0.500000in}}%
\pgfpathlineto{\pgfqpoint{0.500000in}{1.500000in}}%
\pgfpathmoveto{\pgfqpoint{-0.416667in}{0.416667in}}%
\pgfpathlineto{\pgfqpoint{0.583333in}{1.416667in}}%
\pgfpathmoveto{\pgfqpoint{-0.333333in}{0.333333in}}%
\pgfpathlineto{\pgfqpoint{0.666667in}{1.333333in}}%
\pgfpathmoveto{\pgfqpoint{-0.250000in}{0.250000in}}%
\pgfpathlineto{\pgfqpoint{0.750000in}{1.250000in}}%
\pgfpathmoveto{\pgfqpoint{-0.166667in}{0.166667in}}%
\pgfpathlineto{\pgfqpoint{0.833333in}{1.166667in}}%
\pgfpathmoveto{\pgfqpoint{-0.083333in}{0.083333in}}%
\pgfpathlineto{\pgfqpoint{0.916667in}{1.083333in}}%
\pgfpathmoveto{\pgfqpoint{0.000000in}{0.000000in}}%
\pgfpathlineto{\pgfqpoint{1.000000in}{1.000000in}}%
\pgfpathmoveto{\pgfqpoint{0.083333in}{-0.083333in}}%
\pgfpathlineto{\pgfqpoint{1.083333in}{0.916667in}}%
\pgfpathmoveto{\pgfqpoint{0.166667in}{-0.166667in}}%
\pgfpathlineto{\pgfqpoint{1.166667in}{0.833333in}}%
\pgfpathmoveto{\pgfqpoint{0.250000in}{-0.250000in}}%
\pgfpathlineto{\pgfqpoint{1.250000in}{0.750000in}}%
\pgfpathmoveto{\pgfqpoint{0.333333in}{-0.333333in}}%
\pgfpathlineto{\pgfqpoint{1.333333in}{0.666667in}}%
\pgfpathmoveto{\pgfqpoint{0.416667in}{-0.416667in}}%
\pgfpathlineto{\pgfqpoint{1.416667in}{0.583333in}}%
\pgfpathmoveto{\pgfqpoint{0.500000in}{-0.500000in}}%
\pgfpathlineto{\pgfqpoint{1.500000in}{0.500000in}}%
\pgfusepath{stroke}%
\end{pgfscope}%
}%
\pgfsys@transformshift{0.773931in}{0.446667in}%
\pgfsys@useobject{currentpattern}{}%
\pgfsys@transformshift{1in}{0in}%
\pgfsys@transformshift{-1in}{0in}%
\pgfsys@transformshift{0in}{1in}%
\pgfsys@useobject{currentpattern}{}%
\pgfsys@transformshift{1in}{0in}%
\pgfsys@transformshift{-1in}{0in}%
\pgfsys@transformshift{0in}{1in}%
\end{pgfscope}%
\begin{pgfscope}%
\pgfpathrectangle{\pgfqpoint{0.386510in}{0.446667in}}{\pgfqpoint{3.162725in}{1.723114in}}%
\pgfusepath{clip}%
\pgfsetbuttcap%
\pgfsetmiterjoin%
\pgfsetlinewidth{0.501875pt}%
\definecolor{currentstroke}{rgb}{0.501961,0.501961,0.501961}%
\pgfsetstrokecolor{currentstroke}%
\pgfsetdash{}{0pt}%
\pgfpathmoveto{\pgfqpoint{1.017593in}{0.446667in}}%
\pgfpathlineto{\pgfqpoint{1.212522in}{0.446667in}}%
\pgfpathlineto{\pgfqpoint{1.212522in}{1.362356in}}%
\pgfpathlineto{\pgfqpoint{1.017593in}{1.362356in}}%
\pgfpathclose%
\pgfusepath{stroke}%
\end{pgfscope}%
\begin{pgfscope}%
\pgfsetbuttcap%
\pgfsetmiterjoin%
\pgfsetlinewidth{0.501875pt}%
\definecolor{currentstroke}{rgb}{0.501961,0.501961,0.501961}%
\pgfsetstrokecolor{currentstroke}%
\pgfsetdash{}{0pt}%
\pgfpathrectangle{\pgfqpoint{0.386510in}{0.446667in}}{\pgfqpoint{3.162725in}{1.723114in}}%
\pgfusepath{clip}%
\pgfpathmoveto{\pgfqpoint{1.017593in}{0.446667in}}%
\pgfpathlineto{\pgfqpoint{1.212522in}{0.446667in}}%
\pgfpathlineto{\pgfqpoint{1.212522in}{1.362356in}}%
\pgfpathlineto{\pgfqpoint{1.017593in}{1.362356in}}%
\pgfpathclose%
\pgfusepath{clip}%
\pgfsys@defobject{currentpattern}{\pgfqpoint{0in}{0in}}{\pgfqpoint{1in}{1in}}{%
\begin{pgfscope}%
\pgfpathrectangle{\pgfqpoint{0in}{0in}}{\pgfqpoint{1in}{1in}}%
\pgfusepath{clip}%
\pgfpathmoveto{\pgfqpoint{-0.500000in}{0.500000in}}%
\pgfpathlineto{\pgfqpoint{0.500000in}{1.500000in}}%
\pgfpathmoveto{\pgfqpoint{-0.416667in}{0.416667in}}%
\pgfpathlineto{\pgfqpoint{0.583333in}{1.416667in}}%
\pgfpathmoveto{\pgfqpoint{-0.333333in}{0.333333in}}%
\pgfpathlineto{\pgfqpoint{0.666667in}{1.333333in}}%
\pgfpathmoveto{\pgfqpoint{-0.250000in}{0.250000in}}%
\pgfpathlineto{\pgfqpoint{0.750000in}{1.250000in}}%
\pgfpathmoveto{\pgfqpoint{-0.166667in}{0.166667in}}%
\pgfpathlineto{\pgfqpoint{0.833333in}{1.166667in}}%
\pgfpathmoveto{\pgfqpoint{-0.083333in}{0.083333in}}%
\pgfpathlineto{\pgfqpoint{0.916667in}{1.083333in}}%
\pgfpathmoveto{\pgfqpoint{0.000000in}{0.000000in}}%
\pgfpathlineto{\pgfqpoint{1.000000in}{1.000000in}}%
\pgfpathmoveto{\pgfqpoint{0.083333in}{-0.083333in}}%
\pgfpathlineto{\pgfqpoint{1.083333in}{0.916667in}}%
\pgfpathmoveto{\pgfqpoint{0.166667in}{-0.166667in}}%
\pgfpathlineto{\pgfqpoint{1.166667in}{0.833333in}}%
\pgfpathmoveto{\pgfqpoint{0.250000in}{-0.250000in}}%
\pgfpathlineto{\pgfqpoint{1.250000in}{0.750000in}}%
\pgfpathmoveto{\pgfqpoint{0.333333in}{-0.333333in}}%
\pgfpathlineto{\pgfqpoint{1.333333in}{0.666667in}}%
\pgfpathmoveto{\pgfqpoint{0.416667in}{-0.416667in}}%
\pgfpathlineto{\pgfqpoint{1.416667in}{0.583333in}}%
\pgfpathmoveto{\pgfqpoint{0.500000in}{-0.500000in}}%
\pgfpathlineto{\pgfqpoint{1.500000in}{0.500000in}}%
\pgfusepath{stroke}%
\end{pgfscope}%
}%
\pgfsys@transformshift{1.017593in}{0.446667in}%
\pgfsys@useobject{currentpattern}{}%
\pgfsys@transformshift{1in}{0in}%
\pgfsys@transformshift{-1in}{0in}%
\pgfsys@transformshift{0in}{1in}%
\end{pgfscope}%
\begin{pgfscope}%
\pgfpathrectangle{\pgfqpoint{0.386510in}{0.446667in}}{\pgfqpoint{3.162725in}{1.723114in}}%
\pgfusepath{clip}%
\pgfsetbuttcap%
\pgfsetmiterjoin%
\pgfsetlinewidth{0.501875pt}%
\definecolor{currentstroke}{rgb}{0.000000,0.000000,0.000000}%
\pgfsetstrokecolor{currentstroke}%
\pgfsetdash{}{0pt}%
\pgfpathmoveto{\pgfqpoint{1.261254in}{0.446667in}}%
\pgfpathlineto{\pgfqpoint{1.456183in}{0.446667in}}%
\pgfpathlineto{\pgfqpoint{1.456183in}{1.176780in}}%
\pgfpathlineto{\pgfqpoint{1.261254in}{1.176780in}}%
\pgfpathclose%
\pgfusepath{stroke}%
\end{pgfscope}%
\begin{pgfscope}%
\pgfpathrectangle{\pgfqpoint{0.386510in}{0.446667in}}{\pgfqpoint{3.162725in}{1.723114in}}%
\pgfusepath{clip}%
\pgfsetbuttcap%
\pgfsetmiterjoin%
\pgfsetlinewidth{0.501875pt}%
\definecolor{currentstroke}{rgb}{0.000000,0.000000,0.000000}%
\pgfsetstrokecolor{currentstroke}%
\pgfsetdash{}{0pt}%
\pgfpathmoveto{\pgfqpoint{1.504916in}{0.446667in}}%
\pgfpathlineto{\pgfqpoint{1.699845in}{0.446667in}}%
\pgfpathlineto{\pgfqpoint{1.699845in}{2.078923in}}%
\pgfpathlineto{\pgfqpoint{1.504916in}{2.078923in}}%
\pgfpathclose%
\pgfusepath{stroke}%
\end{pgfscope}%
\begin{pgfscope}%
\pgfpathrectangle{\pgfqpoint{0.386510in}{0.446667in}}{\pgfqpoint{3.162725in}{1.723114in}}%
\pgfusepath{clip}%
\pgfsetbuttcap%
\pgfsetmiterjoin%
\pgfsetlinewidth{0.501875pt}%
\definecolor{currentstroke}{rgb}{0.000000,0.000000,0.000000}%
\pgfsetstrokecolor{currentstroke}%
\pgfsetdash{}{0pt}%
\pgfpathmoveto{\pgfqpoint{1.748577in}{0.446667in}}%
\pgfpathlineto{\pgfqpoint{1.943506in}{0.446667in}}%
\pgfpathlineto{\pgfqpoint{1.943506in}{1.093474in}}%
\pgfpathlineto{\pgfqpoint{1.748577in}{1.093474in}}%
\pgfpathclose%
\pgfusepath{stroke}%
\end{pgfscope}%
\begin{pgfscope}%
\pgfpathrectangle{\pgfqpoint{0.386510in}{0.446667in}}{\pgfqpoint{3.162725in}{1.723114in}}%
\pgfusepath{clip}%
\pgfsetbuttcap%
\pgfsetmiterjoin%
\pgfsetlinewidth{0.501875pt}%
\definecolor{currentstroke}{rgb}{0.000000,0.000000,0.000000}%
\pgfsetstrokecolor{currentstroke}%
\pgfsetdash{}{0pt}%
\pgfpathmoveto{\pgfqpoint{1.992238in}{0.446667in}}%
\pgfpathlineto{\pgfqpoint{2.187167in}{0.446667in}}%
\pgfpathlineto{\pgfqpoint{2.187167in}{1.356260in}}%
\pgfpathlineto{\pgfqpoint{1.992238in}{1.356260in}}%
\pgfpathclose%
\pgfusepath{stroke}%
\end{pgfscope}%
\begin{pgfscope}%
\pgfpathrectangle{\pgfqpoint{0.386510in}{0.446667in}}{\pgfqpoint{3.162725in}{1.723114in}}%
\pgfusepath{clip}%
\pgfsetbuttcap%
\pgfsetmiterjoin%
\pgfsetlinewidth{0.501875pt}%
\definecolor{currentstroke}{rgb}{0.000000,0.000000,0.000000}%
\pgfsetstrokecolor{currentstroke}%
\pgfsetdash{}{0pt}%
\pgfpathmoveto{\pgfqpoint{2.235900in}{0.446667in}}%
\pgfpathlineto{\pgfqpoint{2.430829in}{0.446667in}}%
\pgfpathlineto{\pgfqpoint{2.430829in}{0.606506in}}%
\pgfpathlineto{\pgfqpoint{2.235900in}{0.606506in}}%
\pgfpathclose%
\pgfusepath{stroke}%
\end{pgfscope}%
\begin{pgfscope}%
\pgfpathrectangle{\pgfqpoint{0.386510in}{0.446667in}}{\pgfqpoint{3.162725in}{1.723114in}}%
\pgfusepath{clip}%
\pgfsetbuttcap%
\pgfsetmiterjoin%
\pgfsetlinewidth{0.501875pt}%
\definecolor{currentstroke}{rgb}{0.000000,0.000000,0.000000}%
\pgfsetstrokecolor{currentstroke}%
\pgfsetdash{}{0pt}%
\pgfpathmoveto{\pgfqpoint{2.479561in}{0.446667in}}%
\pgfpathlineto{\pgfqpoint{2.674490in}{0.446667in}}%
\pgfpathlineto{\pgfqpoint{2.674490in}{0.842201in}}%
\pgfpathlineto{\pgfqpoint{2.479561in}{0.842201in}}%
\pgfpathclose%
\pgfusepath{stroke}%
\end{pgfscope}%
\begin{pgfscope}%
\pgfpathrectangle{\pgfqpoint{0.386510in}{0.446667in}}{\pgfqpoint{3.162725in}{1.723114in}}%
\pgfusepath{clip}%
\pgfsetbuttcap%
\pgfsetmiterjoin%
\pgfsetlinewidth{0.501875pt}%
\definecolor{currentstroke}{rgb}{0.000000,0.000000,0.000000}%
\pgfsetstrokecolor{currentstroke}%
\pgfsetdash{}{0pt}%
\pgfpathmoveto{\pgfqpoint{2.723223in}{0.446667in}}%
\pgfpathlineto{\pgfqpoint{2.918152in}{0.446667in}}%
\pgfpathlineto{\pgfqpoint{2.918152in}{0.574674in}}%
\pgfpathlineto{\pgfqpoint{2.723223in}{0.574674in}}%
\pgfpathclose%
\pgfusepath{stroke}%
\end{pgfscope}%
\begin{pgfscope}%
\pgfpathrectangle{\pgfqpoint{0.386510in}{0.446667in}}{\pgfqpoint{3.162725in}{1.723114in}}%
\pgfusepath{clip}%
\pgfsetbuttcap%
\pgfsetmiterjoin%
\pgfsetlinewidth{0.501875pt}%
\definecolor{currentstroke}{rgb}{0.000000,0.000000,0.000000}%
\pgfsetstrokecolor{currentstroke}%
\pgfsetdash{}{0pt}%
\pgfpathmoveto{\pgfqpoint{2.966884in}{0.446667in}}%
\pgfpathlineto{\pgfqpoint{3.161813in}{0.446667in}}%
\pgfpathlineto{\pgfqpoint{3.161813in}{0.582801in}}%
\pgfpathlineto{\pgfqpoint{2.966884in}{0.582801in}}%
\pgfpathclose%
\pgfusepath{stroke}%
\end{pgfscope}%
\begin{pgfscope}%
\pgfpathrectangle{\pgfqpoint{0.386510in}{0.446667in}}{\pgfqpoint{3.162725in}{1.723114in}}%
\pgfusepath{clip}%
\pgfsetbuttcap%
\pgfsetmiterjoin%
\pgfsetlinewidth{0.501875pt}%
\definecolor{currentstroke}{rgb}{0.000000,0.000000,0.000000}%
\pgfsetstrokecolor{currentstroke}%
\pgfsetdash{}{0pt}%
\pgfpathmoveto{\pgfqpoint{3.210545in}{0.446667in}}%
\pgfpathlineto{\pgfqpoint{3.405475in}{0.446667in}}%
\pgfpathlineto{\pgfqpoint{3.405475in}{1.266181in}}%
\pgfpathlineto{\pgfqpoint{3.210545in}{1.266181in}}%
\pgfpathclose%
\pgfusepath{stroke}%
\end{pgfscope}%
\begin{pgfscope}%
\pgfsetrectcap%
\pgfsetmiterjoin%
\pgfsetlinewidth{0.501875pt}%
\definecolor{currentstroke}{rgb}{0.000000,0.000000,0.000000}%
\pgfsetstrokecolor{currentstroke}%
\pgfsetdash{}{0pt}%
\pgfpathmoveto{\pgfqpoint{0.386510in}{0.446667in}}%
\pgfpathlineto{\pgfqpoint{0.386510in}{2.169781in}}%
\pgfusepath{stroke}%
\end{pgfscope}%
\begin{pgfscope}%
\pgfsetrectcap%
\pgfsetmiterjoin%
\pgfsetlinewidth{0.501875pt}%
\definecolor{currentstroke}{rgb}{0.000000,0.000000,0.000000}%
\pgfsetstrokecolor{currentstroke}%
\pgfsetdash{}{0pt}%
\pgfpathmoveto{\pgfqpoint{0.386510in}{0.446667in}}%
\pgfpathlineto{\pgfqpoint{3.549235in}{0.446667in}}%
\pgfusepath{stroke}%
\end{pgfscope}%
\begin{pgfscope}%
\pgftext[x=3.391099in,y=2.083625in,right,base]{\sffamily\fontsize{8.000000}{9.600000}\selectfont \texttt{DETECTIONS\,=\,1,2,3}}%
\end{pgfscope}%
\end{pgfpicture}%
\makeatother%
\endgroup%

%% file: bad_pixel_filter.pgf
\begingroup%
\makeatletter%
\begin{pgfpicture}%
\pgfpathrectangle{\pgfpointorigin}{\pgfqpoint{3.611457in}{2.232003in}}%
\pgfusepath{use as bounding box, clip}%
\begin{pgfscope}%
\pgfsetbuttcap%
\pgfsetmiterjoin%
\definecolor{currentfill}{rgb}{1.000000,1.000000,1.000000}%
\pgfsetfillcolor{currentfill}%
\pgfsetlinewidth{0.000000pt}%
\definecolor{currentstroke}{rgb}{1.000000,1.000000,1.000000}%
\pgfsetstrokecolor{currentstroke}%
\pgfsetdash{}{0pt}%
\pgfpathmoveto{\pgfqpoint{0.000000in}{0.000000in}}%
\pgfpathlineto{\pgfqpoint{3.611457in}{0.000000in}}%
\pgfpathlineto{\pgfqpoint{3.611457in}{2.232003in}}%
\pgfpathlineto{\pgfqpoint{0.000000in}{2.232003in}}%
\pgfpathclose%
\pgfusepath{fill}%
\end{pgfscope}%
\begin{pgfscope}%
\pgfsetbuttcap%
\pgfsetmiterjoin%
\definecolor{currentfill}{rgb}{1.000000,1.000000,1.000000}%
\pgfsetfillcolor{currentfill}%
\pgfsetlinewidth{0.000000pt}%
\definecolor{currentstroke}{rgb}{0.000000,0.000000,0.000000}%
\pgfsetstrokecolor{currentstroke}%
\pgfsetstrokeopacity{0.000000}%
\pgfsetdash{}{0pt}%
\pgfpathmoveto{\pgfqpoint{0.301111in}{0.301111in}}%
\pgfpathlineto{\pgfqpoint{3.589235in}{0.301111in}}%
\pgfpathlineto{\pgfqpoint{3.589235in}{2.209781in}}%
\pgfpathlineto{\pgfqpoint{0.301111in}{2.209781in}}%
\pgfpathclose%
\pgfusepath{fill}%
\end{pgfscope}%
\begin{pgfscope}%
\pgfpathrectangle{\pgfqpoint{0.301111in}{0.301111in}}{\pgfqpoint{3.288124in}{1.908670in}}%
\pgfusepath{clip}%
\pgfsetbuttcap%
\pgfsetroundjoin%
\pgfsetlinewidth{0.200750pt}%
\definecolor{currentstroke}{rgb}{0.501961,0.501961,0.501961}%
\pgfsetstrokecolor{currentstroke}%
\pgfsetdash{{0.200000pt}{0.330000pt}}{0.000000pt}%
\pgfpathmoveto{\pgfqpoint{0.301111in}{0.301111in}}%
\pgfpathlineto{\pgfqpoint{0.301111in}{2.209781in}}%
\pgfusepath{stroke}%
\end{pgfscope}%
\begin{pgfscope}%
\pgfpathrectangle{\pgfqpoint{0.301111in}{0.301111in}}{\pgfqpoint{3.288124in}{1.908670in}}%
\pgfusepath{clip}%
\pgfsetbuttcap%
\pgfsetroundjoin%
\pgfsetlinewidth{0.200750pt}%
\definecolor{currentstroke}{rgb}{0.501961,0.501961,0.501961}%
\pgfsetstrokecolor{currentstroke}%
\pgfsetdash{{0.200000pt}{0.330000pt}}{0.000000pt}%
\pgfpathmoveto{\pgfqpoint{1.204299in}{0.301111in}}%
\pgfpathlineto{\pgfqpoint{1.204299in}{2.209781in}}%
\pgfusepath{stroke}%
\end{pgfscope}%
\begin{pgfscope}%
\pgfpathrectangle{\pgfqpoint{0.301111in}{0.301111in}}{\pgfqpoint{3.288124in}{1.908670in}}%
\pgfusepath{clip}%
\pgfsetbuttcap%
\pgfsetroundjoin%
\pgfsetlinewidth{0.200750pt}%
\definecolor{currentstroke}{rgb}{0.501961,0.501961,0.501961}%
\pgfsetstrokecolor{currentstroke}%
\pgfsetdash{{0.200000pt}{0.330000pt}}{0.000000pt}%
\pgfpathmoveto{\pgfqpoint{2.107486in}{0.301111in}}%
\pgfpathlineto{\pgfqpoint{2.107486in}{2.209781in}}%
\pgfusepath{stroke}%
\end{pgfscope}%
\begin{pgfscope}%
\pgfpathrectangle{\pgfqpoint{0.301111in}{0.301111in}}{\pgfqpoint{3.288124in}{1.908670in}}%
\pgfusepath{clip}%
\pgfsetbuttcap%
\pgfsetroundjoin%
\pgfsetlinewidth{0.200750pt}%
\definecolor{currentstroke}{rgb}{0.501961,0.501961,0.501961}%
\pgfsetstrokecolor{currentstroke}%
\pgfsetdash{{0.200000pt}{0.330000pt}}{0.000000pt}%
\pgfpathmoveto{\pgfqpoint{3.010674in}{0.301111in}}%
\pgfpathlineto{\pgfqpoint{3.010674in}{2.209781in}}%
\pgfusepath{stroke}%
\end{pgfscope}%
\begin{pgfscope}%
\pgfpathrectangle{\pgfqpoint{0.301111in}{0.301111in}}{\pgfqpoint{3.288124in}{1.908670in}}%
\pgfusepath{clip}%
\pgfsetbuttcap%
\pgfsetroundjoin%
\pgfsetlinewidth{0.200750pt}%
\definecolor{currentstroke}{rgb}{0.501961,0.501961,0.501961}%
\pgfsetstrokecolor{currentstroke}%
\pgfsetdash{{0.200000pt}{0.330000pt}}{0.000000pt}%
\pgfpathmoveto{\pgfqpoint{0.301111in}{0.301111in}}%
\pgfpathlineto{\pgfqpoint{0.301111in}{2.209781in}}%
\pgfusepath{stroke}%
\end{pgfscope}%
\begin{pgfscope}%
\pgfpathrectangle{\pgfqpoint{0.301111in}{0.301111in}}{\pgfqpoint{3.288124in}{1.908670in}}%
\pgfusepath{clip}%
\pgfsetbuttcap%
\pgfsetroundjoin%
\pgfsetlinewidth{0.200750pt}%
\definecolor{currentstroke}{rgb}{0.501961,0.501961,0.501961}%
\pgfsetstrokecolor{currentstroke}%
\pgfsetdash{{0.200000pt}{0.330000pt}}{0.000000pt}%
\pgfpathmoveto{\pgfqpoint{0.526908in}{0.301111in}}%
\pgfpathlineto{\pgfqpoint{0.526908in}{2.209781in}}%
\pgfusepath{stroke}%
\end{pgfscope}%
\begin{pgfscope}%
\pgfpathrectangle{\pgfqpoint{0.301111in}{0.301111in}}{\pgfqpoint{3.288124in}{1.908670in}}%
\pgfusepath{clip}%
\pgfsetbuttcap%
\pgfsetroundjoin%
\pgfsetlinewidth{0.200750pt}%
\definecolor{currentstroke}{rgb}{0.501961,0.501961,0.501961}%
\pgfsetstrokecolor{currentstroke}%
\pgfsetdash{{0.200000pt}{0.330000pt}}{0.000000pt}%
\pgfpathmoveto{\pgfqpoint{0.752705in}{0.301111in}}%
\pgfpathlineto{\pgfqpoint{0.752705in}{2.209781in}}%
\pgfusepath{stroke}%
\end{pgfscope}%
\begin{pgfscope}%
\pgfpathrectangle{\pgfqpoint{0.301111in}{0.301111in}}{\pgfqpoint{3.288124in}{1.908670in}}%
\pgfusepath{clip}%
\pgfsetbuttcap%
\pgfsetroundjoin%
\pgfsetlinewidth{0.200750pt}%
\definecolor{currentstroke}{rgb}{0.501961,0.501961,0.501961}%
\pgfsetstrokecolor{currentstroke}%
\pgfsetdash{{0.200000pt}{0.330000pt}}{0.000000pt}%
\pgfpathmoveto{\pgfqpoint{0.978502in}{0.301111in}}%
\pgfpathlineto{\pgfqpoint{0.978502in}{2.209781in}}%
\pgfusepath{stroke}%
\end{pgfscope}%
\begin{pgfscope}%
\pgfpathrectangle{\pgfqpoint{0.301111in}{0.301111in}}{\pgfqpoint{3.288124in}{1.908670in}}%
\pgfusepath{clip}%
\pgfsetbuttcap%
\pgfsetroundjoin%
\pgfsetlinewidth{0.200750pt}%
\definecolor{currentstroke}{rgb}{0.501961,0.501961,0.501961}%
\pgfsetstrokecolor{currentstroke}%
\pgfsetdash{{0.200000pt}{0.330000pt}}{0.000000pt}%
\pgfpathmoveto{\pgfqpoint{1.204299in}{0.301111in}}%
\pgfpathlineto{\pgfqpoint{1.204299in}{2.209781in}}%
\pgfusepath{stroke}%
\end{pgfscope}%
\begin{pgfscope}%
\pgfpathrectangle{\pgfqpoint{0.301111in}{0.301111in}}{\pgfqpoint{3.288124in}{1.908670in}}%
\pgfusepath{clip}%
\pgfsetbuttcap%
\pgfsetroundjoin%
\pgfsetlinewidth{0.200750pt}%
\definecolor{currentstroke}{rgb}{0.501961,0.501961,0.501961}%
\pgfsetstrokecolor{currentstroke}%
\pgfsetdash{{0.200000pt}{0.330000pt}}{0.000000pt}%
\pgfpathmoveto{\pgfqpoint{1.430096in}{0.301111in}}%
\pgfpathlineto{\pgfqpoint{1.430096in}{2.209781in}}%
\pgfusepath{stroke}%
\end{pgfscope}%
\begin{pgfscope}%
\pgfpathrectangle{\pgfqpoint{0.301111in}{0.301111in}}{\pgfqpoint{3.288124in}{1.908670in}}%
\pgfusepath{clip}%
\pgfsetbuttcap%
\pgfsetroundjoin%
\pgfsetlinewidth{0.200750pt}%
\definecolor{currentstroke}{rgb}{0.501961,0.501961,0.501961}%
\pgfsetstrokecolor{currentstroke}%
\pgfsetdash{{0.200000pt}{0.330000pt}}{0.000000pt}%
\pgfpathmoveto{\pgfqpoint{1.655893in}{0.301111in}}%
\pgfpathlineto{\pgfqpoint{1.655893in}{2.209781in}}%
\pgfusepath{stroke}%
\end{pgfscope}%
\begin{pgfscope}%
\pgfpathrectangle{\pgfqpoint{0.301111in}{0.301111in}}{\pgfqpoint{3.288124in}{1.908670in}}%
\pgfusepath{clip}%
\pgfsetbuttcap%
\pgfsetroundjoin%
\pgfsetlinewidth{0.200750pt}%
\definecolor{currentstroke}{rgb}{0.501961,0.501961,0.501961}%
\pgfsetstrokecolor{currentstroke}%
\pgfsetdash{{0.200000pt}{0.330000pt}}{0.000000pt}%
\pgfpathmoveto{\pgfqpoint{1.881689in}{0.301111in}}%
\pgfpathlineto{\pgfqpoint{1.881689in}{2.209781in}}%
\pgfusepath{stroke}%
\end{pgfscope}%
\begin{pgfscope}%
\pgfpathrectangle{\pgfqpoint{0.301111in}{0.301111in}}{\pgfqpoint{3.288124in}{1.908670in}}%
\pgfusepath{clip}%
\pgfsetbuttcap%
\pgfsetroundjoin%
\pgfsetlinewidth{0.200750pt}%
\definecolor{currentstroke}{rgb}{0.501961,0.501961,0.501961}%
\pgfsetstrokecolor{currentstroke}%
\pgfsetdash{{0.200000pt}{0.330000pt}}{0.000000pt}%
\pgfpathmoveto{\pgfqpoint{2.107486in}{0.301111in}}%
\pgfpathlineto{\pgfqpoint{2.107486in}{2.209781in}}%
\pgfusepath{stroke}%
\end{pgfscope}%
\begin{pgfscope}%
\pgfpathrectangle{\pgfqpoint{0.301111in}{0.301111in}}{\pgfqpoint{3.288124in}{1.908670in}}%
\pgfusepath{clip}%
\pgfsetbuttcap%
\pgfsetroundjoin%
\pgfsetlinewidth{0.200750pt}%
\definecolor{currentstroke}{rgb}{0.501961,0.501961,0.501961}%
\pgfsetstrokecolor{currentstroke}%
\pgfsetdash{{0.200000pt}{0.330000pt}}{0.000000pt}%
\pgfpathmoveto{\pgfqpoint{2.333283in}{0.301111in}}%
\pgfpathlineto{\pgfqpoint{2.333283in}{2.209781in}}%
\pgfusepath{stroke}%
\end{pgfscope}%
\begin{pgfscope}%
\pgfpathrectangle{\pgfqpoint{0.301111in}{0.301111in}}{\pgfqpoint{3.288124in}{1.908670in}}%
\pgfusepath{clip}%
\pgfsetbuttcap%
\pgfsetroundjoin%
\pgfsetlinewidth{0.200750pt}%
\definecolor{currentstroke}{rgb}{0.501961,0.501961,0.501961}%
\pgfsetstrokecolor{currentstroke}%
\pgfsetdash{{0.200000pt}{0.330000pt}}{0.000000pt}%
\pgfpathmoveto{\pgfqpoint{2.559080in}{0.301111in}}%
\pgfpathlineto{\pgfqpoint{2.559080in}{2.209781in}}%
\pgfusepath{stroke}%
\end{pgfscope}%
\begin{pgfscope}%
\pgfpathrectangle{\pgfqpoint{0.301111in}{0.301111in}}{\pgfqpoint{3.288124in}{1.908670in}}%
\pgfusepath{clip}%
\pgfsetbuttcap%
\pgfsetroundjoin%
\pgfsetlinewidth{0.200750pt}%
\definecolor{currentstroke}{rgb}{0.501961,0.501961,0.501961}%
\pgfsetstrokecolor{currentstroke}%
\pgfsetdash{{0.200000pt}{0.330000pt}}{0.000000pt}%
\pgfpathmoveto{\pgfqpoint{2.784877in}{0.301111in}}%
\pgfpathlineto{\pgfqpoint{2.784877in}{2.209781in}}%
\pgfusepath{stroke}%
\end{pgfscope}%
\begin{pgfscope}%
\pgfpathrectangle{\pgfqpoint{0.301111in}{0.301111in}}{\pgfqpoint{3.288124in}{1.908670in}}%
\pgfusepath{clip}%
\pgfsetbuttcap%
\pgfsetroundjoin%
\pgfsetlinewidth{0.200750pt}%
\definecolor{currentstroke}{rgb}{0.501961,0.501961,0.501961}%
\pgfsetstrokecolor{currentstroke}%
\pgfsetdash{{0.200000pt}{0.330000pt}}{0.000000pt}%
\pgfpathmoveto{\pgfqpoint{3.010674in}{0.301111in}}%
\pgfpathlineto{\pgfqpoint{3.010674in}{2.209781in}}%
\pgfusepath{stroke}%
\end{pgfscope}%
\begin{pgfscope}%
\pgfpathrectangle{\pgfqpoint{0.301111in}{0.301111in}}{\pgfqpoint{3.288124in}{1.908670in}}%
\pgfusepath{clip}%
\pgfsetbuttcap%
\pgfsetroundjoin%
\pgfsetlinewidth{0.200750pt}%
\definecolor{currentstroke}{rgb}{0.501961,0.501961,0.501961}%
\pgfsetstrokecolor{currentstroke}%
\pgfsetdash{{0.200000pt}{0.330000pt}}{0.000000pt}%
\pgfpathmoveto{\pgfqpoint{3.236471in}{0.301111in}}%
\pgfpathlineto{\pgfqpoint{3.236471in}{2.209781in}}%
\pgfusepath{stroke}%
\end{pgfscope}%
\begin{pgfscope}%
\pgfpathrectangle{\pgfqpoint{0.301111in}{0.301111in}}{\pgfqpoint{3.288124in}{1.908670in}}%
\pgfusepath{clip}%
\pgfsetbuttcap%
\pgfsetroundjoin%
\pgfsetlinewidth{0.200750pt}%
\definecolor{currentstroke}{rgb}{0.501961,0.501961,0.501961}%
\pgfsetstrokecolor{currentstroke}%
\pgfsetdash{{0.200000pt}{0.330000pt}}{0.000000pt}%
\pgfpathmoveto{\pgfqpoint{3.462268in}{0.301111in}}%
\pgfpathlineto{\pgfqpoint{3.462268in}{2.209781in}}%
\pgfusepath{stroke}%
\end{pgfscope}%
\begin{pgfscope}%
\pgftext[x=1.945173in,y=0.162222in,,top]{\sffamily\fontsize{9.600000}{11.520000}\selectfont \texttt{XWIN\_IMAGE}}%
\end{pgfscope}%
\begin{pgfscope}%
\pgfpathrectangle{\pgfqpoint{0.301111in}{0.301111in}}{\pgfqpoint{3.288124in}{1.908670in}}%
\pgfusepath{clip}%
\pgfsetbuttcap%
\pgfsetroundjoin%
\pgfsetlinewidth{0.200750pt}%
\definecolor{currentstroke}{rgb}{0.501961,0.501961,0.501961}%
\pgfsetstrokecolor{currentstroke}%
\pgfsetdash{{0.200000pt}{0.330000pt}}{0.000000pt}%
\pgfpathmoveto{\pgfqpoint{0.301111in}{0.301111in}}%
\pgfpathlineto{\pgfqpoint{3.589235in}{0.301111in}}%
\pgfusepath{stroke}%
\end{pgfscope}%
\begin{pgfscope}%
\pgfpathrectangle{\pgfqpoint{0.301111in}{0.301111in}}{\pgfqpoint{3.288124in}{1.908670in}}%
\pgfusepath{clip}%
\pgfsetbuttcap%
\pgfsetroundjoin%
\pgfsetlinewidth{0.200750pt}%
\definecolor{currentstroke}{rgb}{0.501961,0.501961,0.501961}%
\pgfsetstrokecolor{currentstroke}%
\pgfsetdash{{0.200000pt}{0.330000pt}}{0.000000pt}%
\pgfpathmoveto{\pgfqpoint{0.301111in}{1.149409in}}%
\pgfpathlineto{\pgfqpoint{3.589235in}{1.149409in}}%
\pgfusepath{stroke}%
\end{pgfscope}%
\begin{pgfscope}%
\pgfpathrectangle{\pgfqpoint{0.301111in}{0.301111in}}{\pgfqpoint{3.288124in}{1.908670in}}%
\pgfusepath{clip}%
\pgfsetbuttcap%
\pgfsetroundjoin%
\pgfsetlinewidth{0.200750pt}%
\definecolor{currentstroke}{rgb}{0.501961,0.501961,0.501961}%
\pgfsetstrokecolor{currentstroke}%
\pgfsetdash{{0.200000pt}{0.330000pt}}{0.000000pt}%
\pgfpathmoveto{\pgfqpoint{0.301111in}{1.997707in}}%
\pgfpathlineto{\pgfqpoint{3.589235in}{1.997707in}}%
\pgfusepath{stroke}%
\end{pgfscope}%
\begin{pgfscope}%
\pgfpathrectangle{\pgfqpoint{0.301111in}{0.301111in}}{\pgfqpoint{3.288124in}{1.908670in}}%
\pgfusepath{clip}%
\pgfsetbuttcap%
\pgfsetroundjoin%
\pgfsetlinewidth{0.200750pt}%
\definecolor{currentstroke}{rgb}{0.501961,0.501961,0.501961}%
\pgfsetstrokecolor{currentstroke}%
\pgfsetdash{{0.200000pt}{0.330000pt}}{0.000000pt}%
\pgfpathmoveto{\pgfqpoint{0.301111in}{0.301111in}}%
\pgfpathlineto{\pgfqpoint{3.589235in}{0.301111in}}%
\pgfusepath{stroke}%
\end{pgfscope}%
\begin{pgfscope}%
\pgfpathrectangle{\pgfqpoint{0.301111in}{0.301111in}}{\pgfqpoint{3.288124in}{1.908670in}}%
\pgfusepath{clip}%
\pgfsetbuttcap%
\pgfsetroundjoin%
\pgfsetlinewidth{0.200750pt}%
\definecolor{currentstroke}{rgb}{0.501961,0.501961,0.501961}%
\pgfsetstrokecolor{currentstroke}%
\pgfsetdash{{0.200000pt}{0.330000pt}}{0.000000pt}%
\pgfpathmoveto{\pgfqpoint{0.301111in}{0.513186in}}%
\pgfpathlineto{\pgfqpoint{3.589235in}{0.513186in}}%
\pgfusepath{stroke}%
\end{pgfscope}%
\begin{pgfscope}%
\pgfpathrectangle{\pgfqpoint{0.301111in}{0.301111in}}{\pgfqpoint{3.288124in}{1.908670in}}%
\pgfusepath{clip}%
\pgfsetbuttcap%
\pgfsetroundjoin%
\pgfsetlinewidth{0.200750pt}%
\definecolor{currentstroke}{rgb}{0.501961,0.501961,0.501961}%
\pgfsetstrokecolor{currentstroke}%
\pgfsetdash{{0.200000pt}{0.330000pt}}{0.000000pt}%
\pgfpathmoveto{\pgfqpoint{0.301111in}{0.725260in}}%
\pgfpathlineto{\pgfqpoint{3.589235in}{0.725260in}}%
\pgfusepath{stroke}%
\end{pgfscope}%
\begin{pgfscope}%
\pgfpathrectangle{\pgfqpoint{0.301111in}{0.301111in}}{\pgfqpoint{3.288124in}{1.908670in}}%
\pgfusepath{clip}%
\pgfsetbuttcap%
\pgfsetroundjoin%
\pgfsetlinewidth{0.200750pt}%
\definecolor{currentstroke}{rgb}{0.501961,0.501961,0.501961}%
\pgfsetstrokecolor{currentstroke}%
\pgfsetdash{{0.200000pt}{0.330000pt}}{0.000000pt}%
\pgfpathmoveto{\pgfqpoint{0.301111in}{0.937334in}}%
\pgfpathlineto{\pgfqpoint{3.589235in}{0.937334in}}%
\pgfusepath{stroke}%
\end{pgfscope}%
\begin{pgfscope}%
\pgfpathrectangle{\pgfqpoint{0.301111in}{0.301111in}}{\pgfqpoint{3.288124in}{1.908670in}}%
\pgfusepath{clip}%
\pgfsetbuttcap%
\pgfsetroundjoin%
\pgfsetlinewidth{0.200750pt}%
\definecolor{currentstroke}{rgb}{0.501961,0.501961,0.501961}%
\pgfsetstrokecolor{currentstroke}%
\pgfsetdash{{0.200000pt}{0.330000pt}}{0.000000pt}%
\pgfpathmoveto{\pgfqpoint{0.301111in}{1.149409in}}%
\pgfpathlineto{\pgfqpoint{3.589235in}{1.149409in}}%
\pgfusepath{stroke}%
\end{pgfscope}%
\begin{pgfscope}%
\pgfpathrectangle{\pgfqpoint{0.301111in}{0.301111in}}{\pgfqpoint{3.288124in}{1.908670in}}%
\pgfusepath{clip}%
\pgfsetbuttcap%
\pgfsetroundjoin%
\pgfsetlinewidth{0.200750pt}%
\definecolor{currentstroke}{rgb}{0.501961,0.501961,0.501961}%
\pgfsetstrokecolor{currentstroke}%
\pgfsetdash{{0.200000pt}{0.330000pt}}{0.000000pt}%
\pgfpathmoveto{\pgfqpoint{0.301111in}{1.361483in}}%
\pgfpathlineto{\pgfqpoint{3.589235in}{1.361483in}}%
\pgfusepath{stroke}%
\end{pgfscope}%
\begin{pgfscope}%
\pgfpathrectangle{\pgfqpoint{0.301111in}{0.301111in}}{\pgfqpoint{3.288124in}{1.908670in}}%
\pgfusepath{clip}%
\pgfsetbuttcap%
\pgfsetroundjoin%
\pgfsetlinewidth{0.200750pt}%
\definecolor{currentstroke}{rgb}{0.501961,0.501961,0.501961}%
\pgfsetstrokecolor{currentstroke}%
\pgfsetdash{{0.200000pt}{0.330000pt}}{0.000000pt}%
\pgfpathmoveto{\pgfqpoint{0.301111in}{1.573558in}}%
\pgfpathlineto{\pgfqpoint{3.589235in}{1.573558in}}%
\pgfusepath{stroke}%
\end{pgfscope}%
\begin{pgfscope}%
\pgfpathrectangle{\pgfqpoint{0.301111in}{0.301111in}}{\pgfqpoint{3.288124in}{1.908670in}}%
\pgfusepath{clip}%
\pgfsetbuttcap%
\pgfsetroundjoin%
\pgfsetlinewidth{0.200750pt}%
\definecolor{currentstroke}{rgb}{0.501961,0.501961,0.501961}%
\pgfsetstrokecolor{currentstroke}%
\pgfsetdash{{0.200000pt}{0.330000pt}}{0.000000pt}%
\pgfpathmoveto{\pgfqpoint{0.301111in}{1.785632in}}%
\pgfpathlineto{\pgfqpoint{3.589235in}{1.785632in}}%
\pgfusepath{stroke}%
\end{pgfscope}%
\begin{pgfscope}%
\pgfpathrectangle{\pgfqpoint{0.301111in}{0.301111in}}{\pgfqpoint{3.288124in}{1.908670in}}%
\pgfusepath{clip}%
\pgfsetbuttcap%
\pgfsetroundjoin%
\pgfsetlinewidth{0.200750pt}%
\definecolor{currentstroke}{rgb}{0.501961,0.501961,0.501961}%
\pgfsetstrokecolor{currentstroke}%
\pgfsetdash{{0.200000pt}{0.330000pt}}{0.000000pt}%
\pgfpathmoveto{\pgfqpoint{0.301111in}{1.997707in}}%
\pgfpathlineto{\pgfqpoint{3.589235in}{1.997707in}}%
\pgfusepath{stroke}%
\end{pgfscope}%
\begin{pgfscope}%
\pgftext[x=0.162222in,y=1.255446in,,bottom,rotate=90.000000]{\sffamily\fontsize{9.600000}{11.520000}\selectfont \texttt{YWIN\_IMAGE}}%
\end{pgfscope}%
\begin{pgfscope}%
\pgfpathrectangle{\pgfqpoint{0.301111in}{0.301111in}}{\pgfqpoint{3.288124in}{1.908670in}}%
\pgfusepath{clip}%
\pgfsetbuttcap%
\pgfsetroundjoin%
\pgfsetlinewidth{0.501875pt}%
\definecolor{currentstroke}{rgb}{0.000000,0.000000,0.000000}%
\pgfsetstrokecolor{currentstroke}%
\pgfsetdash{{1.850000pt}{0.800000pt}}{0.000000pt}%
\pgfpathmoveto{\pgfqpoint{0.913785in}{1.698990in}}%
\pgfpathlineto{\pgfqpoint{0.913785in}{2.123139in}}%
\pgfpathlineto{\pgfqpoint{1.365379in}{2.123139in}}%
\pgfpathlineto{\pgfqpoint{1.365379in}{1.698990in}}%
\pgfpathlineto{\pgfqpoint{0.913785in}{1.698990in}}%
\pgfusepath{stroke}%
\end{pgfscope}%
\begin{pgfscope}%
\pgfpathrectangle{\pgfqpoint{0.301111in}{0.301111in}}{\pgfqpoint{3.288124in}{1.908670in}}%
\pgfusepath{clip}%
\pgfsetbuttcap%
\pgfsetroundjoin%
\pgfsetlinewidth{0.501875pt}%
\definecolor{currentstroke}{rgb}{0.000000,0.000000,0.000000}%
\pgfsetstrokecolor{currentstroke}%
\pgfsetdash{{1.850000pt}{0.800000pt}}{0.000000pt}%
\pgfpathmoveto{\pgfqpoint{1.925140in}{0.662547in}}%
\pgfpathlineto{\pgfqpoint{1.925140in}{1.086696in}}%
\pgfpathlineto{\pgfqpoint{2.376734in}{1.086696in}}%
\pgfpathlineto{\pgfqpoint{2.376734in}{0.662547in}}%
\pgfpathlineto{\pgfqpoint{1.925140in}{0.662547in}}%
\pgfusepath{stroke}%
\end{pgfscope}%
\begin{pgfscope}%
\pgfpathrectangle{\pgfqpoint{0.301111in}{0.301111in}}{\pgfqpoint{3.288124in}{1.908670in}}%
\pgfusepath{clip}%
\pgfsetbuttcap%
\pgfsetroundjoin%
\definecolor{currentfill}{rgb}{0.501961,0.501961,0.501961}%
\pgfsetfillcolor{currentfill}%
\pgfsetlinewidth{1.003750pt}%
\definecolor{currentstroke}{rgb}{0.501961,0.501961,0.501961}%
\pgfsetstrokecolor{currentstroke}%
\pgfsetdash{}{0pt}%
\pgfsys@defobject{currentmarker}{\pgfqpoint{-0.020833in}{-0.020833in}}{\pgfqpoint{0.020833in}{0.020833in}}{%
\pgfpathmoveto{\pgfqpoint{-0.020833in}{-0.020833in}}%
\pgfpathlineto{\pgfqpoint{0.020833in}{0.020833in}}%
\pgfpathmoveto{\pgfqpoint{-0.020833in}{0.020833in}}%
\pgfpathlineto{\pgfqpoint{0.020833in}{-0.020833in}}%
\pgfusepath{stroke,fill}%
}%
\begin{pgfscope}%
\pgfsys@transformshift{1.172089in}{1.838779in}%
\pgfsys@useobject{currentmarker}{}%
\end{pgfscope}%
\begin{pgfscope}%
\pgfsys@transformshift{1.151021in}{1.851701in}%
\pgfsys@useobject{currentmarker}{}%
\end{pgfscope}%
\begin{pgfscope}%
\pgfsys@transformshift{1.148059in}{1.905190in}%
\pgfsys@useobject{currentmarker}{}%
\end{pgfscope}%
\begin{pgfscope}%
\pgfsys@transformshift{1.127126in}{1.964687in}%
\pgfsys@useobject{currentmarker}{}%
\end{pgfscope}%
\begin{pgfscope}%
\pgfsys@transformshift{1.099615in}{1.994967in}%
\pgfsys@useobject{currentmarker}{}%
\end{pgfscope}%
\end{pgfscope}%
\begin{pgfscope}%
\pgfpathrectangle{\pgfqpoint{0.301111in}{0.301111in}}{\pgfqpoint{3.288124in}{1.908670in}}%
\pgfusepath{clip}%
\pgfsetbuttcap%
\pgfsetroundjoin%
\definecolor{currentfill}{rgb}{0.000000,0.000000,0.000000}%
\pgfsetfillcolor{currentfill}%
\pgfsetlinewidth{1.003750pt}%
\definecolor{currentstroke}{rgb}{0.000000,0.000000,0.000000}%
\pgfsetstrokecolor{currentstroke}%
\pgfsetdash{}{0pt}%
\pgfsys@defobject{currentmarker}{\pgfqpoint{-0.010417in}{-0.010417in}}{\pgfqpoint{0.010417in}{0.010417in}}{%
\pgfpathmoveto{\pgfqpoint{0.000000in}{-0.010417in}}%
\pgfpathcurveto{\pgfqpoint{0.002763in}{-0.010417in}}{\pgfqpoint{0.005412in}{-0.009319in}}{\pgfqpoint{0.007366in}{-0.007366in}}%
\pgfpathcurveto{\pgfqpoint{0.009319in}{-0.005412in}}{\pgfqpoint{0.010417in}{-0.002763in}}{\pgfqpoint{0.010417in}{0.000000in}}%
\pgfpathcurveto{\pgfqpoint{0.010417in}{0.002763in}}{\pgfqpoint{0.009319in}{0.005412in}}{\pgfqpoint{0.007366in}{0.007366in}}%
\pgfpathcurveto{\pgfqpoint{0.005412in}{0.009319in}}{\pgfqpoint{0.002763in}{0.010417in}}{\pgfqpoint{0.000000in}{0.010417in}}%
\pgfpathcurveto{\pgfqpoint{-0.002763in}{0.010417in}}{\pgfqpoint{-0.005412in}{0.009319in}}{\pgfqpoint{-0.007366in}{0.007366in}}%
\pgfpathcurveto{\pgfqpoint{-0.009319in}{0.005412in}}{\pgfqpoint{-0.010417in}{0.002763in}}{\pgfqpoint{-0.010417in}{0.000000in}}%
\pgfpathcurveto{\pgfqpoint{-0.010417in}{-0.002763in}}{\pgfqpoint{-0.009319in}{-0.005412in}}{\pgfqpoint{-0.007366in}{-0.007366in}}%
\pgfpathcurveto{\pgfqpoint{-0.005412in}{-0.009319in}}{\pgfqpoint{-0.002763in}{-0.010417in}}{\pgfqpoint{0.000000in}{-0.010417in}}%
\pgfpathclose%
\pgfusepath{stroke,fill}%
}%
\begin{pgfscope}%
\pgfsys@transformshift{1.675077in}{1.147784in}%
\pgfsys@useobject{currentmarker}{}%
\end{pgfscope}%
\begin{pgfscope}%
\pgfsys@transformshift{1.796726in}{1.044477in}%
\pgfsys@useobject{currentmarker}{}%
\end{pgfscope}%
\begin{pgfscope}%
\pgfsys@transformshift{2.104459in}{0.886087in}%
\pgfsys@useobject{currentmarker}{}%
\end{pgfscope}%
\begin{pgfscope}%
\pgfsys@transformshift{2.338424in}{0.761193in}%
\pgfsys@useobject{currentmarker}{}%
\end{pgfscope}%
\begin{pgfscope}%
\pgfsys@transformshift{2.840002in}{0.533566in}%
\pgfsys@useobject{currentmarker}{}%
\end{pgfscope}%
\end{pgfscope}%
\begin{pgfscope}%
\pgfsetrectcap%
\pgfsetmiterjoin%
\pgfsetlinewidth{0.501875pt}%
\definecolor{currentstroke}{rgb}{0.000000,0.000000,0.000000}%
\pgfsetstrokecolor{currentstroke}%
\pgfsetdash{}{0pt}%
\pgfpathmoveto{\pgfqpoint{0.301111in}{0.301111in}}%
\pgfpathlineto{\pgfqpoint{0.301111in}{2.209781in}}%
\pgfusepath{stroke}%
\end{pgfscope}%
\begin{pgfscope}%
\pgfsetrectcap%
\pgfsetmiterjoin%
\pgfsetlinewidth{0.501875pt}%
\definecolor{currentstroke}{rgb}{0.000000,0.000000,0.000000}%
\pgfsetstrokecolor{currentstroke}%
\pgfsetdash{}{0pt}%
\pgfpathmoveto{\pgfqpoint{0.301111in}{0.301111in}}%
\pgfpathlineto{\pgfqpoint{3.589235in}{0.301111in}}%
\pgfusepath{stroke}%
\end{pgfscope}%
\begin{pgfscope}%
\pgftext[x=3.424829in,y=2.114347in,right,base]{\sffamily\fontsize{8.000000}{9.600000}\selectfont \texttt{DELTA\_PIXEL\,=\,2}}%
\end{pgfscope}%
\begin{pgfscope}%
\pgfsetbuttcap%
\pgfsetroundjoin%
\definecolor{currentfill}{rgb}{0.501961,0.501961,0.501961}%
\pgfsetfillcolor{currentfill}%
\pgfsetlinewidth{1.003750pt}%
\definecolor{currentstroke}{rgb}{0.501961,0.501961,0.501961}%
\pgfsetstrokecolor{currentstroke}%
\pgfsetdash{}{0pt}%
\pgfsys@defobject{currentmarker}{\pgfqpoint{-0.020833in}{-0.020833in}}{\pgfqpoint{0.020833in}{0.020833in}}{%
\pgfpathmoveto{\pgfqpoint{-0.020833in}{-0.020833in}}%
\pgfpathlineto{\pgfqpoint{0.020833in}{0.020833in}}%
\pgfpathmoveto{\pgfqpoint{-0.020833in}{0.020833in}}%
\pgfpathlineto{\pgfqpoint{0.020833in}{-0.020833in}}%
\pgfusepath{stroke,fill}%
}%
\begin{pgfscope}%
\pgfsys@transformshift{2.656662in}{1.981514in}%
\pgfsys@useobject{currentmarker}{}%
\end{pgfscope}%
\end{pgfscope}%
\begin{pgfscope}%
\pgftext[x=2.856662in,y=1.942625in,left,base]{\sffamily\fontsize{8.000000}{9.600000}\selectfont Bad Pixel}%
\end{pgfscope}%
\begin{pgfscope}%
\pgfsetbuttcap%
\pgfsetroundjoin%
\definecolor{currentfill}{rgb}{0.000000,0.000000,0.000000}%
\pgfsetfillcolor{currentfill}%
\pgfsetlinewidth{1.003750pt}%
\definecolor{currentstroke}{rgb}{0.000000,0.000000,0.000000}%
\pgfsetstrokecolor{currentstroke}%
\pgfsetdash{}{0pt}%
\pgfsys@defobject{currentmarker}{\pgfqpoint{-0.010417in}{-0.010417in}}{\pgfqpoint{0.010417in}{0.010417in}}{%
\pgfpathmoveto{\pgfqpoint{0.000000in}{-0.010417in}}%
\pgfpathcurveto{\pgfqpoint{0.002763in}{-0.010417in}}{\pgfqpoint{0.005412in}{-0.009319in}}{\pgfqpoint{0.007366in}{-0.007366in}}%
\pgfpathcurveto{\pgfqpoint{0.009319in}{-0.005412in}}{\pgfqpoint{0.010417in}{-0.002763in}}{\pgfqpoint{0.010417in}{0.000000in}}%
\pgfpathcurveto{\pgfqpoint{0.010417in}{0.002763in}}{\pgfqpoint{0.009319in}{0.005412in}}{\pgfqpoint{0.007366in}{0.007366in}}%
\pgfpathcurveto{\pgfqpoint{0.005412in}{0.009319in}}{\pgfqpoint{0.002763in}{0.010417in}}{\pgfqpoint{0.000000in}{0.010417in}}%
\pgfpathcurveto{\pgfqpoint{-0.002763in}{0.010417in}}{\pgfqpoint{-0.005412in}{0.009319in}}{\pgfqpoint{-0.007366in}{0.007366in}}%
\pgfpathcurveto{\pgfqpoint{-0.009319in}{0.005412in}}{\pgfqpoint{-0.010417in}{0.002763in}}{\pgfqpoint{-0.010417in}{0.000000in}}%
\pgfpathcurveto{\pgfqpoint{-0.010417in}{-0.002763in}}{\pgfqpoint{-0.009319in}{-0.005412in}}{\pgfqpoint{-0.007366in}{-0.007366in}}%
\pgfpathcurveto{\pgfqpoint{-0.005412in}{-0.009319in}}{\pgfqpoint{-0.002763in}{-0.010417in}}{\pgfqpoint{0.000000in}{-0.010417in}}%
\pgfpathclose%
\pgfusepath{stroke,fill}%
}%
\begin{pgfscope}%
\pgfsys@transformshift{2.656662in}{1.826625in}%
\pgfsys@useobject{currentmarker}{}%
\end{pgfscope}%
\end{pgfscope}%
\begin{pgfscope}%
\pgftext[x=2.856662in,y=1.787737in,left,base]{\sffamily\fontsize{8.000000}{9.600000}\selectfont SSO}%
\end{pgfscope}%
\end{pgfpicture}%
\makeatother%
\endgroup%

%% file: outlier_motion_filter.pgf
\begingroup%
\makeatletter%
\begin{pgfpicture}%
\pgfpathrectangle{\pgfpointorigin}{\pgfqpoint{3.611457in}{2.232003in}}%
\pgfusepath{use as bounding box, clip}%
\begin{pgfscope}%
\pgfsetbuttcap%
\pgfsetmiterjoin%
\definecolor{currentfill}{rgb}{1.000000,1.000000,1.000000}%
\pgfsetfillcolor{currentfill}%
\pgfsetlinewidth{0.000000pt}%
\definecolor{currentstroke}{rgb}{1.000000,1.000000,1.000000}%
\pgfsetstrokecolor{currentstroke}%
\pgfsetdash{}{0pt}%
\pgfpathmoveto{\pgfqpoint{0.000000in}{0.000000in}}%
\pgfpathlineto{\pgfqpoint{3.611457in}{0.000000in}}%
\pgfpathlineto{\pgfqpoint{3.611457in}{2.232003in}}%
\pgfpathlineto{\pgfqpoint{0.000000in}{2.232003in}}%
\pgfpathclose%
\pgfusepath{fill}%
\end{pgfscope}%
\begin{pgfscope}%
\pgfsetbuttcap%
\pgfsetmiterjoin%
\definecolor{currentfill}{rgb}{1.000000,1.000000,1.000000}%
\pgfsetfillcolor{currentfill}%
\pgfsetlinewidth{0.000000pt}%
\definecolor{currentstroke}{rgb}{0.000000,0.000000,0.000000}%
\pgfsetstrokecolor{currentstroke}%
\pgfsetstrokeopacity{0.000000}%
\pgfsetdash{}{0pt}%
\pgfpathmoveto{\pgfqpoint{0.301111in}{0.370556in}}%
\pgfpathlineto{\pgfqpoint{3.589235in}{0.370556in}}%
\pgfpathlineto{\pgfqpoint{3.589235in}{2.209781in}}%
\pgfpathlineto{\pgfqpoint{0.301111in}{2.209781in}}%
\pgfpathclose%
\pgfusepath{fill}%
\end{pgfscope}%
\begin{pgfscope}%
\pgftext[x=1.945173in,y=0.162222in,,top]{\sffamily\fontsize{9.600000}{11.520000}\selectfont Epoch / MJD}%
\end{pgfscope}%
\begin{pgfscope}%
\pgftext[x=0.162222in,y=1.290168in,,bottom,rotate=90.000000]{\sffamily\fontsize{9.600000}{11.520000}\selectfont Right Ascension / deg}%
\end{pgfscope}%
\begin{pgfscope}%
\pgfpathrectangle{\pgfqpoint{0.301111in}{0.370556in}}{\pgfqpoint{3.288124in}{1.839225in}}%
\pgfusepath{clip}%
\pgfsetbuttcap%
\pgfsetroundjoin%
\pgfsetlinewidth{0.803000pt}%
\definecolor{currentstroke}{rgb}{0.501961,0.501961,0.501961}%
\pgfsetstrokecolor{currentstroke}%
\pgfsetdash{{0.800000pt}{1.320000pt}}{0.000000pt}%
\pgfpathmoveto{\pgfqpoint{0.876809in}{2.212281in}}%
\pgfpathlineto{\pgfqpoint{0.892973in}{2.200183in}}%
\pgfpathlineto{\pgfqpoint{0.925855in}{2.175574in}}%
\pgfpathlineto{\pgfqpoint{0.958736in}{2.150965in}}%
\pgfpathlineto{\pgfqpoint{0.991617in}{2.126355in}}%
\pgfpathlineto{\pgfqpoint{1.024498in}{2.101746in}}%
\pgfpathlineto{\pgfqpoint{1.057380in}{2.077137in}}%
\pgfpathlineto{\pgfqpoint{1.090261in}{2.052528in}}%
\pgfpathlineto{\pgfqpoint{1.123142in}{2.027918in}}%
\pgfpathlineto{\pgfqpoint{1.156023in}{2.003309in}}%
\pgfpathlineto{\pgfqpoint{1.188905in}{1.978700in}}%
\pgfpathlineto{\pgfqpoint{1.221786in}{1.954091in}}%
\pgfpathlineto{\pgfqpoint{1.254667in}{1.929481in}}%
\pgfpathlineto{\pgfqpoint{1.287548in}{1.904872in}}%
\pgfpathlineto{\pgfqpoint{1.320429in}{1.880263in}}%
\pgfpathlineto{\pgfqpoint{1.353311in}{1.855654in}}%
\pgfpathlineto{\pgfqpoint{1.386192in}{1.831045in}}%
\pgfpathlineto{\pgfqpoint{1.419073in}{1.806435in}}%
\pgfpathlineto{\pgfqpoint{1.451954in}{1.781826in}}%
\pgfpathlineto{\pgfqpoint{1.484836in}{1.757217in}}%
\pgfpathlineto{\pgfqpoint{1.517717in}{1.732608in}}%
\pgfpathlineto{\pgfqpoint{1.550598in}{1.707998in}}%
\pgfpathlineto{\pgfqpoint{1.583479in}{1.683389in}}%
\pgfpathlineto{\pgfqpoint{1.616361in}{1.658780in}}%
\pgfpathlineto{\pgfqpoint{1.649242in}{1.634171in}}%
\pgfpathlineto{\pgfqpoint{1.682123in}{1.609562in}}%
\pgfpathlineto{\pgfqpoint{1.715004in}{1.584952in}}%
\pgfpathlineto{\pgfqpoint{1.747886in}{1.560343in}}%
\pgfpathlineto{\pgfqpoint{1.780767in}{1.535734in}}%
\pgfpathlineto{\pgfqpoint{1.813648in}{1.511125in}}%
\pgfpathlineto{\pgfqpoint{1.846529in}{1.486515in}}%
\pgfpathlineto{\pgfqpoint{1.879410in}{1.461906in}}%
\pgfpathlineto{\pgfqpoint{1.912292in}{1.437297in}}%
\pgfpathlineto{\pgfqpoint{1.945173in}{1.412688in}}%
\pgfpathlineto{\pgfqpoint{1.978054in}{1.388079in}}%
\pgfpathlineto{\pgfqpoint{2.010935in}{1.363469in}}%
\pgfpathlineto{\pgfqpoint{2.043817in}{1.338860in}}%
\pgfpathlineto{\pgfqpoint{2.076698in}{1.314251in}}%
\pgfpathlineto{\pgfqpoint{2.109579in}{1.289642in}}%
\pgfpathlineto{\pgfqpoint{2.142460in}{1.265032in}}%
\pgfpathlineto{\pgfqpoint{2.175342in}{1.240423in}}%
\pgfpathlineto{\pgfqpoint{2.208223in}{1.215814in}}%
\pgfpathlineto{\pgfqpoint{2.241104in}{1.191205in}}%
\pgfpathlineto{\pgfqpoint{2.273985in}{1.166595in}}%
\pgfpathlineto{\pgfqpoint{2.306867in}{1.141986in}}%
\pgfpathlineto{\pgfqpoint{2.339748in}{1.117377in}}%
\pgfpathlineto{\pgfqpoint{2.372629in}{1.092768in}}%
\pgfpathlineto{\pgfqpoint{2.405510in}{1.068159in}}%
\pgfpathlineto{\pgfqpoint{2.438392in}{1.043549in}}%
\pgfpathlineto{\pgfqpoint{2.471273in}{1.018940in}}%
\pgfpathlineto{\pgfqpoint{2.504154in}{0.994331in}}%
\pgfpathlineto{\pgfqpoint{2.537035in}{0.969722in}}%
\pgfpathlineto{\pgfqpoint{2.569916in}{0.945112in}}%
\pgfpathlineto{\pgfqpoint{2.602798in}{0.920503in}}%
\pgfpathlineto{\pgfqpoint{2.635679in}{0.895894in}}%
\pgfpathlineto{\pgfqpoint{2.668560in}{0.871285in}}%
\pgfpathlineto{\pgfqpoint{2.701441in}{0.846676in}}%
\pgfpathlineto{\pgfqpoint{2.734323in}{0.822066in}}%
\pgfpathlineto{\pgfqpoint{2.767204in}{0.797457in}}%
\pgfpathlineto{\pgfqpoint{2.800085in}{0.772848in}}%
\pgfpathlineto{\pgfqpoint{2.832966in}{0.748239in}}%
\pgfpathlineto{\pgfqpoint{2.865848in}{0.723629in}}%
\pgfpathlineto{\pgfqpoint{2.898729in}{0.699020in}}%
\pgfpathlineto{\pgfqpoint{2.931610in}{0.674411in}}%
\pgfpathlineto{\pgfqpoint{2.964491in}{0.649802in}}%
\pgfpathlineto{\pgfqpoint{2.997373in}{0.625192in}}%
\pgfpathlineto{\pgfqpoint{3.030254in}{0.600583in}}%
\pgfpathlineto{\pgfqpoint{3.063135in}{0.575974in}}%
\pgfpathlineto{\pgfqpoint{3.096016in}{0.551365in}}%
\pgfpathlineto{\pgfqpoint{3.128897in}{0.526756in}}%
\pgfpathlineto{\pgfqpoint{3.161779in}{0.502146in}}%
\pgfpathlineto{\pgfqpoint{3.194660in}{0.477537in}}%
\pgfpathlineto{\pgfqpoint{3.227541in}{0.452928in}}%
\pgfpathlineto{\pgfqpoint{3.260422in}{0.428319in}}%
\pgfpathlineto{\pgfqpoint{3.293304in}{0.403709in}}%
\pgfpathlineto{\pgfqpoint{3.326185in}{0.379100in}}%
\pgfpathlineto{\pgfqpoint{3.340942in}{0.368056in}}%
\pgfusepath{stroke}%
\end{pgfscope}%
\begin{pgfscope}%
\pgfpathrectangle{\pgfqpoint{0.301111in}{0.370556in}}{\pgfqpoint{3.288124in}{1.839225in}}%
\pgfusepath{clip}%
\pgfsetrectcap%
\pgfsetroundjoin%
\pgfsetlinewidth{0.803000pt}%
\definecolor{currentstroke}{rgb}{0.000000,0.000000,0.000000}%
\pgfsetstrokecolor{currentstroke}%
\pgfsetdash{}{0pt}%
\pgfpathmoveto{\pgfqpoint{1.121199in}{2.212281in}}%
\pgfpathlineto{\pgfqpoint{1.123142in}{2.210533in}}%
\pgfpathlineto{\pgfqpoint{1.156023in}{2.180952in}}%
\pgfpathlineto{\pgfqpoint{1.188905in}{2.151371in}}%
\pgfpathlineto{\pgfqpoint{1.221786in}{2.121789in}}%
\pgfpathlineto{\pgfqpoint{1.254667in}{2.092208in}}%
\pgfpathlineto{\pgfqpoint{1.287548in}{2.062626in}}%
\pgfpathlineto{\pgfqpoint{1.320429in}{2.033045in}}%
\pgfpathlineto{\pgfqpoint{1.353311in}{2.003464in}}%
\pgfpathlineto{\pgfqpoint{1.386192in}{1.973882in}}%
\pgfpathlineto{\pgfqpoint{1.419073in}{1.944301in}}%
\pgfpathlineto{\pgfqpoint{1.451954in}{1.914720in}}%
\pgfpathlineto{\pgfqpoint{1.484836in}{1.885138in}}%
\pgfpathlineto{\pgfqpoint{1.517717in}{1.855557in}}%
\pgfpathlineto{\pgfqpoint{1.550598in}{1.825975in}}%
\pgfpathlineto{\pgfqpoint{1.583479in}{1.796394in}}%
\pgfpathlineto{\pgfqpoint{1.616361in}{1.766813in}}%
\pgfpathlineto{\pgfqpoint{1.649242in}{1.737231in}}%
\pgfpathlineto{\pgfqpoint{1.682123in}{1.707650in}}%
\pgfpathlineto{\pgfqpoint{1.715004in}{1.678068in}}%
\pgfpathlineto{\pgfqpoint{1.747886in}{1.648487in}}%
\pgfpathlineto{\pgfqpoint{1.780767in}{1.618906in}}%
\pgfpathlineto{\pgfqpoint{1.813648in}{1.589324in}}%
\pgfpathlineto{\pgfqpoint{1.846529in}{1.559743in}}%
\pgfpathlineto{\pgfqpoint{1.879410in}{1.530162in}}%
\pgfpathlineto{\pgfqpoint{1.912292in}{1.500580in}}%
\pgfpathlineto{\pgfqpoint{1.945173in}{1.470999in}}%
\pgfpathlineto{\pgfqpoint{1.978054in}{1.441417in}}%
\pgfpathlineto{\pgfqpoint{2.010935in}{1.411836in}}%
\pgfpathlineto{\pgfqpoint{2.043817in}{1.382255in}}%
\pgfpathlineto{\pgfqpoint{2.076698in}{1.352673in}}%
\pgfpathlineto{\pgfqpoint{2.109579in}{1.323092in}}%
\pgfpathlineto{\pgfqpoint{2.142460in}{1.293510in}}%
\pgfpathlineto{\pgfqpoint{2.175342in}{1.263929in}}%
\pgfpathlineto{\pgfqpoint{2.208223in}{1.234348in}}%
\pgfpathlineto{\pgfqpoint{2.241104in}{1.204766in}}%
\pgfpathlineto{\pgfqpoint{2.273985in}{1.175185in}}%
\pgfpathlineto{\pgfqpoint{2.306867in}{1.145604in}}%
\pgfpathlineto{\pgfqpoint{2.339748in}{1.116022in}}%
\pgfpathlineto{\pgfqpoint{2.372629in}{1.086441in}}%
\pgfpathlineto{\pgfqpoint{2.405510in}{1.056859in}}%
\pgfpathlineto{\pgfqpoint{2.438392in}{1.027278in}}%
\pgfpathlineto{\pgfqpoint{2.471273in}{0.997697in}}%
\pgfpathlineto{\pgfqpoint{2.504154in}{0.968115in}}%
\pgfpathlineto{\pgfqpoint{2.537035in}{0.938534in}}%
\pgfpathlineto{\pgfqpoint{2.569916in}{0.908952in}}%
\pgfpathlineto{\pgfqpoint{2.602798in}{0.879371in}}%
\pgfpathlineto{\pgfqpoint{2.635679in}{0.849790in}}%
\pgfpathlineto{\pgfqpoint{2.668560in}{0.820208in}}%
\pgfpathlineto{\pgfqpoint{2.701441in}{0.790627in}}%
\pgfpathlineto{\pgfqpoint{2.734323in}{0.761046in}}%
\pgfpathlineto{\pgfqpoint{2.767204in}{0.731464in}}%
\pgfpathlineto{\pgfqpoint{2.800085in}{0.701883in}}%
\pgfpathlineto{\pgfqpoint{2.832966in}{0.672301in}}%
\pgfpathlineto{\pgfqpoint{2.865848in}{0.642720in}}%
\pgfpathlineto{\pgfqpoint{2.898729in}{0.613139in}}%
\pgfpathlineto{\pgfqpoint{2.931610in}{0.583557in}}%
\pgfpathlineto{\pgfqpoint{2.964491in}{0.553976in}}%
\pgfpathlineto{\pgfqpoint{2.997373in}{0.524395in}}%
\pgfpathlineto{\pgfqpoint{3.030254in}{0.494813in}}%
\pgfpathlineto{\pgfqpoint{3.063135in}{0.465232in}}%
\pgfpathlineto{\pgfqpoint{3.096016in}{0.435650in}}%
\pgfpathlineto{\pgfqpoint{3.128897in}{0.406069in}}%
\pgfpathlineto{\pgfqpoint{3.161779in}{0.376488in}}%
\pgfpathlineto{\pgfqpoint{3.171151in}{0.368056in}}%
\pgfusepath{stroke}%
\end{pgfscope}%
\begin{pgfscope}%
\pgfpathrectangle{\pgfqpoint{0.301111in}{0.370556in}}{\pgfqpoint{3.288124in}{1.839225in}}%
\pgfusepath{clip}%
\pgfsetbuttcap%
\pgfsetroundjoin%
\definecolor{currentfill}{rgb}{0.000000,0.000000,0.000000}%
\pgfsetfillcolor{currentfill}%
\pgfsetlinewidth{1.003750pt}%
\definecolor{currentstroke}{rgb}{0.000000,0.000000,0.000000}%
\pgfsetstrokecolor{currentstroke}%
\pgfsetdash{}{0pt}%
\pgfsys@defobject{currentmarker}{\pgfqpoint{-0.010417in}{-0.010417in}}{\pgfqpoint{0.010417in}{0.010417in}}{%
\pgfpathmoveto{\pgfqpoint{0.000000in}{-0.010417in}}%
\pgfpathcurveto{\pgfqpoint{0.002763in}{-0.010417in}}{\pgfqpoint{0.005412in}{-0.009319in}}{\pgfqpoint{0.007366in}{-0.007366in}}%
\pgfpathcurveto{\pgfqpoint{0.009319in}{-0.005412in}}{\pgfqpoint{0.010417in}{-0.002763in}}{\pgfqpoint{0.010417in}{0.000000in}}%
\pgfpathcurveto{\pgfqpoint{0.010417in}{0.002763in}}{\pgfqpoint{0.009319in}{0.005412in}}{\pgfqpoint{0.007366in}{0.007366in}}%
\pgfpathcurveto{\pgfqpoint{0.005412in}{0.009319in}}{\pgfqpoint{0.002763in}{0.010417in}}{\pgfqpoint{0.000000in}{0.010417in}}%
\pgfpathcurveto{\pgfqpoint{-0.002763in}{0.010417in}}{\pgfqpoint{-0.005412in}{0.009319in}}{\pgfqpoint{-0.007366in}{0.007366in}}%
\pgfpathcurveto{\pgfqpoint{-0.009319in}{0.005412in}}{\pgfqpoint{-0.010417in}{0.002763in}}{\pgfqpoint{-0.010417in}{0.000000in}}%
\pgfpathcurveto{\pgfqpoint{-0.010417in}{-0.002763in}}{\pgfqpoint{-0.009319in}{-0.005412in}}{\pgfqpoint{-0.007366in}{-0.007366in}}%
\pgfpathcurveto{\pgfqpoint{-0.005412in}{-0.009319in}}{\pgfqpoint{-0.002763in}{-0.010417in}}{\pgfqpoint{0.000000in}{-0.010417in}}%
\pgfpathclose%
\pgfusepath{stroke,fill}%
}%
\begin{pgfscope}%
\pgfsys@transformshift{2.728460in}{0.770959in}%
\pgfsys@useobject{currentmarker}{}%
\end{pgfscope}%
\begin{pgfscope}%
\pgfsys@transformshift{2.474146in}{1.010287in}%
\pgfsys@useobject{currentmarker}{}%
\end{pgfscope}%
\begin{pgfscope}%
\pgfsys@transformshift{2.286105in}{1.137891in}%
\pgfsys@useobject{currentmarker}{}%
\end{pgfscope}%
\begin{pgfscope}%
\pgfsys@transformshift{1.905558in}{1.521289in}%
\pgfsys@useobject{currentmarker}{}%
\end{pgfscope}%
\begin{pgfscope}%
\pgfsys@transformshift{1.759894in}{1.614337in}%
\pgfsys@useobject{currentmarker}{}%
\end{pgfscope}%
\end{pgfscope}%
\begin{pgfscope}%
\pgfpathrectangle{\pgfqpoint{0.301111in}{0.370556in}}{\pgfqpoint{3.288124in}{1.839225in}}%
\pgfusepath{clip}%
\pgfsetbuttcap%
\pgfsetroundjoin%
\definecolor{currentfill}{rgb}{0.501961,0.501961,0.501961}%
\pgfsetfillcolor{currentfill}%
\pgfsetlinewidth{1.003750pt}%
\definecolor{currentstroke}{rgb}{0.501961,0.501961,0.501961}%
\pgfsetstrokecolor{currentstroke}%
\pgfsetdash{}{0pt}%
\pgfsys@defobject{currentmarker}{\pgfqpoint{-0.020833in}{-0.020833in}}{\pgfqpoint{0.020833in}{0.020833in}}{%
\pgfpathmoveto{\pgfqpoint{-0.020833in}{-0.020833in}}%
\pgfpathlineto{\pgfqpoint{0.020833in}{0.020833in}}%
\pgfpathmoveto{\pgfqpoint{-0.020833in}{0.020833in}}%
\pgfpathlineto{\pgfqpoint{0.020833in}{-0.020833in}}%
\pgfusepath{stroke,fill}%
}%
\begin{pgfscope}%
\pgfsys@transformshift{1.011312in}{1.514660in}%
\pgfsys@useobject{currentmarker}{}%
\end{pgfscope}%
\end{pgfscope}%
\begin{pgfscope}%
\pgfsetrectcap%
\pgfsetmiterjoin%
\pgfsetlinewidth{0.501875pt}%
\definecolor{currentstroke}{rgb}{0.000000,0.000000,0.000000}%
\pgfsetstrokecolor{currentstroke}%
\pgfsetdash{}{0pt}%
\pgfpathmoveto{\pgfqpoint{0.301111in}{0.370556in}}%
\pgfpathlineto{\pgfqpoint{0.301111in}{2.209781in}}%
\pgfusepath{stroke}%
\end{pgfscope}%
\begin{pgfscope}%
\pgfsetrectcap%
\pgfsetmiterjoin%
\pgfsetlinewidth{0.501875pt}%
\definecolor{currentstroke}{rgb}{0.501961,0.501961,0.501961}%
\pgfsetstrokecolor{currentstroke}%
\pgfsetdash{}{0pt}%
\pgfpathmoveto{\pgfqpoint{0.301111in}{0.370556in}}%
\pgfpathlineto{\pgfqpoint{3.589235in}{0.370556in}}%
\pgfusepath{stroke}%
\end{pgfscope}%
\begin{pgfscope}%
\definecolor{textcolor}{rgb}{0.501961,0.501961,0.501961}%
\pgfsetstrokecolor{textcolor}%
\pgfsetfillcolor{textcolor}%
\pgftext[x=0.498399in,y=2.025858in,left,base]{\color{textcolor}\sffamily\fontsize{8.000000}{9.600000}\selectfont \(\displaystyle R^2\)\,=\,0.36}%
\end{pgfscope}%
\begin{pgfscope}%
\pgftext[x=1.385603in,y=0.255069in,,base]{\sffamily\fontsize{6.000000}{7.200000}\selectfont 1\,MAD}%
\end{pgfscope}%
\begin{pgfscope}%
\pgfsetroundcap%
\pgfsetroundjoin%
\pgfsetlinewidth{0.501875pt}%
\definecolor{currentstroke}{rgb}{0.501961,0.501961,0.501961}%
\pgfsetstrokecolor{currentstroke}%
\pgfsetdash{}{0pt}%
\pgfpathmoveto{\pgfqpoint{1.759894in}{0.335833in}}%
\pgfpathlineto{\pgfqpoint{1.759894in}{0.335833in}}%
\pgfpathlineto{\pgfqpoint{1.759894in}{0.405278in}}%
\pgfpathlineto{\pgfqpoint{1.759894in}{0.405278in}}%
\pgfpathmoveto{\pgfqpoint{1.759894in}{0.370556in}}%
\pgfpathquadraticcurveto{\pgfqpoint{1.385603in}{0.370556in}}{\pgfqpoint{1.011312in}{0.370556in}}%
\pgfpathmoveto{\pgfqpoint{1.011312in}{0.405278in}}%
\pgfpathlineto{\pgfqpoint{1.011312in}{0.405278in}}%
\pgfpathlineto{\pgfqpoint{1.011312in}{0.335833in}}%
\pgfpathlineto{\pgfqpoint{1.011312in}{0.335833in}}%
\pgfusepath{stroke}%
\end{pgfscope}%
\begin{pgfscope}%
\pgfsetroundcap%
\pgfsetroundjoin%
\pgfsetlinewidth{0.501875pt}%
\definecolor{currentstroke}{rgb}{0.000000,0.000000,0.000000}%
\pgfsetstrokecolor{currentstroke}%
\pgfsetdash{}{0pt}%
\pgfpathmoveto{\pgfqpoint{1.650341in}{0.335833in}}%
\pgfpathlineto{\pgfqpoint{1.650341in}{0.335833in}}%
\pgfpathlineto{\pgfqpoint{1.650341in}{0.405278in}}%
\pgfpathlineto{\pgfqpoint{1.650341in}{0.405278in}}%
\pgfpathmoveto{\pgfqpoint{1.650341in}{0.370556in}}%
\pgfpathquadraticcurveto{\pgfqpoint{1.385603in}{0.370556in}}{\pgfqpoint{1.120865in}{0.370556in}}%
\pgfpathmoveto{\pgfqpoint{1.120865in}{0.405278in}}%
\pgfpathlineto{\pgfqpoint{1.120865in}{0.405278in}}%
\pgfpathlineto{\pgfqpoint{1.120865in}{0.335833in}}%
\pgfpathlineto{\pgfqpoint{1.120865in}{0.335833in}}%
\pgfusepath{stroke}%
\end{pgfscope}%
\begin{pgfscope}%
\pgftext[x=1.419073in,y=2.025858in,left,base]{\sffamily\fontsize{8.000000}{9.600000}\selectfont \(\displaystyle R^2\)\,=\,1.00}%
\end{pgfscope}%
\begin{pgfscope}%
\pgfsetbuttcap%
\pgfsetmiterjoin%
\definecolor{currentfill}{rgb}{1.000000,1.000000,1.000000}%
\pgfsetfillcolor{currentfill}%
\pgfsetlinewidth{0.000000pt}%
\definecolor{currentstroke}{rgb}{0.000000,0.000000,0.000000}%
\pgfsetstrokecolor{currentstroke}%
\pgfsetdash{}{0pt}%
\pgfpathmoveto{\pgfqpoint{2.174183in}{1.948747in}}%
\pgfpathlineto{\pgfqpoint{3.480384in}{1.948747in}}%
\pgfpathlineto{\pgfqpoint{3.480384in}{2.272614in}}%
\pgfpathlineto{\pgfqpoint{2.174183in}{2.272614in}}%
\pgfpathclose%
\pgfusepath{fill}%
\end{pgfscope}%
\begin{pgfscope}%
\pgftext[x=2.229739in,y=2.139947in,left,base]{\sffamily\fontsize{8.000000}{9.600000}\selectfont \texttt{OUTLIER\_THRESHOLD\,\,=\,\,1}}%
\end{pgfscope}%
\begin{pgfscope}%
\pgftext[x=2.679783in,y=2.025858in,left,base]{\sffamily\fontsize{8.000000}{9.600000}\selectfont \texttt{R\_SQU\_M\,=\,0.95}}%
\end{pgfscope}%
\begin{pgfscope}%
\pgfsetbuttcap%
\pgfsetroundjoin%
\definecolor{currentfill}{rgb}{0.000000,0.000000,0.000000}%
\pgfsetfillcolor{currentfill}%
\pgfsetlinewidth{1.003750pt}%
\definecolor{currentstroke}{rgb}{0.000000,0.000000,0.000000}%
\pgfsetstrokecolor{currentstroke}%
\pgfsetdash{}{0pt}%
\pgfsys@defobject{currentmarker}{\pgfqpoint{-0.010417in}{-0.010417in}}{\pgfqpoint{0.010417in}{0.010417in}}{%
\pgfpathmoveto{\pgfqpoint{0.000000in}{-0.010417in}}%
\pgfpathcurveto{\pgfqpoint{0.002763in}{-0.010417in}}{\pgfqpoint{0.005412in}{-0.009319in}}{\pgfqpoint{0.007366in}{-0.007366in}}%
\pgfpathcurveto{\pgfqpoint{0.009319in}{-0.005412in}}{\pgfqpoint{0.010417in}{-0.002763in}}{\pgfqpoint{0.010417in}{0.000000in}}%
\pgfpathcurveto{\pgfqpoint{0.010417in}{0.002763in}}{\pgfqpoint{0.009319in}{0.005412in}}{\pgfqpoint{0.007366in}{0.007366in}}%
\pgfpathcurveto{\pgfqpoint{0.005412in}{0.009319in}}{\pgfqpoint{0.002763in}{0.010417in}}{\pgfqpoint{0.000000in}{0.010417in}}%
\pgfpathcurveto{\pgfqpoint{-0.002763in}{0.010417in}}{\pgfqpoint{-0.005412in}{0.009319in}}{\pgfqpoint{-0.007366in}{0.007366in}}%
\pgfpathcurveto{\pgfqpoint{-0.009319in}{0.005412in}}{\pgfqpoint{-0.010417in}{0.002763in}}{\pgfqpoint{-0.010417in}{0.000000in}}%
\pgfpathcurveto{\pgfqpoint{-0.010417in}{-0.002763in}}{\pgfqpoint{-0.009319in}{-0.005412in}}{\pgfqpoint{-0.007366in}{-0.007366in}}%
\pgfpathcurveto{\pgfqpoint{-0.005412in}{-0.009319in}}{\pgfqpoint{-0.002763in}{-0.010417in}}{\pgfqpoint{0.000000in}{-0.010417in}}%
\pgfpathclose%
\pgfusepath{stroke,fill}%
}%
\begin{pgfscope}%
\pgfsys@transformshift{0.512222in}{0.685889in}%
\pgfsys@useobject{currentmarker}{}%
\end{pgfscope}%
\end{pgfscope}%
\begin{pgfscope}%
\pgftext[x=0.712222in,y=0.647000in,left,base]{\sffamily\fontsize{8.000000}{9.600000}\selectfont SSO detections}%
\end{pgfscope}%
\begin{pgfscope}%
\pgfsetbuttcap%
\pgfsetroundjoin%
\definecolor{currentfill}{rgb}{0.501961,0.501961,0.501961}%
\pgfsetfillcolor{currentfill}%
\pgfsetlinewidth{1.003750pt}%
\definecolor{currentstroke}{rgb}{0.501961,0.501961,0.501961}%
\pgfsetstrokecolor{currentstroke}%
\pgfsetdash{}{0pt}%
\pgfsys@defobject{currentmarker}{\pgfqpoint{-0.020833in}{-0.020833in}}{\pgfqpoint{0.020833in}{0.020833in}}{%
\pgfpathmoveto{\pgfqpoint{-0.020833in}{-0.020833in}}%
\pgfpathlineto{\pgfqpoint{0.020833in}{0.020833in}}%
\pgfpathmoveto{\pgfqpoint{-0.020833in}{0.020833in}}%
\pgfpathlineto{\pgfqpoint{0.020833in}{-0.020833in}}%
\pgfusepath{stroke,fill}%
}%
\begin{pgfscope}%
\pgfsys@transformshift{0.512222in}{0.531000in}%
\pgfsys@useobject{currentmarker}{}%
\end{pgfscope}%
\end{pgfscope}%
\begin{pgfscope}%
\pgftext[x=0.712222in,y=0.492111in,left,base]{\sffamily\fontsize{8.000000}{9.600000}\selectfont Random association}%
\end{pgfscope}%
\end{pgfpicture}%
\makeatother%
\endgroup%

%% file: stellar_catalogue.pgf
\begingroup%
\makeatletter%
\begin{pgfpicture}%
\pgfpathrectangle{\pgfpointorigin}{\pgfqpoint{3.611457in}{2.232003in}}%
\pgfusepath{use as bounding box, clip}%
\begin{pgfscope}%
\pgfsetbuttcap%
\pgfsetmiterjoin%
\definecolor{currentfill}{rgb}{1.000000,1.000000,1.000000}%
\pgfsetfillcolor{currentfill}%
\pgfsetlinewidth{0.000000pt}%
\definecolor{currentstroke}{rgb}{1.000000,1.000000,1.000000}%
\pgfsetstrokecolor{currentstroke}%
\pgfsetdash{}{0pt}%
\pgfpathmoveto{\pgfqpoint{0.000000in}{0.000000in}}%
\pgfpathlineto{\pgfqpoint{3.611457in}{0.000000in}}%
\pgfpathlineto{\pgfqpoint{3.611457in}{2.232003in}}%
\pgfpathlineto{\pgfqpoint{0.000000in}{2.232003in}}%
\pgfpathclose%
\pgfusepath{fill}%
\end{pgfscope}%
\begin{pgfscope}%
\pgfsetbuttcap%
\pgfsetmiterjoin%
\definecolor{currentfill}{rgb}{1.000000,1.000000,1.000000}%
\pgfsetfillcolor{currentfill}%
\pgfsetlinewidth{0.000000pt}%
\definecolor{currentstroke}{rgb}{0.000000,0.000000,0.000000}%
\pgfsetstrokecolor{currentstroke}%
\pgfsetstrokeopacity{0.000000}%
\pgfsetdash{}{0pt}%
\pgfpathmoveto{\pgfqpoint{0.705864in}{0.446667in}}%
\pgfpathlineto{\pgfqpoint{3.459033in}{0.446667in}}%
\pgfpathlineto{\pgfqpoint{3.459033in}{2.169781in}}%
\pgfpathlineto{\pgfqpoint{0.705864in}{2.169781in}}%
\pgfpathclose%
\pgfusepath{fill}%
\end{pgfscope}%
\begin{pgfscope}%
\pgfsetbuttcap%
\pgfsetroundjoin%
\definecolor{currentfill}{rgb}{0.000000,0.000000,0.000000}%
\pgfsetfillcolor{currentfill}%
\pgfsetlinewidth{0.501875pt}%
\definecolor{currentstroke}{rgb}{0.000000,0.000000,0.000000}%
\pgfsetstrokecolor{currentstroke}%
\pgfsetdash{}{0pt}%
\pgfsys@defobject{currentmarker}{\pgfqpoint{0.000000in}{-0.055556in}}{\pgfqpoint{0.000000in}{0.000000in}}{%
\pgfpathmoveto{\pgfqpoint{0.000000in}{0.000000in}}%
\pgfpathlineto{\pgfqpoint{0.000000in}{-0.055556in}}%
\pgfusepath{stroke,fill}%
}%
\begin{pgfscope}%
\pgfsys@transformshift{1.190297in}{0.446667in}%
\pgfsys@useobject{currentmarker}{}%
\end{pgfscope}%
\end{pgfscope}%
\begin{pgfscope}%
\pgftext[x=1.190297in,y=0.307778in,,top]{\sffamily\fontsize{6.664000}{7.996800}\selectfont \(\displaystyle 210.5\)}%
\end{pgfscope}%
\begin{pgfscope}%
\pgfsetbuttcap%
\pgfsetroundjoin%
\definecolor{currentfill}{rgb}{0.000000,0.000000,0.000000}%
\pgfsetfillcolor{currentfill}%
\pgfsetlinewidth{0.501875pt}%
\definecolor{currentstroke}{rgb}{0.000000,0.000000,0.000000}%
\pgfsetstrokecolor{currentstroke}%
\pgfsetdash{}{0pt}%
\pgfsys@defobject{currentmarker}{\pgfqpoint{0.000000in}{-0.055556in}}{\pgfqpoint{0.000000in}{0.000000in}}{%
\pgfpathmoveto{\pgfqpoint{0.000000in}{0.000000in}}%
\pgfpathlineto{\pgfqpoint{0.000000in}{-0.055556in}}%
\pgfusepath{stroke,fill}%
}%
\begin{pgfscope}%
\pgfsys@transformshift{1.757481in}{0.446667in}%
\pgfsys@useobject{currentmarker}{}%
\end{pgfscope}%
\end{pgfscope}%
\begin{pgfscope}%
\pgftext[x=1.757481in,y=0.307778in,,top]{\sffamily\fontsize{6.664000}{7.996800}\selectfont \(\displaystyle 210.6\)}%
\end{pgfscope}%
\begin{pgfscope}%
\pgfsetbuttcap%
\pgfsetroundjoin%
\definecolor{currentfill}{rgb}{0.000000,0.000000,0.000000}%
\pgfsetfillcolor{currentfill}%
\pgfsetlinewidth{0.501875pt}%
\definecolor{currentstroke}{rgb}{0.000000,0.000000,0.000000}%
\pgfsetstrokecolor{currentstroke}%
\pgfsetdash{}{0pt}%
\pgfsys@defobject{currentmarker}{\pgfqpoint{0.000000in}{-0.055556in}}{\pgfqpoint{0.000000in}{0.000000in}}{%
\pgfpathmoveto{\pgfqpoint{0.000000in}{0.000000in}}%
\pgfpathlineto{\pgfqpoint{0.000000in}{-0.055556in}}%
\pgfusepath{stroke,fill}%
}%
\begin{pgfscope}%
\pgfsys@transformshift{2.324665in}{0.446667in}%
\pgfsys@useobject{currentmarker}{}%
\end{pgfscope}%
\end{pgfscope}%
\begin{pgfscope}%
\pgftext[x=2.324665in,y=0.307778in,,top]{\sffamily\fontsize{6.664000}{7.996800}\selectfont \(\displaystyle 210.7\)}%
\end{pgfscope}%
\begin{pgfscope}%
\pgfsetbuttcap%
\pgfsetroundjoin%
\definecolor{currentfill}{rgb}{0.000000,0.000000,0.000000}%
\pgfsetfillcolor{currentfill}%
\pgfsetlinewidth{0.501875pt}%
\definecolor{currentstroke}{rgb}{0.000000,0.000000,0.000000}%
\pgfsetstrokecolor{currentstroke}%
\pgfsetdash{}{0pt}%
\pgfsys@defobject{currentmarker}{\pgfqpoint{0.000000in}{-0.055556in}}{\pgfqpoint{0.000000in}{0.000000in}}{%
\pgfpathmoveto{\pgfqpoint{0.000000in}{0.000000in}}%
\pgfpathlineto{\pgfqpoint{0.000000in}{-0.055556in}}%
\pgfusepath{stroke,fill}%
}%
\begin{pgfscope}%
\pgfsys@transformshift{2.891849in}{0.446667in}%
\pgfsys@useobject{currentmarker}{}%
\end{pgfscope}%
\end{pgfscope}%
\begin{pgfscope}%
\pgftext[x=2.891849in,y=0.307778in,,top]{\sffamily\fontsize{6.664000}{7.996800}\selectfont \(\displaystyle 210.8\)}%
\end{pgfscope}%
\begin{pgfscope}%
\pgfsetbuttcap%
\pgfsetroundjoin%
\definecolor{currentfill}{rgb}{0.000000,0.000000,0.000000}%
\pgfsetfillcolor{currentfill}%
\pgfsetlinewidth{0.501875pt}%
\definecolor{currentstroke}{rgb}{0.000000,0.000000,0.000000}%
\pgfsetstrokecolor{currentstroke}%
\pgfsetdash{}{0pt}%
\pgfsys@defobject{currentmarker}{\pgfqpoint{0.000000in}{-0.055556in}}{\pgfqpoint{0.000000in}{0.000000in}}{%
\pgfpathmoveto{\pgfqpoint{0.000000in}{0.000000in}}%
\pgfpathlineto{\pgfqpoint{0.000000in}{-0.055556in}}%
\pgfusepath{stroke,fill}%
}%
\begin{pgfscope}%
\pgfsys@transformshift{3.459033in}{0.446667in}%
\pgfsys@useobject{currentmarker}{}%
\end{pgfscope}%
\end{pgfscope}%
\begin{pgfscope}%
\pgftext[x=3.459033in,y=0.307778in,,top]{\sffamily\fontsize{6.664000}{7.996800}\selectfont \(\displaystyle 210.9\)}%
\end{pgfscope}%
\begin{pgfscope}%
\pgftext[x=2.082449in,y=0.170033in,,top]{\sffamily\fontsize{9.600000}{11.520000}\selectfont Right Ascension / deg}%
\end{pgfscope}%
\begin{pgfscope}%
\pgfsetbuttcap%
\pgfsetroundjoin%
\definecolor{currentfill}{rgb}{0.000000,0.000000,0.000000}%
\pgfsetfillcolor{currentfill}%
\pgfsetlinewidth{0.501875pt}%
\definecolor{currentstroke}{rgb}{0.000000,0.000000,0.000000}%
\pgfsetstrokecolor{currentstroke}%
\pgfsetdash{}{0pt}%
\pgfsys@defobject{currentmarker}{\pgfqpoint{-0.055556in}{0.000000in}}{\pgfqpoint{0.000000in}{0.000000in}}{%
\pgfpathmoveto{\pgfqpoint{0.000000in}{0.000000in}}%
\pgfpathlineto{\pgfqpoint{-0.055556in}{0.000000in}}%
\pgfusepath{stroke,fill}%
}%
\begin{pgfscope}%
\pgfsys@transformshift{0.705864in}{1.997470in}%
\pgfsys@useobject{currentmarker}{}%
\end{pgfscope}%
\end{pgfscope}%
\begin{pgfscope}%
\pgftext[x=0.263520in,y=1.965353in,left,base]{\sffamily\fontsize{6.664000}{7.996800}\selectfont \(\displaystyle -60.95\)}%
\end{pgfscope}%
\begin{pgfscope}%
\pgfsetbuttcap%
\pgfsetroundjoin%
\definecolor{currentfill}{rgb}{0.000000,0.000000,0.000000}%
\pgfsetfillcolor{currentfill}%
\pgfsetlinewidth{0.501875pt}%
\definecolor{currentstroke}{rgb}{0.000000,0.000000,0.000000}%
\pgfsetstrokecolor{currentstroke}%
\pgfsetdash{}{0pt}%
\pgfsys@defobject{currentmarker}{\pgfqpoint{-0.055556in}{0.000000in}}{\pgfqpoint{0.000000in}{0.000000in}}{%
\pgfpathmoveto{\pgfqpoint{0.000000in}{0.000000in}}%
\pgfpathlineto{\pgfqpoint{-0.055556in}{0.000000in}}%
\pgfusepath{stroke,fill}%
}%
\begin{pgfscope}%
\pgfsys@transformshift{0.705864in}{1.710284in}%
\pgfsys@useobject{currentmarker}{}%
\end{pgfscope}%
\end{pgfscope}%
\begin{pgfscope}%
\pgftext[x=0.263520in,y=1.678167in,left,base]{\sffamily\fontsize{6.664000}{7.996800}\selectfont \(\displaystyle -60.90\)}%
\end{pgfscope}%
\begin{pgfscope}%
\pgfsetbuttcap%
\pgfsetroundjoin%
\definecolor{currentfill}{rgb}{0.000000,0.000000,0.000000}%
\pgfsetfillcolor{currentfill}%
\pgfsetlinewidth{0.501875pt}%
\definecolor{currentstroke}{rgb}{0.000000,0.000000,0.000000}%
\pgfsetstrokecolor{currentstroke}%
\pgfsetdash{}{0pt}%
\pgfsys@defobject{currentmarker}{\pgfqpoint{-0.055556in}{0.000000in}}{\pgfqpoint{0.000000in}{0.000000in}}{%
\pgfpathmoveto{\pgfqpoint{0.000000in}{0.000000in}}%
\pgfpathlineto{\pgfqpoint{-0.055556in}{0.000000in}}%
\pgfusepath{stroke,fill}%
}%
\begin{pgfscope}%
\pgfsys@transformshift{0.705864in}{1.423098in}%
\pgfsys@useobject{currentmarker}{}%
\end{pgfscope}%
\end{pgfscope}%
\begin{pgfscope}%
\pgftext[x=0.263520in,y=1.390981in,left,base]{\sffamily\fontsize{6.664000}{7.996800}\selectfont \(\displaystyle -60.85\)}%
\end{pgfscope}%
\begin{pgfscope}%
\pgfsetbuttcap%
\pgfsetroundjoin%
\definecolor{currentfill}{rgb}{0.000000,0.000000,0.000000}%
\pgfsetfillcolor{currentfill}%
\pgfsetlinewidth{0.501875pt}%
\definecolor{currentstroke}{rgb}{0.000000,0.000000,0.000000}%
\pgfsetstrokecolor{currentstroke}%
\pgfsetdash{}{0pt}%
\pgfsys@defobject{currentmarker}{\pgfqpoint{-0.055556in}{0.000000in}}{\pgfqpoint{0.000000in}{0.000000in}}{%
\pgfpathmoveto{\pgfqpoint{0.000000in}{0.000000in}}%
\pgfpathlineto{\pgfqpoint{-0.055556in}{0.000000in}}%
\pgfusepath{stroke,fill}%
}%
\begin{pgfscope}%
\pgfsys@transformshift{0.705864in}{1.135912in}%
\pgfsys@useobject{currentmarker}{}%
\end{pgfscope}%
\end{pgfscope}%
\begin{pgfscope}%
\pgftext[x=0.263520in,y=1.103796in,left,base]{\sffamily\fontsize{6.664000}{7.996800}\selectfont \(\displaystyle -60.80\)}%
\end{pgfscope}%
\begin{pgfscope}%
\pgfsetbuttcap%
\pgfsetroundjoin%
\definecolor{currentfill}{rgb}{0.000000,0.000000,0.000000}%
\pgfsetfillcolor{currentfill}%
\pgfsetlinewidth{0.501875pt}%
\definecolor{currentstroke}{rgb}{0.000000,0.000000,0.000000}%
\pgfsetstrokecolor{currentstroke}%
\pgfsetdash{}{0pt}%
\pgfsys@defobject{currentmarker}{\pgfqpoint{-0.055556in}{0.000000in}}{\pgfqpoint{0.000000in}{0.000000in}}{%
\pgfpathmoveto{\pgfqpoint{0.000000in}{0.000000in}}%
\pgfpathlineto{\pgfqpoint{-0.055556in}{0.000000in}}%
\pgfusepath{stroke,fill}%
}%
\begin{pgfscope}%
\pgfsys@transformshift{0.705864in}{0.848727in}%
\pgfsys@useobject{currentmarker}{}%
\end{pgfscope}%
\end{pgfscope}%
\begin{pgfscope}%
\pgftext[x=0.263520in,y=0.816610in,left,base]{\sffamily\fontsize{6.664000}{7.996800}\selectfont \(\displaystyle -60.75\)}%
\end{pgfscope}%
\begin{pgfscope}%
\pgfsetbuttcap%
\pgfsetroundjoin%
\definecolor{currentfill}{rgb}{0.000000,0.000000,0.000000}%
\pgfsetfillcolor{currentfill}%
\pgfsetlinewidth{0.501875pt}%
\definecolor{currentstroke}{rgb}{0.000000,0.000000,0.000000}%
\pgfsetstrokecolor{currentstroke}%
\pgfsetdash{}{0pt}%
\pgfsys@defobject{currentmarker}{\pgfqpoint{-0.055556in}{0.000000in}}{\pgfqpoint{0.000000in}{0.000000in}}{%
\pgfpathmoveto{\pgfqpoint{0.000000in}{0.000000in}}%
\pgfpathlineto{\pgfqpoint{-0.055556in}{0.000000in}}%
\pgfusepath{stroke,fill}%
}%
\begin{pgfscope}%
\pgfsys@transformshift{0.705864in}{0.561541in}%
\pgfsys@useobject{currentmarker}{}%
\end{pgfscope}%
\end{pgfscope}%
\begin{pgfscope}%
\pgftext[x=0.263520in,y=0.529424in,left,base]{\sffamily\fontsize{6.664000}{7.996800}\selectfont \(\displaystyle -60.70\)}%
\end{pgfscope}%
\begin{pgfscope}%
\pgftext[x=0.207964in,y=1.308224in,,bottom,rotate=90.000000]{\sffamily\fontsize{9.600000}{11.520000}\selectfont Declination / deg}%
\end{pgfscope}%
\begin{pgfscope}%
\pgfpathrectangle{\pgfqpoint{0.705864in}{0.446667in}}{\pgfqpoint{2.753169in}{1.723114in}}%
\pgfusepath{clip}%
\pgfsetbuttcap%
\pgfsetbeveljoin%
\definecolor{currentfill}{rgb}{0.000000,0.000000,0.000000}%
\pgfsetfillcolor{currentfill}%
\pgfsetlinewidth{1.003750pt}%
\definecolor{currentstroke}{rgb}{0.000000,0.000000,0.000000}%
\pgfsetstrokecolor{currentstroke}%
\pgfsetdash{}{0pt}%
\pgfsys@defobject{currentmarker}{\pgfqpoint{-0.033023in}{-0.028091in}}{\pgfqpoint{0.033023in}{0.034722in}}{%
\pgfpathmoveto{\pgfqpoint{0.000000in}{0.034722in}}%
\pgfpathlineto{\pgfqpoint{-0.007796in}{0.010730in}}%
\pgfpathlineto{\pgfqpoint{-0.033023in}{0.010730in}}%
\pgfpathlineto{\pgfqpoint{-0.012614in}{-0.004098in}}%
\pgfpathlineto{\pgfqpoint{-0.020409in}{-0.028091in}}%
\pgfpathlineto{\pgfqpoint{-0.000000in}{-0.013263in}}%
\pgfpathlineto{\pgfqpoint{0.020409in}{-0.028091in}}%
\pgfpathlineto{\pgfqpoint{0.012614in}{-0.004098in}}%
\pgfpathlineto{\pgfqpoint{0.033023in}{0.010730in}}%
\pgfpathlineto{\pgfqpoint{0.007796in}{0.010730in}}%
\pgfpathclose%
\pgfusepath{stroke,fill}%
}%
\begin{pgfscope}%
\pgfsys@transformshift{2.102130in}{1.331084in}%
\pgfsys@useobject{currentmarker}{}%
\end{pgfscope}%
\end{pgfscope}%
\begin{pgfscope}%
\pgfpathrectangle{\pgfqpoint{0.705864in}{0.446667in}}{\pgfqpoint{2.753169in}{1.723114in}}%
\pgfusepath{clip}%
\pgfsetbuttcap%
\pgfsetroundjoin%
\definecolor{currentfill}{rgb}{0.000000,0.000000,0.000000}%
\pgfsetfillcolor{currentfill}%
\pgfsetlinewidth{1.003750pt}%
\definecolor{currentstroke}{rgb}{0.000000,0.000000,0.000000}%
\pgfsetstrokecolor{currentstroke}%
\pgfsetdash{}{0pt}%
\pgfsys@defobject{currentmarker}{\pgfqpoint{-0.010417in}{-0.010417in}}{\pgfqpoint{0.010417in}{0.010417in}}{%
\pgfpathmoveto{\pgfqpoint{0.000000in}{-0.010417in}}%
\pgfpathcurveto{\pgfqpoint{0.002763in}{-0.010417in}}{\pgfqpoint{0.005412in}{-0.009319in}}{\pgfqpoint{0.007366in}{-0.007366in}}%
\pgfpathcurveto{\pgfqpoint{0.009319in}{-0.005412in}}{\pgfqpoint{0.010417in}{-0.002763in}}{\pgfqpoint{0.010417in}{0.000000in}}%
\pgfpathcurveto{\pgfqpoint{0.010417in}{0.002763in}}{\pgfqpoint{0.009319in}{0.005412in}}{\pgfqpoint{0.007366in}{0.007366in}}%
\pgfpathcurveto{\pgfqpoint{0.005412in}{0.009319in}}{\pgfqpoint{0.002763in}{0.010417in}}{\pgfqpoint{0.000000in}{0.010417in}}%
\pgfpathcurveto{\pgfqpoint{-0.002763in}{0.010417in}}{\pgfqpoint{-0.005412in}{0.009319in}}{\pgfqpoint{-0.007366in}{0.007366in}}%
\pgfpathcurveto{\pgfqpoint{-0.009319in}{0.005412in}}{\pgfqpoint{-0.010417in}{0.002763in}}{\pgfqpoint{-0.010417in}{0.000000in}}%
\pgfpathcurveto{\pgfqpoint{-0.010417in}{-0.002763in}}{\pgfqpoint{-0.009319in}{-0.005412in}}{\pgfqpoint{-0.007366in}{-0.007366in}}%
\pgfpathcurveto{\pgfqpoint{-0.005412in}{-0.009319in}}{\pgfqpoint{-0.002763in}{-0.010417in}}{\pgfqpoint{0.000000in}{-0.010417in}}%
\pgfpathclose%
\pgfusepath{stroke,fill}%
}%
\begin{pgfscope}%
\pgfsys@transformshift{2.858312in}{1.327525in}%
\pgfsys@useobject{currentmarker}{}%
\end{pgfscope}%
\end{pgfscope}%
\begin{pgfscope}%
\pgfpathrectangle{\pgfqpoint{0.705864in}{0.446667in}}{\pgfqpoint{2.753169in}{1.723114in}}%
\pgfusepath{clip}%
\pgfsetbuttcap%
\pgfsetroundjoin%
\definecolor{currentfill}{rgb}{0.501961,0.501961,0.501961}%
\pgfsetfillcolor{currentfill}%
\pgfsetlinewidth{1.003750pt}%
\definecolor{currentstroke}{rgb}{0.501961,0.501961,0.501961}%
\pgfsetstrokecolor{currentstroke}%
\pgfsetdash{}{0pt}%
\pgfsys@defobject{currentmarker}{\pgfqpoint{-0.020833in}{-0.020833in}}{\pgfqpoint{0.020833in}{0.020833in}}{%
\pgfpathmoveto{\pgfqpoint{-0.020833in}{-0.020833in}}%
\pgfpathlineto{\pgfqpoint{0.020833in}{0.020833in}}%
\pgfpathmoveto{\pgfqpoint{-0.020833in}{0.020833in}}%
\pgfpathlineto{\pgfqpoint{0.020833in}{-0.020833in}}%
\pgfusepath{stroke,fill}%
}%
\begin{pgfscope}%
\pgfsys@transformshift{2.102130in}{0.965013in}%
\pgfsys@useobject{currentmarker}{}%
\end{pgfscope}%
\end{pgfscope}%
\begin{pgfscope}%
\pgfpathrectangle{\pgfqpoint{0.705864in}{0.446667in}}{\pgfqpoint{2.753169in}{1.723114in}}%
\pgfusepath{clip}%
\pgfsetbuttcap%
\pgfsetroundjoin%
\definecolor{currentfill}{rgb}{0.000000,0.000000,0.000000}%
\pgfsetfillcolor{currentfill}%
\pgfsetlinewidth{1.003750pt}%
\definecolor{currentstroke}{rgb}{0.000000,0.000000,0.000000}%
\pgfsetstrokecolor{currentstroke}%
\pgfsetdash{}{0pt}%
\pgfsys@defobject{currentmarker}{\pgfqpoint{-0.010417in}{-0.010417in}}{\pgfqpoint{0.010417in}{0.010417in}}{%
\pgfpathmoveto{\pgfqpoint{0.000000in}{-0.010417in}}%
\pgfpathcurveto{\pgfqpoint{0.002763in}{-0.010417in}}{\pgfqpoint{0.005412in}{-0.009319in}}{\pgfqpoint{0.007366in}{-0.007366in}}%
\pgfpathcurveto{\pgfqpoint{0.009319in}{-0.005412in}}{\pgfqpoint{0.010417in}{-0.002763in}}{\pgfqpoint{0.010417in}{0.000000in}}%
\pgfpathcurveto{\pgfqpoint{0.010417in}{0.002763in}}{\pgfqpoint{0.009319in}{0.005412in}}{\pgfqpoint{0.007366in}{0.007366in}}%
\pgfpathcurveto{\pgfqpoint{0.005412in}{0.009319in}}{\pgfqpoint{0.002763in}{0.010417in}}{\pgfqpoint{0.000000in}{0.010417in}}%
\pgfpathcurveto{\pgfqpoint{-0.002763in}{0.010417in}}{\pgfqpoint{-0.005412in}{0.009319in}}{\pgfqpoint{-0.007366in}{0.007366in}}%
\pgfpathcurveto{\pgfqpoint{-0.009319in}{0.005412in}}{\pgfqpoint{-0.010417in}{0.002763in}}{\pgfqpoint{-0.010417in}{0.000000in}}%
\pgfpathcurveto{\pgfqpoint{-0.010417in}{-0.002763in}}{\pgfqpoint{-0.009319in}{-0.005412in}}{\pgfqpoint{-0.007366in}{-0.007366in}}%
\pgfpathcurveto{\pgfqpoint{-0.005412in}{-0.009319in}}{\pgfqpoint{-0.002763in}{-0.010417in}}{\pgfqpoint{0.000000in}{-0.010417in}}%
\pgfpathclose%
\pgfusepath{stroke,fill}%
}%
\begin{pgfscope}%
\pgfsys@transformshift{3.121278in}{0.884065in}%
\pgfsys@useobject{currentmarker}{}%
\end{pgfscope}%
\end{pgfscope}%
\begin{pgfscope}%
\pgfpathrectangle{\pgfqpoint{0.705864in}{0.446667in}}{\pgfqpoint{2.753169in}{1.723114in}}%
\pgfusepath{clip}%
\pgfsetbuttcap%
\pgfsetroundjoin%
\definecolor{currentfill}{rgb}{0.000000,0.000000,0.000000}%
\pgfsetfillcolor{currentfill}%
\pgfsetlinewidth{1.003750pt}%
\definecolor{currentstroke}{rgb}{0.000000,0.000000,0.000000}%
\pgfsetstrokecolor{currentstroke}%
\pgfsetdash{}{0pt}%
\pgfsys@defobject{currentmarker}{\pgfqpoint{-0.010417in}{-0.010417in}}{\pgfqpoint{0.010417in}{0.010417in}}{%
\pgfpathmoveto{\pgfqpoint{0.000000in}{-0.010417in}}%
\pgfpathcurveto{\pgfqpoint{0.002763in}{-0.010417in}}{\pgfqpoint{0.005412in}{-0.009319in}}{\pgfqpoint{0.007366in}{-0.007366in}}%
\pgfpathcurveto{\pgfqpoint{0.009319in}{-0.005412in}}{\pgfqpoint{0.010417in}{-0.002763in}}{\pgfqpoint{0.010417in}{0.000000in}}%
\pgfpathcurveto{\pgfqpoint{0.010417in}{0.002763in}}{\pgfqpoint{0.009319in}{0.005412in}}{\pgfqpoint{0.007366in}{0.007366in}}%
\pgfpathcurveto{\pgfqpoint{0.005412in}{0.009319in}}{\pgfqpoint{0.002763in}{0.010417in}}{\pgfqpoint{0.000000in}{0.010417in}}%
\pgfpathcurveto{\pgfqpoint{-0.002763in}{0.010417in}}{\pgfqpoint{-0.005412in}{0.009319in}}{\pgfqpoint{-0.007366in}{0.007366in}}%
\pgfpathcurveto{\pgfqpoint{-0.009319in}{0.005412in}}{\pgfqpoint{-0.010417in}{0.002763in}}{\pgfqpoint{-0.010417in}{0.000000in}}%
\pgfpathcurveto{\pgfqpoint{-0.010417in}{-0.002763in}}{\pgfqpoint{-0.009319in}{-0.005412in}}{\pgfqpoint{-0.007366in}{-0.007366in}}%
\pgfpathcurveto{\pgfqpoint{-0.005412in}{-0.009319in}}{\pgfqpoint{-0.002763in}{-0.010417in}}{\pgfqpoint{0.000000in}{-0.010417in}}%
\pgfpathclose%
\pgfusepath{stroke,fill}%
}%
\begin{pgfscope}%
\pgfsys@transformshift{2.555574in}{0.706823in}%
\pgfsys@useobject{currentmarker}{}%
\end{pgfscope}%
\end{pgfscope}%
\begin{pgfscope}%
\pgfpathrectangle{\pgfqpoint{0.705864in}{0.446667in}}{\pgfqpoint{2.753169in}{1.723114in}}%
\pgfusepath{clip}%
\pgfsetbuttcap%
\pgfsetroundjoin%
\definecolor{currentfill}{rgb}{0.000000,0.000000,0.000000}%
\pgfsetfillcolor{currentfill}%
\pgfsetlinewidth{1.003750pt}%
\definecolor{currentstroke}{rgb}{0.000000,0.000000,0.000000}%
\pgfsetstrokecolor{currentstroke}%
\pgfsetdash{}{0pt}%
\pgfsys@defobject{currentmarker}{\pgfqpoint{-0.010417in}{-0.010417in}}{\pgfqpoint{0.010417in}{0.010417in}}{%
\pgfpathmoveto{\pgfqpoint{0.000000in}{-0.010417in}}%
\pgfpathcurveto{\pgfqpoint{0.002763in}{-0.010417in}}{\pgfqpoint{0.005412in}{-0.009319in}}{\pgfqpoint{0.007366in}{-0.007366in}}%
\pgfpathcurveto{\pgfqpoint{0.009319in}{-0.005412in}}{\pgfqpoint{0.010417in}{-0.002763in}}{\pgfqpoint{0.010417in}{0.000000in}}%
\pgfpathcurveto{\pgfqpoint{0.010417in}{0.002763in}}{\pgfqpoint{0.009319in}{0.005412in}}{\pgfqpoint{0.007366in}{0.007366in}}%
\pgfpathcurveto{\pgfqpoint{0.005412in}{0.009319in}}{\pgfqpoint{0.002763in}{0.010417in}}{\pgfqpoint{0.000000in}{0.010417in}}%
\pgfpathcurveto{\pgfqpoint{-0.002763in}{0.010417in}}{\pgfqpoint{-0.005412in}{0.009319in}}{\pgfqpoint{-0.007366in}{0.007366in}}%
\pgfpathcurveto{\pgfqpoint{-0.009319in}{0.005412in}}{\pgfqpoint{-0.010417in}{0.002763in}}{\pgfqpoint{-0.010417in}{0.000000in}}%
\pgfpathcurveto{\pgfqpoint{-0.010417in}{-0.002763in}}{\pgfqpoint{-0.009319in}{-0.005412in}}{\pgfqpoint{-0.007366in}{-0.007366in}}%
\pgfpathcurveto{\pgfqpoint{-0.005412in}{-0.009319in}}{\pgfqpoint{-0.002763in}{-0.010417in}}{\pgfqpoint{0.000000in}{-0.010417in}}%
\pgfpathclose%
\pgfusepath{stroke,fill}%
}%
\begin{pgfscope}%
\pgfsys@transformshift{2.672751in}{0.104115in}%
\pgfsys@useobject{currentmarker}{}%
\end{pgfscope}%
\end{pgfscope}%
\begin{pgfscope}%
\pgfpathrectangle{\pgfqpoint{0.705864in}{0.446667in}}{\pgfqpoint{2.753169in}{1.723114in}}%
\pgfusepath{clip}%
\pgfsetbuttcap%
\pgfsetroundjoin%
\pgfsetlinewidth{0.501875pt}%
\definecolor{currentstroke}{rgb}{0.501961,0.501961,0.501961}%
\pgfsetstrokecolor{currentstroke}%
\pgfsetdash{{1.850000pt}{0.800000pt}}{0.000000pt}%
\pgfpathmoveto{\pgfqpoint{1.708253in}{1.331084in}}%
\pgfpathlineto{\pgfqpoint{1.709041in}{1.305857in}}%
\pgfpathlineto{\pgfqpoint{1.710618in}{1.287434in}}%
\pgfpathlineto{\pgfqpoint{1.712984in}{1.269446in}}%
\pgfpathlineto{\pgfqpoint{1.715350in}{1.255708in}}%
\pgfpathlineto{\pgfqpoint{1.718504in}{1.240676in}}%
\pgfpathlineto{\pgfqpoint{1.722446in}{1.224972in}}%
\pgfpathlineto{\pgfqpoint{1.726389in}{1.211442in}}%
\pgfpathlineto{\pgfqpoint{1.731120in}{1.197153in}}%
\pgfpathlineto{\pgfqpoint{1.736640in}{1.182402in}}%
\pgfpathlineto{\pgfqpoint{1.742160in}{1.169180in}}%
\pgfpathlineto{\pgfqpoint{1.748468in}{1.155500in}}%
\pgfpathlineto{\pgfqpoint{1.755565in}{1.141543in}}%
\pgfpathlineto{\pgfqpoint{1.762662in}{1.128800in}}%
\pgfpathlineto{\pgfqpoint{1.770548in}{1.115804in}}%
\pgfpathlineto{\pgfqpoint{1.778433in}{1.103830in}}%
\pgfpathlineto{\pgfqpoint{1.787107in}{1.091656in}}%
\pgfpathlineto{\pgfqpoint{1.796570in}{1.079394in}}%
\pgfpathlineto{\pgfqpoint{1.806032in}{1.068051in}}%
\pgfpathlineto{\pgfqpoint{1.816283in}{1.056668in}}%
\pgfpathlineto{\pgfqpoint{1.826534in}{1.046117in}}%
\pgfpathlineto{\pgfqpoint{1.837574in}{1.035582in}}%
\pgfpathlineto{\pgfqpoint{1.848614in}{1.025819in}}%
\pgfpathlineto{\pgfqpoint{1.860442in}{1.016134in}}%
\pgfpathlineto{\pgfqpoint{1.872270in}{1.007181in}}%
\pgfpathlineto{\pgfqpoint{1.884098in}{0.998900in}}%
\pgfpathlineto{\pgfqpoint{1.896715in}{0.990754in}}%
\pgfpathlineto{\pgfqpoint{1.909331in}{0.983266in}}%
\pgfpathlineto{\pgfqpoint{1.921948in}{0.976397in}}%
\pgfpathlineto{\pgfqpoint{1.935353in}{0.969736in}}%
\pgfpathlineto{\pgfqpoint{1.948759in}{0.963696in}}%
\pgfpathlineto{\pgfqpoint{1.962164in}{0.958248in}}%
\pgfpathlineto{\pgfqpoint{1.976358in}{0.953097in}}%
\pgfpathlineto{\pgfqpoint{1.990551in}{0.948554in}}%
\pgfpathlineto{\pgfqpoint{2.004745in}{0.944599in}}%
\pgfpathlineto{\pgfqpoint{2.018939in}{0.941213in}}%
\pgfpathlineto{\pgfqpoint{2.033133in}{0.938382in}}%
\pgfpathlineto{\pgfqpoint{2.047327in}{0.936095in}}%
\pgfpathlineto{\pgfqpoint{2.062309in}{0.934258in}}%
\pgfpathlineto{\pgfqpoint{2.077291in}{0.933009in}}%
\pgfpathlineto{\pgfqpoint{2.092274in}{0.932340in}}%
\pgfpathlineto{\pgfqpoint{2.107256in}{0.932249in}}%
\pgfpathlineto{\pgfqpoint{2.122238in}{0.932735in}}%
\pgfpathlineto{\pgfqpoint{2.137221in}{0.933801in}}%
\pgfpathlineto{\pgfqpoint{2.152203in}{0.935451in}}%
\pgfpathlineto{\pgfqpoint{2.166397in}{0.937560in}}%
\pgfpathlineto{\pgfqpoint{2.180591in}{0.940208in}}%
\pgfpathlineto{\pgfqpoint{2.194784in}{0.943408in}}%
\pgfpathlineto{\pgfqpoint{2.208978in}{0.947171in}}%
\pgfpathlineto{\pgfqpoint{2.223172in}{0.951516in}}%
\pgfpathlineto{\pgfqpoint{2.237366in}{0.956462in}}%
\pgfpathlineto{\pgfqpoint{2.250771in}{0.961707in}}%
\pgfpathlineto{\pgfqpoint{2.264176in}{0.967535in}}%
\pgfpathlineto{\pgfqpoint{2.277581in}{0.973973in}}%
\pgfpathlineto{\pgfqpoint{2.290987in}{0.981055in}}%
\pgfpathlineto{\pgfqpoint{2.303603in}{0.988345in}}%
\pgfpathlineto{\pgfqpoint{2.316220in}{0.996281in}}%
\pgfpathlineto{\pgfqpoint{2.328048in}{1.004349in}}%
\pgfpathlineto{\pgfqpoint{2.339876in}{1.013072in}}%
\pgfpathlineto{\pgfqpoint{2.351705in}{1.022505in}}%
\pgfpathlineto{\pgfqpoint{2.362744in}{1.032011in}}%
\pgfpathlineto{\pgfqpoint{2.373784in}{1.042261in}}%
\pgfpathlineto{\pgfqpoint{2.384035in}{1.052517in}}%
\pgfpathlineto{\pgfqpoint{2.394286in}{1.063568in}}%
\pgfpathlineto{\pgfqpoint{2.403749in}{1.074563in}}%
\pgfpathlineto{\pgfqpoint{2.413211in}{1.086426in}}%
\pgfpathlineto{\pgfqpoint{2.421885in}{1.098177in}}%
\pgfpathlineto{\pgfqpoint{2.430559in}{1.110902in}}%
\pgfpathlineto{\pgfqpoint{2.438444in}{1.123467in}}%
\pgfpathlineto{\pgfqpoint{2.445541in}{1.135744in}}%
\pgfpathlineto{\pgfqpoint{2.452638in}{1.149130in}}%
\pgfpathlineto{\pgfqpoint{2.458947in}{1.162171in}}%
\pgfpathlineto{\pgfqpoint{2.465255in}{1.176572in}}%
\pgfpathlineto{\pgfqpoint{2.470775in}{1.190614in}}%
\pgfpathlineto{\pgfqpoint{2.475506in}{1.204074in}}%
\pgfpathlineto{\pgfqpoint{2.480237in}{1.219346in}}%
\pgfpathlineto{\pgfqpoint{2.484180in}{1.234069in}}%
\pgfpathlineto{\pgfqpoint{2.487334in}{1.247837in}}%
\pgfpathlineto{\pgfqpoint{2.490488in}{1.264541in}}%
\pgfpathlineto{\pgfqpoint{2.492854in}{1.280706in}}%
\pgfpathlineto{\pgfqpoint{2.494431in}{1.295426in}}%
\pgfpathlineto{\pgfqpoint{2.495220in}{1.305857in}}%
\pgfpathlineto{\pgfqpoint{2.496008in}{1.331084in}}%
\pgfpathlineto{\pgfqpoint{2.496008in}{1.331084in}}%
\pgfusepath{stroke}%
\end{pgfscope}%
\begin{pgfscope}%
\pgfpathrectangle{\pgfqpoint{0.705864in}{0.446667in}}{\pgfqpoint{2.753169in}{1.723114in}}%
\pgfusepath{clip}%
\pgfsetbuttcap%
\pgfsetroundjoin%
\pgfsetlinewidth{0.501875pt}%
\definecolor{currentstroke}{rgb}{0.501961,0.501961,0.501961}%
\pgfsetstrokecolor{currentstroke}%
\pgfsetdash{{1.850000pt}{0.800000pt}}{0.000000pt}%
\pgfpathmoveto{\pgfqpoint{1.708253in}{1.331084in}}%
\pgfpathlineto{\pgfqpoint{1.709041in}{1.356310in}}%
\pgfpathlineto{\pgfqpoint{1.710618in}{1.374734in}}%
\pgfpathlineto{\pgfqpoint{1.712984in}{1.392721in}}%
\pgfpathlineto{\pgfqpoint{1.715350in}{1.406460in}}%
\pgfpathlineto{\pgfqpoint{1.718504in}{1.421491in}}%
\pgfpathlineto{\pgfqpoint{1.722446in}{1.437196in}}%
\pgfpathlineto{\pgfqpoint{1.726389in}{1.450726in}}%
\pgfpathlineto{\pgfqpoint{1.731120in}{1.465014in}}%
\pgfpathlineto{\pgfqpoint{1.736640in}{1.479766in}}%
\pgfpathlineto{\pgfqpoint{1.742160in}{1.492988in}}%
\pgfpathlineto{\pgfqpoint{1.748468in}{1.506667in}}%
\pgfpathlineto{\pgfqpoint{1.755565in}{1.520625in}}%
\pgfpathlineto{\pgfqpoint{1.762662in}{1.533367in}}%
\pgfpathlineto{\pgfqpoint{1.770548in}{1.546363in}}%
\pgfpathlineto{\pgfqpoint{1.778433in}{1.558337in}}%
\pgfpathlineto{\pgfqpoint{1.787107in}{1.570512in}}%
\pgfpathlineto{\pgfqpoint{1.796570in}{1.582774in}}%
\pgfpathlineto{\pgfqpoint{1.806032in}{1.594116in}}%
\pgfpathlineto{\pgfqpoint{1.816283in}{1.605499in}}%
\pgfpathlineto{\pgfqpoint{1.826534in}{1.616051in}}%
\pgfpathlineto{\pgfqpoint{1.837574in}{1.626586in}}%
\pgfpathlineto{\pgfqpoint{1.848614in}{1.636348in}}%
\pgfpathlineto{\pgfqpoint{1.860442in}{1.646033in}}%
\pgfpathlineto{\pgfqpoint{1.872270in}{1.654987in}}%
\pgfpathlineto{\pgfqpoint{1.884098in}{1.663267in}}%
\pgfpathlineto{\pgfqpoint{1.896715in}{1.671414in}}%
\pgfpathlineto{\pgfqpoint{1.909331in}{1.678901in}}%
\pgfpathlineto{\pgfqpoint{1.921948in}{1.685771in}}%
\pgfpathlineto{\pgfqpoint{1.935353in}{1.692432in}}%
\pgfpathlineto{\pgfqpoint{1.948759in}{1.698471in}}%
\pgfpathlineto{\pgfqpoint{1.962164in}{1.703919in}}%
\pgfpathlineto{\pgfqpoint{1.976358in}{1.709071in}}%
\pgfpathlineto{\pgfqpoint{1.990551in}{1.713614in}}%
\pgfpathlineto{\pgfqpoint{2.004745in}{1.717569in}}%
\pgfpathlineto{\pgfqpoint{2.018939in}{1.720955in}}%
\pgfpathlineto{\pgfqpoint{2.033133in}{1.723785in}}%
\pgfpathlineto{\pgfqpoint{2.047327in}{1.726073in}}%
\pgfpathlineto{\pgfqpoint{2.062309in}{1.727909in}}%
\pgfpathlineto{\pgfqpoint{2.077291in}{1.729159in}}%
\pgfpathlineto{\pgfqpoint{2.092274in}{1.729828in}}%
\pgfpathlineto{\pgfqpoint{2.107256in}{1.729919in}}%
\pgfpathlineto{\pgfqpoint{2.122238in}{1.729433in}}%
\pgfpathlineto{\pgfqpoint{2.137221in}{1.728367in}}%
\pgfpathlineto{\pgfqpoint{2.152203in}{1.726717in}}%
\pgfpathlineto{\pgfqpoint{2.166397in}{1.724608in}}%
\pgfpathlineto{\pgfqpoint{2.180591in}{1.721959in}}%
\pgfpathlineto{\pgfqpoint{2.194784in}{1.718760in}}%
\pgfpathlineto{\pgfqpoint{2.208978in}{1.714996in}}%
\pgfpathlineto{\pgfqpoint{2.223172in}{1.710652in}}%
\pgfpathlineto{\pgfqpoint{2.237366in}{1.705706in}}%
\pgfpathlineto{\pgfqpoint{2.250771in}{1.700460in}}%
\pgfpathlineto{\pgfqpoint{2.264176in}{1.694633in}}%
\pgfpathlineto{\pgfqpoint{2.277581in}{1.688195in}}%
\pgfpathlineto{\pgfqpoint{2.290987in}{1.681113in}}%
\pgfpathlineto{\pgfqpoint{2.303603in}{1.673823in}}%
\pgfpathlineto{\pgfqpoint{2.316220in}{1.665887in}}%
\pgfpathlineto{\pgfqpoint{2.328048in}{1.657819in}}%
\pgfpathlineto{\pgfqpoint{2.339876in}{1.649096in}}%
\pgfpathlineto{\pgfqpoint{2.351705in}{1.639662in}}%
\pgfpathlineto{\pgfqpoint{2.362744in}{1.630157in}}%
\pgfpathlineto{\pgfqpoint{2.373784in}{1.619907in}}%
\pgfpathlineto{\pgfqpoint{2.384035in}{1.609651in}}%
\pgfpathlineto{\pgfqpoint{2.394286in}{1.598600in}}%
\pgfpathlineto{\pgfqpoint{2.403749in}{1.587604in}}%
\pgfpathlineto{\pgfqpoint{2.413211in}{1.575741in}}%
\pgfpathlineto{\pgfqpoint{2.421885in}{1.563991in}}%
\pgfpathlineto{\pgfqpoint{2.430559in}{1.551266in}}%
\pgfpathlineto{\pgfqpoint{2.438444in}{1.538700in}}%
\pgfpathlineto{\pgfqpoint{2.445541in}{1.526424in}}%
\pgfpathlineto{\pgfqpoint{2.452638in}{1.513038in}}%
\pgfpathlineto{\pgfqpoint{2.458947in}{1.499996in}}%
\pgfpathlineto{\pgfqpoint{2.465255in}{1.485596in}}%
\pgfpathlineto{\pgfqpoint{2.470775in}{1.471554in}}%
\pgfpathlineto{\pgfqpoint{2.475506in}{1.458094in}}%
\pgfpathlineto{\pgfqpoint{2.480237in}{1.442822in}}%
\pgfpathlineto{\pgfqpoint{2.484180in}{1.428099in}}%
\pgfpathlineto{\pgfqpoint{2.487334in}{1.414331in}}%
\pgfpathlineto{\pgfqpoint{2.490488in}{1.397626in}}%
\pgfpathlineto{\pgfqpoint{2.492854in}{1.381461in}}%
\pgfpathlineto{\pgfqpoint{2.494431in}{1.366742in}}%
\pgfpathlineto{\pgfqpoint{2.495220in}{1.356310in}}%
\pgfpathlineto{\pgfqpoint{2.496008in}{1.331084in}}%
\pgfpathlineto{\pgfqpoint{2.496008in}{1.331084in}}%
\pgfusepath{stroke}%
\end{pgfscope}%
\begin{pgfscope}%
\pgfsetrectcap%
\pgfsetmiterjoin%
\pgfsetlinewidth{0.501875pt}%
\definecolor{currentstroke}{rgb}{0.000000,0.000000,0.000000}%
\pgfsetstrokecolor{currentstroke}%
\pgfsetdash{}{0pt}%
\pgfpathmoveto{\pgfqpoint{0.705864in}{0.446667in}}%
\pgfpathlineto{\pgfqpoint{0.705864in}{2.169781in}}%
\pgfusepath{stroke}%
\end{pgfscope}%
\begin{pgfscope}%
\pgfsetrectcap%
\pgfsetmiterjoin%
\pgfsetlinewidth{0.501875pt}%
\definecolor{currentstroke}{rgb}{0.000000,0.000000,0.000000}%
\pgfsetstrokecolor{currentstroke}%
\pgfsetdash{}{0pt}%
\pgfpathmoveto{\pgfqpoint{0.705864in}{0.446667in}}%
\pgfpathlineto{\pgfqpoint{3.459033in}{0.446667in}}%
\pgfusepath{stroke}%
\end{pgfscope}%
\begin{pgfscope}%
\pgftext[x=3.321374in,y=1.997470in,right,base]{\sffamily\fontsize{8.000000}{9.600000}\selectfont \texttt{DISTANCE\,=\,200}}%
\end{pgfscope}%
\begin{pgfscope}%
\pgfsetbuttcap%
\pgfsetbeveljoin%
\definecolor{currentfill}{rgb}{0.000000,0.000000,0.000000}%
\pgfsetfillcolor{currentfill}%
\pgfsetlinewidth{1.003750pt}%
\definecolor{currentstroke}{rgb}{0.000000,0.000000,0.000000}%
\pgfsetstrokecolor{currentstroke}%
\pgfsetdash{}{0pt}%
\pgfsys@defobject{currentmarker}{\pgfqpoint{-0.033023in}{-0.028091in}}{\pgfqpoint{0.033023in}{0.034722in}}{%
\pgfpathmoveto{\pgfqpoint{0.000000in}{0.034722in}}%
\pgfpathlineto{\pgfqpoint{-0.007796in}{0.010730in}}%
\pgfpathlineto{\pgfqpoint{-0.033023in}{0.010730in}}%
\pgfpathlineto{\pgfqpoint{-0.012614in}{-0.004098in}}%
\pgfpathlineto{\pgfqpoint{-0.020409in}{-0.028091in}}%
\pgfpathlineto{\pgfqpoint{-0.000000in}{-0.013263in}}%
\pgfpathlineto{\pgfqpoint{0.020409in}{-0.028091in}}%
\pgfpathlineto{\pgfqpoint{0.012614in}{-0.004098in}}%
\pgfpathlineto{\pgfqpoint{0.033023in}{0.010730in}}%
\pgfpathlineto{\pgfqpoint{0.007796in}{0.010730in}}%
\pgfpathclose%
\pgfusepath{stroke,fill}%
}%
\begin{pgfscope}%
\pgfsys@transformshift{0.916975in}{0.917333in}%
\pgfsys@useobject{currentmarker}{}%
\end{pgfscope}%
\end{pgfscope}%
\begin{pgfscope}%
\pgftext[x=1.116975in,y=0.878444in,left,base]{\sffamily\fontsize{8.000000}{9.600000}\selectfont Star}%
\end{pgfscope}%
\begin{pgfscope}%
\pgfsetbuttcap%
\pgfsetroundjoin%
\definecolor{currentfill}{rgb}{0.000000,0.000000,0.000000}%
\pgfsetfillcolor{currentfill}%
\pgfsetlinewidth{1.003750pt}%
\definecolor{currentstroke}{rgb}{0.000000,0.000000,0.000000}%
\pgfsetstrokecolor{currentstroke}%
\pgfsetdash{}{0pt}%
\pgfsys@defobject{currentmarker}{\pgfqpoint{-0.010417in}{-0.010417in}}{\pgfqpoint{0.010417in}{0.010417in}}{%
\pgfpathmoveto{\pgfqpoint{0.000000in}{-0.010417in}}%
\pgfpathcurveto{\pgfqpoint{0.002763in}{-0.010417in}}{\pgfqpoint{0.005412in}{-0.009319in}}{\pgfqpoint{0.007366in}{-0.007366in}}%
\pgfpathcurveto{\pgfqpoint{0.009319in}{-0.005412in}}{\pgfqpoint{0.010417in}{-0.002763in}}{\pgfqpoint{0.010417in}{0.000000in}}%
\pgfpathcurveto{\pgfqpoint{0.010417in}{0.002763in}}{\pgfqpoint{0.009319in}{0.005412in}}{\pgfqpoint{0.007366in}{0.007366in}}%
\pgfpathcurveto{\pgfqpoint{0.005412in}{0.009319in}}{\pgfqpoint{0.002763in}{0.010417in}}{\pgfqpoint{0.000000in}{0.010417in}}%
\pgfpathcurveto{\pgfqpoint{-0.002763in}{0.010417in}}{\pgfqpoint{-0.005412in}{0.009319in}}{\pgfqpoint{-0.007366in}{0.007366in}}%
\pgfpathcurveto{\pgfqpoint{-0.009319in}{0.005412in}}{\pgfqpoint{-0.010417in}{0.002763in}}{\pgfqpoint{-0.010417in}{0.000000in}}%
\pgfpathcurveto{\pgfqpoint{-0.010417in}{-0.002763in}}{\pgfqpoint{-0.009319in}{-0.005412in}}{\pgfqpoint{-0.007366in}{-0.007366in}}%
\pgfpathcurveto{\pgfqpoint{-0.005412in}{-0.009319in}}{\pgfqpoint{-0.002763in}{-0.010417in}}{\pgfqpoint{0.000000in}{-0.010417in}}%
\pgfpathclose%
\pgfusepath{stroke,fill}%
}%
\begin{pgfscope}%
\pgfsys@transformshift{0.916975in}{0.762444in}%
\pgfsys@useobject{currentmarker}{}%
\end{pgfscope}%
\end{pgfscope}%
\begin{pgfscope}%
\pgftext[x=1.116975in,y=0.723555in,left,base]{\sffamily\fontsize{8.000000}{9.600000}\selectfont SSOs}%
\end{pgfscope}%
\begin{pgfscope}%
\pgfsetbuttcap%
\pgfsetroundjoin%
\definecolor{currentfill}{rgb}{0.501961,0.501961,0.501961}%
\pgfsetfillcolor{currentfill}%
\pgfsetlinewidth{1.003750pt}%
\definecolor{currentstroke}{rgb}{0.501961,0.501961,0.501961}%
\pgfsetstrokecolor{currentstroke}%
\pgfsetdash{}{0pt}%
\pgfsys@defobject{currentmarker}{\pgfqpoint{-0.020833in}{-0.020833in}}{\pgfqpoint{0.020833in}{0.020833in}}{%
\pgfpathmoveto{\pgfqpoint{-0.020833in}{-0.020833in}}%
\pgfpathlineto{\pgfqpoint{0.020833in}{0.020833in}}%
\pgfpathmoveto{\pgfqpoint{-0.020833in}{0.020833in}}%
\pgfpathlineto{\pgfqpoint{0.020833in}{-0.020833in}}%
\pgfusepath{stroke,fill}%
}%
\begin{pgfscope}%
\pgfsys@transformshift{0.916975in}{0.607111in}%
\pgfsys@useobject{currentmarker}{}%
\end{pgfscope}%
\end{pgfscope}%
\begin{pgfscope}%
\pgftext[x=1.116975in,y=0.568222in,left,base]{\sffamily\fontsize{8.000000}{9.600000}\selectfont Refraction spike}%
\end{pgfscope}%
\end{pgfpicture}%
\makeatother%
\endgroup%

%% file: gtc_runs.pgf
\begingroup%
\makeatletter%
\begin{pgfpicture}%
\pgfpathrectangle{\pgfpointorigin}{\pgfqpoint{3.611457in}{2.232003in}}%
\pgfusepath{use as bounding box, clip}%
\begin{pgfscope}%
\pgfsetbuttcap%
\pgfsetmiterjoin%
\definecolor{currentfill}{rgb}{1.000000,1.000000,1.000000}%
\pgfsetfillcolor{currentfill}%
\pgfsetlinewidth{0.000000pt}%
\definecolor{currentstroke}{rgb}{1.000000,1.000000,1.000000}%
\pgfsetstrokecolor{currentstroke}%
\pgfsetdash{}{0pt}%
\pgfpathmoveto{\pgfqpoint{0.000000in}{0.000000in}}%
\pgfpathlineto{\pgfqpoint{3.611457in}{0.000000in}}%
\pgfpathlineto{\pgfqpoint{3.611457in}{2.232003in}}%
\pgfpathlineto{\pgfqpoint{0.000000in}{2.232003in}}%
\pgfpathclose%
\pgfusepath{fill}%
\end{pgfscope}%
\begin{pgfscope}%
\pgfsetbuttcap%
\pgfsetmiterjoin%
\definecolor{currentfill}{rgb}{1.000000,1.000000,1.000000}%
\pgfsetfillcolor{currentfill}%
\pgfsetlinewidth{0.000000pt}%
\definecolor{currentstroke}{rgb}{0.000000,0.000000,0.000000}%
\pgfsetstrokecolor{currentstroke}%
\pgfsetstrokeopacity{0.000000}%
\pgfsetdash{}{0pt}%
\pgfpathmoveto{\pgfqpoint{0.330160in}{0.448876in}}%
\pgfpathlineto{\pgfqpoint{3.549235in}{0.448876in}}%
\pgfpathlineto{\pgfqpoint{3.549235in}{2.169781in}}%
\pgfpathlineto{\pgfqpoint{0.330160in}{2.169781in}}%
\pgfpathclose%
\pgfusepath{fill}%
\end{pgfscope}%
\begin{pgfscope}%
\pgfpathrectangle{\pgfqpoint{0.330160in}{0.448876in}}{\pgfqpoint{3.219075in}{1.720905in}}%
\pgfusepath{clip}%
\pgfsetbuttcap%
\pgfsetmiterjoin%
\pgfsetlinewidth{0.501875pt}%
\definecolor{currentstroke}{rgb}{0.501961,0.501961,0.501961}%
\pgfsetstrokecolor{currentstroke}%
\pgfsetdash{{1.850000pt}{0.800000pt}}{0.000000pt}%
\pgfpathmoveto{\pgfqpoint{0.625242in}{1.669376in}}%
\pgfpathlineto{\pgfqpoint{0.786196in}{1.669376in}}%
\pgfpathlineto{\pgfqpoint{0.786196in}{1.899527in}}%
\pgfpathlineto{\pgfqpoint{0.625242in}{1.899527in}}%
\pgfpathclose%
\pgfusepath{stroke}%
\end{pgfscope}%
\begin{pgfscope}%
\pgfpathrectangle{\pgfqpoint{0.330160in}{0.448876in}}{\pgfqpoint{3.219075in}{1.720905in}}%
\pgfusepath{clip}%
\pgfsetbuttcap%
\pgfsetmiterjoin%
\pgfsetlinewidth{0.501875pt}%
\definecolor{currentstroke}{rgb}{0.501961,0.501961,0.501961}%
\pgfsetstrokecolor{currentstroke}%
\pgfsetdash{{1.850000pt}{0.800000pt}}{0.000000pt}%
\pgfpathmoveto{\pgfqpoint{0.786196in}{1.006819in}}%
\pgfpathlineto{\pgfqpoint{0.947149in}{1.006819in}}%
\pgfpathlineto{\pgfqpoint{0.947149in}{1.034716in}}%
\pgfpathlineto{\pgfqpoint{0.786196in}{1.034716in}}%
\pgfpathclose%
\pgfusepath{stroke}%
\end{pgfscope}%
\begin{pgfscope}%
\pgfpathrectangle{\pgfqpoint{0.330160in}{0.448876in}}{\pgfqpoint{3.219075in}{1.720905in}}%
\pgfusepath{clip}%
\pgfsetbuttcap%
\pgfsetmiterjoin%
\pgfsetlinewidth{0.501875pt}%
\definecolor{currentstroke}{rgb}{0.501961,0.501961,0.501961}%
\pgfsetstrokecolor{currentstroke}%
\pgfsetdash{{1.850000pt}{0.800000pt}}{0.000000pt}%
\pgfpathmoveto{\pgfqpoint{1.698267in}{1.732145in}}%
\pgfpathlineto{\pgfqpoint{1.859221in}{1.732145in}}%
\pgfpathlineto{\pgfqpoint{1.859221in}{2.059936in}}%
\pgfpathlineto{\pgfqpoint{1.698267in}{2.059936in}}%
\pgfpathclose%
\pgfusepath{stroke}%
\end{pgfscope}%
\begin{pgfscope}%
\pgfpathrectangle{\pgfqpoint{0.330160in}{0.448876in}}{\pgfqpoint{3.219075in}{1.720905in}}%
\pgfusepath{clip}%
\pgfsetbuttcap%
\pgfsetmiterjoin%
\pgfsetlinewidth{0.501875pt}%
\definecolor{currentstroke}{rgb}{0.501961,0.501961,0.501961}%
\pgfsetstrokecolor{currentstroke}%
\pgfsetdash{{1.850000pt}{0.800000pt}}{0.000000pt}%
\pgfpathmoveto{\pgfqpoint{1.859221in}{0.951025in}}%
\pgfpathlineto{\pgfqpoint{2.020174in}{0.951025in}}%
\pgfpathlineto{\pgfqpoint{2.020174in}{0.978922in}}%
\pgfpathlineto{\pgfqpoint{1.859221in}{0.978922in}}%
\pgfpathclose%
\pgfusepath{stroke}%
\end{pgfscope}%
\begin{pgfscope}%
\pgfpathrectangle{\pgfqpoint{0.330160in}{0.448876in}}{\pgfqpoint{3.219075in}{1.720905in}}%
\pgfusepath{clip}%
\pgfsetbuttcap%
\pgfsetmiterjoin%
\pgfsetlinewidth{0.501875pt}%
\definecolor{currentstroke}{rgb}{0.501961,0.501961,0.501961}%
\pgfsetstrokecolor{currentstroke}%
\pgfsetdash{{1.850000pt}{0.800000pt}}{0.000000pt}%
\pgfpathmoveto{\pgfqpoint{2.771292in}{1.711222in}}%
\pgfpathlineto{\pgfqpoint{2.932245in}{1.711222in}}%
\pgfpathlineto{\pgfqpoint{2.932245in}{2.087833in}}%
\pgfpathlineto{\pgfqpoint{2.771292in}{2.087833in}}%
\pgfpathclose%
\pgfusepath{stroke}%
\end{pgfscope}%
\begin{pgfscope}%
\pgfpathrectangle{\pgfqpoint{0.330160in}{0.448876in}}{\pgfqpoint{3.219075in}{1.720905in}}%
\pgfusepath{clip}%
\pgfsetbuttcap%
\pgfsetmiterjoin%
\pgfsetlinewidth{0.501875pt}%
\definecolor{currentstroke}{rgb}{0.501961,0.501961,0.501961}%
\pgfsetstrokecolor{currentstroke}%
\pgfsetdash{{1.850000pt}{0.800000pt}}{0.000000pt}%
\pgfpathmoveto{\pgfqpoint{2.932245in}{0.992871in}}%
\pgfpathlineto{\pgfqpoint{3.093199in}{0.992871in}}%
\pgfpathlineto{\pgfqpoint{3.093199in}{1.041691in}}%
\pgfpathlineto{\pgfqpoint{2.932245in}{1.041691in}}%
\pgfpathclose%
\pgfusepath{stroke}%
\end{pgfscope}%
\begin{pgfscope}%
\pgfsetbuttcap%
\pgfsetroundjoin%
\definecolor{currentfill}{rgb}{0.000000,0.000000,0.000000}%
\pgfsetfillcolor{currentfill}%
\pgfsetlinewidth{0.501875pt}%
\definecolor{currentstroke}{rgb}{0.000000,0.000000,0.000000}%
\pgfsetstrokecolor{currentstroke}%
\pgfsetdash{}{0pt}%
\pgfsys@defobject{currentmarker}{\pgfqpoint{0.000000in}{-0.055556in}}{\pgfqpoint{0.000000in}{0.000000in}}{%
\pgfpathmoveto{\pgfqpoint{0.000000in}{0.000000in}}%
\pgfpathlineto{\pgfqpoint{0.000000in}{-0.055556in}}%
\pgfusepath{stroke,fill}%
}%
\begin{pgfscope}%
\pgfsys@transformshift{0.866672in}{0.448876in}%
\pgfsys@useobject{currentmarker}{}%
\end{pgfscope}%
\end{pgfscope}%
\begin{pgfscope}%
\pgftext[x=0.866672in,y=0.309988in,,top]{\sffamily\fontsize{6.664000}{7.996800}\selectfont 50 / ("/h)}%
\end{pgfscope}%
\begin{pgfscope}%
\pgfsetbuttcap%
\pgfsetroundjoin%
\definecolor{currentfill}{rgb}{0.000000,0.000000,0.000000}%
\pgfsetfillcolor{currentfill}%
\pgfsetlinewidth{0.501875pt}%
\definecolor{currentstroke}{rgb}{0.000000,0.000000,0.000000}%
\pgfsetstrokecolor{currentstroke}%
\pgfsetdash{}{0pt}%
\pgfsys@defobject{currentmarker}{\pgfqpoint{0.000000in}{-0.055556in}}{\pgfqpoint{0.000000in}{0.000000in}}{%
\pgfpathmoveto{\pgfqpoint{0.000000in}{0.000000in}}%
\pgfpathlineto{\pgfqpoint{0.000000in}{-0.055556in}}%
\pgfusepath{stroke,fill}%
}%
\begin{pgfscope}%
\pgfsys@transformshift{1.939697in}{0.448876in}%
\pgfsys@useobject{currentmarker}{}%
\end{pgfscope}%
\end{pgfscope}%
\begin{pgfscope}%
\pgftext[x=1.939697in,y=0.309988in,,top]{\sffamily\fontsize{6.664000}{7.996800}\selectfont 100 / ("/h)}%
\end{pgfscope}%
\begin{pgfscope}%
\pgfsetbuttcap%
\pgfsetroundjoin%
\definecolor{currentfill}{rgb}{0.000000,0.000000,0.000000}%
\pgfsetfillcolor{currentfill}%
\pgfsetlinewidth{0.501875pt}%
\definecolor{currentstroke}{rgb}{0.000000,0.000000,0.000000}%
\pgfsetstrokecolor{currentstroke}%
\pgfsetdash{}{0pt}%
\pgfsys@defobject{currentmarker}{\pgfqpoint{0.000000in}{-0.055556in}}{\pgfqpoint{0.000000in}{0.000000in}}{%
\pgfpathmoveto{\pgfqpoint{0.000000in}{0.000000in}}%
\pgfpathlineto{\pgfqpoint{0.000000in}{-0.055556in}}%
\pgfusepath{stroke,fill}%
}%
\begin{pgfscope}%
\pgfsys@transformshift{3.012722in}{0.448876in}%
\pgfsys@useobject{currentmarker}{}%
\end{pgfscope}%
\end{pgfscope}%
\begin{pgfscope}%
\pgftext[x=3.012722in,y=0.309988in,,top]{\sffamily\fontsize{6.664000}{7.996800}\selectfont 150 / ("/h)}%
\end{pgfscope}%
\begin{pgfscope}%
\pgftext[x=1.939697in,y=0.161877in,,top]{\sffamily\fontsize{9.600000}{11.520000}\selectfont Equivalent \texttt{CROSSID\_RADIUS}}%
\end{pgfscope}%
\begin{pgfscope}%
\pgfsetbuttcap%
\pgfsetroundjoin%
\definecolor{currentfill}{rgb}{0.000000,0.000000,0.000000}%
\pgfsetfillcolor{currentfill}%
\pgfsetlinewidth{0.501875pt}%
\definecolor{currentstroke}{rgb}{0.000000,0.000000,0.000000}%
\pgfsetstrokecolor{currentstroke}%
\pgfsetdash{}{0pt}%
\pgfsys@defobject{currentmarker}{\pgfqpoint{-0.055556in}{0.000000in}}{\pgfqpoint{0.000000in}{0.000000in}}{%
\pgfpathmoveto{\pgfqpoint{0.000000in}{0.000000in}}%
\pgfpathlineto{\pgfqpoint{-0.055556in}{0.000000in}}%
\pgfusepath{stroke,fill}%
}%
\begin{pgfscope}%
\pgfsys@transformshift{0.330160in}{0.448876in}%
\pgfsys@useobject{currentmarker}{}%
\end{pgfscope}%
\end{pgfscope}%
\begin{pgfscope}%
\pgftext[x=0.135908in,y=0.416760in,left,base]{\sffamily\fontsize{6.664000}{7.996800}\selectfont \(\displaystyle 0\)}%
\end{pgfscope}%
\begin{pgfscope}%
\pgfsetbuttcap%
\pgfsetroundjoin%
\definecolor{currentfill}{rgb}{0.000000,0.000000,0.000000}%
\pgfsetfillcolor{currentfill}%
\pgfsetlinewidth{0.501875pt}%
\definecolor{currentstroke}{rgb}{0.000000,0.000000,0.000000}%
\pgfsetstrokecolor{currentstroke}%
\pgfsetdash{}{0pt}%
\pgfsys@defobject{currentmarker}{\pgfqpoint{-0.055556in}{0.000000in}}{\pgfqpoint{0.000000in}{0.000000in}}{%
\pgfpathmoveto{\pgfqpoint{0.000000in}{0.000000in}}%
\pgfpathlineto{\pgfqpoint{-0.055556in}{0.000000in}}%
\pgfusepath{stroke,fill}%
}%
\begin{pgfscope}%
\pgfsys@transformshift{0.330160in}{0.797591in}%
\pgfsys@useobject{currentmarker}{}%
\end{pgfscope}%
\end{pgfscope}%
\begin{pgfscope}%
\pgftext[x=0.080545in,y=0.765474in,left,base]{\sffamily\fontsize{6.664000}{7.996800}\selectfont \(\displaystyle 50\)}%
\end{pgfscope}%
\begin{pgfscope}%
\pgfsetbuttcap%
\pgfsetroundjoin%
\definecolor{currentfill}{rgb}{0.000000,0.000000,0.000000}%
\pgfsetfillcolor{currentfill}%
\pgfsetlinewidth{0.501875pt}%
\definecolor{currentstroke}{rgb}{0.000000,0.000000,0.000000}%
\pgfsetstrokecolor{currentstroke}%
\pgfsetdash{}{0pt}%
\pgfsys@defobject{currentmarker}{\pgfqpoint{-0.055556in}{0.000000in}}{\pgfqpoint{0.000000in}{0.000000in}}{%
\pgfpathmoveto{\pgfqpoint{0.000000in}{0.000000in}}%
\pgfpathlineto{\pgfqpoint{-0.055556in}{0.000000in}}%
\pgfusepath{stroke,fill}%
}%
\begin{pgfscope}%
\pgfsys@transformshift{0.330160in}{1.146305in}%
\pgfsys@useobject{currentmarker}{}%
\end{pgfscope}%
\end{pgfscope}%
\begin{pgfscope}%
\pgftext[x=0.025182in,y=1.114188in,left,base]{\sffamily\fontsize{6.664000}{7.996800}\selectfont \(\displaystyle 100\)}%
\end{pgfscope}%
\begin{pgfscope}%
\pgfsetbuttcap%
\pgfsetroundjoin%
\definecolor{currentfill}{rgb}{0.000000,0.000000,0.000000}%
\pgfsetfillcolor{currentfill}%
\pgfsetlinewidth{0.501875pt}%
\definecolor{currentstroke}{rgb}{0.000000,0.000000,0.000000}%
\pgfsetstrokecolor{currentstroke}%
\pgfsetdash{}{0pt}%
\pgfsys@defobject{currentmarker}{\pgfqpoint{-0.055556in}{0.000000in}}{\pgfqpoint{0.000000in}{0.000000in}}{%
\pgfpathmoveto{\pgfqpoint{0.000000in}{0.000000in}}%
\pgfpathlineto{\pgfqpoint{-0.055556in}{0.000000in}}%
\pgfusepath{stroke,fill}%
}%
\begin{pgfscope}%
\pgfsys@transformshift{0.330160in}{1.495019in}%
\pgfsys@useobject{currentmarker}{}%
\end{pgfscope}%
\end{pgfscope}%
\begin{pgfscope}%
\pgftext[x=0.025182in,y=1.462902in,left,base]{\sffamily\fontsize{6.664000}{7.996800}\selectfont \(\displaystyle 150\)}%
\end{pgfscope}%
\begin{pgfscope}%
\pgfsetbuttcap%
\pgfsetroundjoin%
\definecolor{currentfill}{rgb}{0.000000,0.000000,0.000000}%
\pgfsetfillcolor{currentfill}%
\pgfsetlinewidth{0.501875pt}%
\definecolor{currentstroke}{rgb}{0.000000,0.000000,0.000000}%
\pgfsetstrokecolor{currentstroke}%
\pgfsetdash{}{0pt}%
\pgfsys@defobject{currentmarker}{\pgfqpoint{-0.055556in}{0.000000in}}{\pgfqpoint{0.000000in}{0.000000in}}{%
\pgfpathmoveto{\pgfqpoint{0.000000in}{0.000000in}}%
\pgfpathlineto{\pgfqpoint{-0.055556in}{0.000000in}}%
\pgfusepath{stroke,fill}%
}%
\begin{pgfscope}%
\pgfsys@transformshift{0.330160in}{1.843733in}%
\pgfsys@useobject{currentmarker}{}%
\end{pgfscope}%
\end{pgfscope}%
\begin{pgfscope}%
\pgftext[x=0.025182in,y=1.811616in,left,base]{\sffamily\fontsize{6.664000}{7.996800}\selectfont \(\displaystyle 200\)}%
\end{pgfscope}%
\begin{pgfscope}%
\pgfpathrectangle{\pgfqpoint{0.330160in}{0.448876in}}{\pgfqpoint{3.219075in}{1.720905in}}%
\pgfusepath{clip}%
\pgfsetbuttcap%
\pgfsetmiterjoin%
\pgfsetlinewidth{0.501875pt}%
\definecolor{currentstroke}{rgb}{0.000000,0.000000,0.000000}%
\pgfsetstrokecolor{currentstroke}%
\pgfsetdash{}{0pt}%
\pgfpathmoveto{\pgfqpoint{0.625242in}{0.448876in}}%
\pgfpathlineto{\pgfqpoint{0.786196in}{0.448876in}}%
\pgfpathlineto{\pgfqpoint{0.786196in}{1.669376in}}%
\pgfpathlineto{\pgfqpoint{0.625242in}{1.669376in}}%
\pgfpathclose%
\pgfusepath{stroke}%
\end{pgfscope}%
\begin{pgfscope}%
\pgfpathrectangle{\pgfqpoint{0.330160in}{0.448876in}}{\pgfqpoint{3.219075in}{1.720905in}}%
\pgfusepath{clip}%
\pgfsetbuttcap%
\pgfsetmiterjoin%
\pgfsetlinewidth{0.501875pt}%
\definecolor{currentstroke}{rgb}{0.000000,0.000000,0.000000}%
\pgfsetstrokecolor{currentstroke}%
\pgfsetdash{}{0pt}%
\pgfpathmoveto{\pgfqpoint{0.786196in}{0.448876in}}%
\pgfpathlineto{\pgfqpoint{0.947149in}{0.448876in}}%
\pgfpathlineto{\pgfqpoint{0.947149in}{1.006819in}}%
\pgfpathlineto{\pgfqpoint{0.786196in}{1.006819in}}%
\pgfpathclose%
\pgfusepath{stroke}%
\end{pgfscope}%
\begin{pgfscope}%
\pgfpathrectangle{\pgfqpoint{0.330160in}{0.448876in}}{\pgfqpoint{3.219075in}{1.720905in}}%
\pgfusepath{clip}%
\pgfsetbuttcap%
\pgfsetmiterjoin%
\pgfsetlinewidth{0.501875pt}%
\definecolor{currentstroke}{rgb}{0.000000,0.000000,0.000000}%
\pgfsetstrokecolor{currentstroke}%
\pgfsetdash{}{0pt}%
\pgfpathmoveto{\pgfqpoint{0.947149in}{0.448876in}}%
\pgfpathlineto{\pgfqpoint{1.108103in}{0.448876in}}%
\pgfpathlineto{\pgfqpoint{1.108103in}{0.783642in}}%
\pgfpathlineto{\pgfqpoint{0.947149in}{0.783642in}}%
\pgfpathclose%
\pgfusepath{stroke}%
\end{pgfscope}%
\begin{pgfscope}%
\pgfpathrectangle{\pgfqpoint{0.330160in}{0.448876in}}{\pgfqpoint{3.219075in}{1.720905in}}%
\pgfusepath{clip}%
\pgfsetbuttcap%
\pgfsetmiterjoin%
\pgfsetlinewidth{0.501875pt}%
\definecolor{currentstroke}{rgb}{0.000000,0.000000,0.000000}%
\pgfsetstrokecolor{currentstroke}%
\pgfsetdash{}{0pt}%
\pgfpathmoveto{\pgfqpoint{1.698267in}{0.448876in}}%
\pgfpathlineto{\pgfqpoint{1.859221in}{0.448876in}}%
\pgfpathlineto{\pgfqpoint{1.859221in}{1.732145in}}%
\pgfpathlineto{\pgfqpoint{1.698267in}{1.732145in}}%
\pgfpathclose%
\pgfusepath{stroke}%
\end{pgfscope}%
\begin{pgfscope}%
\pgfpathrectangle{\pgfqpoint{0.330160in}{0.448876in}}{\pgfqpoint{3.219075in}{1.720905in}}%
\pgfusepath{clip}%
\pgfsetbuttcap%
\pgfsetmiterjoin%
\pgfsetlinewidth{0.501875pt}%
\definecolor{currentstroke}{rgb}{0.000000,0.000000,0.000000}%
\pgfsetstrokecolor{currentstroke}%
\pgfsetdash{}{0pt}%
\pgfpathmoveto{\pgfqpoint{1.859221in}{0.448876in}}%
\pgfpathlineto{\pgfqpoint{2.020174in}{0.448876in}}%
\pgfpathlineto{\pgfqpoint{2.020174in}{0.951025in}}%
\pgfpathlineto{\pgfqpoint{1.859221in}{0.951025in}}%
\pgfpathclose%
\pgfusepath{stroke}%
\end{pgfscope}%
\begin{pgfscope}%
\pgfpathrectangle{\pgfqpoint{0.330160in}{0.448876in}}{\pgfqpoint{3.219075in}{1.720905in}}%
\pgfusepath{clip}%
\pgfsetbuttcap%
\pgfsetmiterjoin%
\pgfsetlinewidth{0.501875pt}%
\definecolor{currentstroke}{rgb}{0.000000,0.000000,0.000000}%
\pgfsetstrokecolor{currentstroke}%
\pgfsetdash{}{0pt}%
\pgfpathmoveto{\pgfqpoint{2.020174in}{0.448876in}}%
\pgfpathlineto{\pgfqpoint{2.181128in}{0.448876in}}%
\pgfpathlineto{\pgfqpoint{2.181128in}{0.783642in}}%
\pgfpathlineto{\pgfqpoint{2.020174in}{0.783642in}}%
\pgfpathclose%
\pgfusepath{stroke}%
\end{pgfscope}%
\begin{pgfscope}%
\pgfpathrectangle{\pgfqpoint{0.330160in}{0.448876in}}{\pgfqpoint{3.219075in}{1.720905in}}%
\pgfusepath{clip}%
\pgfsetbuttcap%
\pgfsetmiterjoin%
\pgfsetlinewidth{0.501875pt}%
\definecolor{currentstroke}{rgb}{0.000000,0.000000,0.000000}%
\pgfsetstrokecolor{currentstroke}%
\pgfsetdash{}{0pt}%
\pgfpathmoveto{\pgfqpoint{2.771292in}{0.448876in}}%
\pgfpathlineto{\pgfqpoint{2.932245in}{0.448876in}}%
\pgfpathlineto{\pgfqpoint{2.932245in}{1.711222in}}%
\pgfpathlineto{\pgfqpoint{2.771292in}{1.711222in}}%
\pgfpathclose%
\pgfusepath{stroke}%
\end{pgfscope}%
\begin{pgfscope}%
\pgfpathrectangle{\pgfqpoint{0.330160in}{0.448876in}}{\pgfqpoint{3.219075in}{1.720905in}}%
\pgfusepath{clip}%
\pgfsetbuttcap%
\pgfsetmiterjoin%
\pgfsetlinewidth{0.501875pt}%
\definecolor{currentstroke}{rgb}{0.000000,0.000000,0.000000}%
\pgfsetstrokecolor{currentstroke}%
\pgfsetdash{}{0pt}%
\pgfpathmoveto{\pgfqpoint{2.932245in}{0.448876in}}%
\pgfpathlineto{\pgfqpoint{3.093199in}{0.448876in}}%
\pgfpathlineto{\pgfqpoint{3.093199in}{0.992871in}}%
\pgfpathlineto{\pgfqpoint{2.932245in}{0.992871in}}%
\pgfpathclose%
\pgfusepath{stroke}%
\end{pgfscope}%
\begin{pgfscope}%
\pgfpathrectangle{\pgfqpoint{0.330160in}{0.448876in}}{\pgfqpoint{3.219075in}{1.720905in}}%
\pgfusepath{clip}%
\pgfsetbuttcap%
\pgfsetmiterjoin%
\pgfsetlinewidth{0.501875pt}%
\definecolor{currentstroke}{rgb}{0.000000,0.000000,0.000000}%
\pgfsetstrokecolor{currentstroke}%
\pgfsetdash{}{0pt}%
\pgfpathmoveto{\pgfqpoint{3.093199in}{0.448876in}}%
\pgfpathlineto{\pgfqpoint{3.254153in}{0.448876in}}%
\pgfpathlineto{\pgfqpoint{3.254153in}{0.776668in}}%
\pgfpathlineto{\pgfqpoint{3.093199in}{0.776668in}}%
\pgfpathclose%
\pgfusepath{stroke}%
\end{pgfscope}%
\begin{pgfscope}%
\pgfsetrectcap%
\pgfsetmiterjoin%
\pgfsetlinewidth{0.501875pt}%
\definecolor{currentstroke}{rgb}{0.000000,0.000000,0.000000}%
\pgfsetstrokecolor{currentstroke}%
\pgfsetdash{}{0pt}%
\pgfpathmoveto{\pgfqpoint{0.330160in}{0.448876in}}%
\pgfpathlineto{\pgfqpoint{0.330160in}{2.169781in}}%
\pgfusepath{stroke}%
\end{pgfscope}%
\begin{pgfscope}%
\pgfsetrectcap%
\pgfsetmiterjoin%
\pgfsetlinewidth{0.501875pt}%
\definecolor{currentstroke}{rgb}{0.000000,0.000000,0.000000}%
\pgfsetstrokecolor{currentstroke}%
\pgfsetdash{}{0pt}%
\pgfpathmoveto{\pgfqpoint{0.330160in}{0.448876in}}%
\pgfpathlineto{\pgfqpoint{3.549235in}{0.448876in}}%
\pgfusepath{stroke}%
\end{pgfscope}%
\begin{pgfscope}%
\pgftext[x=0.813021in,y=1.669376in,left,]{\sffamily\fontsize{6.000000}{7.200000}\selectfont 175}%
\end{pgfscope}%
\begin{pgfscope}%
\definecolor{textcolor}{rgb}{0.501961,0.501961,0.501961}%
\pgfsetstrokecolor{textcolor}%
\pgfsetfillcolor{textcolor}%
\pgftext[x=0.813021in,y=1.899527in,left,]{\color{textcolor}\sffamily\fontsize{6.000000}{7.200000}\selectfont 33}%
\end{pgfscope}%
\begin{pgfscope}%
\pgftext[x=0.973975in,y=1.006819in,left,top]{\sffamily\fontsize{6.000000}{7.200000}\selectfont 80}%
\end{pgfscope}%
\begin{pgfscope}%
\definecolor{textcolor}{rgb}{0.501961,0.501961,0.501961}%
\pgfsetstrokecolor{textcolor}%
\pgfsetfillcolor{textcolor}%
\pgftext[x=0.973975in,y=1.034716in,left,bottom]{\color{textcolor}\sffamily\fontsize{6.000000}{7.200000}\selectfont 4}%
\end{pgfscope}%
\begin{pgfscope}%
\pgftext[x=1.134929in,y=0.783642in,left,]{\sffamily\fontsize{6.000000}{7.200000}\selectfont 48}%
\end{pgfscope}%
\begin{pgfscope}%
\pgftext[x=1.886046in,y=1.732145in,left,]{\sffamily\fontsize{6.000000}{7.200000}\selectfont 184}%
\end{pgfscope}%
\begin{pgfscope}%
\definecolor{textcolor}{rgb}{0.501961,0.501961,0.501961}%
\pgfsetstrokecolor{textcolor}%
\pgfsetfillcolor{textcolor}%
\pgftext[x=1.886046in,y=2.059936in,left,]{\color{textcolor}\sffamily\fontsize{6.000000}{7.200000}\selectfont 47}%
\end{pgfscope}%
\begin{pgfscope}%
\pgftext[x=2.047000in,y=0.951025in,left,top]{\sffamily\fontsize{6.000000}{7.200000}\selectfont 72}%
\end{pgfscope}%
\begin{pgfscope}%
\definecolor{textcolor}{rgb}{0.501961,0.501961,0.501961}%
\pgfsetstrokecolor{textcolor}%
\pgfsetfillcolor{textcolor}%
\pgftext[x=2.047000in,y=0.978922in,left,bottom]{\color{textcolor}\sffamily\fontsize{6.000000}{7.200000}\selectfont 4}%
\end{pgfscope}%
\begin{pgfscope}%
\pgftext[x=2.207954in,y=0.783642in,left,]{\sffamily\fontsize{6.000000}{7.200000}\selectfont 48}%
\end{pgfscope}%
\begin{pgfscope}%
\pgftext[x=2.959071in,y=1.711222in,left,]{\sffamily\fontsize{6.000000}{7.200000}\selectfont 181}%
\end{pgfscope}%
\begin{pgfscope}%
\definecolor{textcolor}{rgb}{0.501961,0.501961,0.501961}%
\pgfsetstrokecolor{textcolor}%
\pgfsetfillcolor{textcolor}%
\pgftext[x=2.959071in,y=2.087833in,left,]{\color{textcolor}\sffamily\fontsize{6.000000}{7.200000}\selectfont 54}%
\end{pgfscope}%
\begin{pgfscope}%
\pgftext[x=3.120025in,y=0.992871in,left,top]{\sffamily\fontsize{6.000000}{7.200000}\selectfont 78}%
\end{pgfscope}%
\begin{pgfscope}%
\definecolor{textcolor}{rgb}{0.501961,0.501961,0.501961}%
\pgfsetstrokecolor{textcolor}%
\pgfsetfillcolor{textcolor}%
\pgftext[x=3.120025in,y=1.041691in,left,bottom]{\color{textcolor}\sffamily\fontsize{6.000000}{7.200000}\selectfont 7}%
\end{pgfscope}%
\begin{pgfscope}%
\pgftext[x=3.280979in,y=0.776668in,left,]{\sffamily\fontsize{6.000000}{7.200000}\selectfont 47}%
\end{pgfscope}%
\begin{pgfscope}%
\pgftext[x=0.726552in,y=0.483748in,left,base,rotate=90.000000]{\sffamily\fontsize{6.000000}{7.200000}\selectfont \textit{Default}}%
\end{pgfscope}%
\begin{pgfscope}%
\pgftext[x=0.887506in,y=0.483748in,left,base,rotate=90.000000]{\sffamily\fontsize{6.000000}{7.200000}\selectfont \textit{Trail}}%
\end{pgfscope}%
\begin{pgfscope}%
\pgftext[x=1.048460in,y=0.483748in,left,base,rotate=90.000000]{\sffamily\fontsize{6.000000}{7.200000}\selectfont \textit{Strict}}%
\end{pgfscope}%
\end{pgfpicture}%
\makeatother%
\endgroup%